\documentclass[traditabstract]{aa}
\usepackage{txfonts} 
\usepackage{graphicx}

\usepackage{txfonts}
\def\la{\raise.5ex\hbox{$<$}\kern-.8em\lower 1mm\hbox{$\sim$}}
\def\ma{\raise.5ex\hbox{$>$}\kern-.8em\lower 1mm\hbox{$\sim$}}

\def\kms{$\rm km\, s^{-1}$}
\def\cm3{$\rm cm^{-3}$}

\def\Vs{$V_{\rm s}$~}
\def\n0{$n_{\rm 0}$}
\def\B0{$B_{0}$}
\def\ne{$n_{\rm e}$~}

\def\erg{$\rm erg\, cm^{-2}\, s^{-1}$}

\def\L12{L$_{12\mu m}$~}
\def\F12{F$_{12\mu m}$~}

\def\Hb{H$\beta$~}

\begin{document}
   \title{Distribution of the heavy elements throughout the\\
  extended narrow line region of the Seyfert galaxy NGC 7212}

   \author{M. Contini \inst{1,2},
   V. Cracco\inst{1},
   S. Ciroi\inst{1},
\and 
G. La Mura\inst{1}
}

   \institute{Dipartimento di Fisica e Astronomia, University of Padova, 
              Vicolo dell'Osservatorio 2, I-35122 Padova, Italy 
                \and  
             School of Physics and Astronomy, Tel-Aviv University,
              Tel-Aviv 69978, Israel\\ 
             } 

   \date{Received }

   \abstract
{The latest observations of line and continuum spectra emitted from the extended narrow line region (ENLR) of the
Seyfert 2 galaxy NGC 7212 are analysed using models accounting for photoionization from the active
nucleus and shocks.  The results show that
relatively high (500--800 \kms) shock velocities appear on the edge of the cone and outside of it.
The model-inferred AGN flux, which is lower than $10^{-11}$ photons cm$^{-2}$ s$^{-1}$ eV$^{-1}$ at the Lyman
limit, is more typical of low-luminosity AGN, and less so for Seyfert 2 galaxies. The preshock densities are
characteristic of the ENLR and range between 80--150 cm$^{-3}$.
Nitrogen and sulphur are found depleted by a factor lower than 2, particularly at the eastern edge.
Oxygen is depleted at several locations. The Fe/H ratio is approximately solar, whereas the Ne/H
relative abundance is unusually high, 1.5--2 times the solar value.
Modelling the continuum spectral energy distribution (SED), we have found radio synchrotron radiation
generated by the Fermi mechanism at the shock front, whereas the X-rays are produced
by the bremsstrahlung from a relatively high temperature plasma.}


\titlerunning{Depletion of heavy elements in NGC 7212}
\authorrunning{Contini et al.}

\maketitle
%
%________________________________________________________________

\section{Introduction}\label{intro}

Morphological and spectroscopic studies of Seyfert galaxies have shed light on the nature of active galactic nuclei (AGN). Specifcally, the spectra of type 2 Seyferts are generally dominated by strong forbidden and permitted lines revealing the physical properties of the extended narrow line region (ENLR). The spectra indicate that radiation from the AGN dominates the radiation field; however, line ratios corresponding to low ionization levels and neutral lines could be explained by underlying shock wave hydrodynamics. The full width at half maximum (FWHM) of the lines reveal velocities up to 1000 \kms. Modelling of line and continuum spectra observed in many single regions throughout the ENLR provided a new dimension to our knowledge. For instance, the complex nature of \object{NGC 7130}, in terms of the intermingled activity of starbursts, shocks, and of an active nucleus and their mutual location, could be traced by Contini et al. (\cite{Con02a}).
In  such objects, e.g. \object{NGC 4388} (Ciroi et al. \cite{C03}) or \object{Mrk 298} (Radovich et al. \cite{Rad05}), the relative importance of the 
radiation flux from the AGN, of shocks, and of radiation from starbursts could be determined in the different regions of the galaxy. This method, which is useful for isolated galaxies, becomes fundamental for the analysis 
of Seyfert nuclei in merging systems originating from collisions, even in case of multiple (generally double) nuclei that are not yet clearly identified as such by the observations. For instance, the biconical structure of the high excitation region which emerges from the toroidal obscuring material, can be fragmented by collision, 
and patches of highly ionized material can appear at the edges of the ENLR, as e.g. in \object{NGC 3393} (Cooke et al. \cite{coo00}). 
The interaction of the ENLR galactic matter with fast shocks, star formation, dust grain destruction etc. are amply analysed and summarised by Ramos Almeida et al. (\cite{R09}) for \object{NGC 7212} and other Seyfert 2 galaxies.

\object{NGC 7212} (z=0.0266) belongs to a compact group of interacting galaxies, with two galaxies in a clear ongoing merger (Wasilewski 1981). Spectropolarimetry studies
show the presence of a hidden broad line region (BLR) (Tran \& Kay 1992), clumpy nuclear morphology,
and irregular diffuse emission. Mu{\~n}oz Mar{\'{\i}}n et al. (\cite{M07}) also claim that
these galaxies manifest some cases in which in addition to the extended emission, the UV light stems from the knots or clumps likely produced by star clusters. They identify the bright clumps to the south as star-forming regions with a certain contribution from the AGN. Exploring whether photoionization or shocks dominate in NGC 7212, Bianchi et al. (\cite{bi06}) found a striking resemblance of the [\ion{O}{III}] structure with soft X-ray emission, a clear indication that shocks are at work.

In an accompanying paper, Cracco et al. (\cite{Cr11}) presented new observations of NGC 7212 and suggested 
that its ENLR gas has an external origin, likely due to gravitational interaction effects in act in this triple system. The optical spectra detailed by Cracco et al. are rich in
number of lines: e.g. oxygen from three ionization levels ([\ion{O}{III}]5007, [\ion{O}{II}]3727 and [\ion{O}{I}]6300), \ion{He}{II} 4686 and \ion{He}{I} 5876,
[\ion{Ne}{III}]3869, [\ion{N}{II}]6548,6583, the two [\ion{S}{II}] lines 6716 and 6731, lines from higher ionization levels, e.g. [\ion{Ar}{IV}]4713 and [\ion{Fe}{VII}]6087, in addition to 
H$\beta$ and H$\alpha$.  The line ratios constrain the physical conditions of the emitting gas and the relative abundances.
These authors pointed out for the first time the presence of an extended ionization cone in NGC 7212, with 
high values of [\ion{O}{III}]/H$\beta$, up to 3.6 kpc. In fact, the Veilleux \& Osterbrock (\cite{VO87}) diagnostic diagrams indicate that the main source of ionization, at least in the entire field of view of the integral field spectra, is the AGN. 
However, high values of [\ion{N}{II}]/H$\alpha$ and [\ion{S}{II}]/H$\alpha$ are observed towards the region 
of gravitational interaction between the  interacting galaxies, suggesting a possible combination of ionization 
by the active nucleus and shocks. On the other hand, no contribution from stellar ionization is required to 
explain the observed emission line ratios. An archival broad-band Hubble Space Telescope (HST) image of NGC 7212 obtained with the F606W filter revelas a structure composed of clouds or filaments (see fig.\,19 in Cracco et al. \cite{Cr11}). The evidence of dust located in the ENLR also supports the idea that the gas in the cone has a filamentary structure.  
In addition, comparing the more significant line ratios 
(e.g. [\ion{O}{III}]/H$\beta$, [\ion{O}{II}]/H$\beta$, [\ion{N}{II}]/H$\alpha$, etc) with diagnostic diagrams, Cracco et al. found depletion of heavy elements such as N and O relative to solar. 

The emission line profiles of [\ion{O}{III}] inside the ionization 
cone display blue wings in the northern side of the cone, and red wings in the southern side, while in the nucleus 
the profiles are symmetric. This suggests the presence of gas in radial motions, which is confirmed also by 
the analysis of high resolution echelle spectra characterized by multiple kinematical components at different velocities. If radial motions are in act in the ENLR of NGC 7212, shock ionization is expected to play a non negligible role.

Moreover, it is well known that the higher the flux from the photoionizing source, the stronger the [\ion{O}{III}] lines are, while emission lines such as [\ion{O}{II}], [\ion{N}{II}], etc., namely low ionization level lines, are significant when the gas is affected by strong shock. 
Consequently, we decided to reproduce the observational data presented by Cracco et al. (\cite{Cr11})
by means of models accounting for the coupled effect of photoionization and shock, improving their interpretation, and adding information concerning the physical and chemical conditions throughout the galaxy.

The structure of the paper is as follows: observations of the optical spectra are described in Sect.\,\ref{obs}. 
In Sect.\,\ref{model} the models and the modelling method are explained.
Modelling results of the line and continuum spectra are presented in Sect.\,\ref{model_result} and in Sect.\,\ref{sed}, respectively. Discussion and concluding remarks follow in Sect.\, \ref{disc} and Sect.\,\ref{result}.

\begin{figure}
\centering
\includegraphics[width=8.5cm]{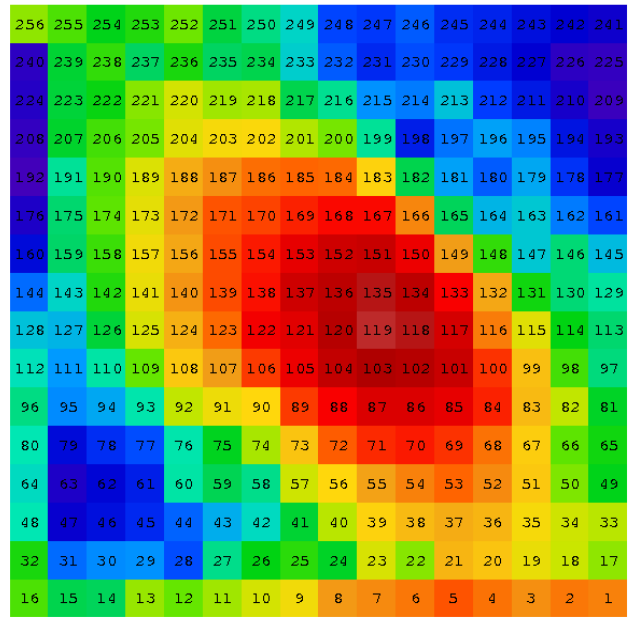}
\caption{2D map of the stellar continuum at 5500 \AA. The numbers identify the spectra in Fig.\,\ref{fig4b}. The image size is $16\times16$ square arcsec. North is up and East is to the left.}
\label{fig1}
\end{figure}

\section{Observations}\label{obs}

Cracco et al. (\cite{Cr11}) present integral field spectra taken with the Multi Pupil Fiber Spectrograph (MPFS; Afanasiev et al. \cite{afanasiev}) at the Special Astrophysical Observatory (SAO RAS).  256 simultaneous spectra arranged in an array of $16\times16$ spatial elements were obtained in the range 3800--7300 \AA, with a resolution of 6 \AA. The field of view was 16$\times$16 square arcsec, seeing was about 1.5\arcsec.
At the distance of the galaxy the spatial scale is 0.513 kpc arcsec$^{-1}$ ($\rm H_0=75$ km s$^{-1}$ Mpc$^{-1}$), yielding field of view of $8.2\times8.2$ kpc$^2$.

The data were reduced with {\sc p3d} (Sandin et al. 2010) and corrected for reddening, atmospheric absorption bands, and atmospheric refraction with {\sc iraf}\footnote{{\sc iraf} is distributed by the National Optical Astronomy Observatories, which are operated by the Association of Universities for Research in Astronomy, Inc., under cooperative agreement with the National Science Foundation.}. 
The stellar component was subtracted with {\sc starlight} (Cid Fernandes et al. 2005, 2007) and finally the spectra were measured with {\sc pan} (Peak Analysis). Spectra are numbered from 1 to 256 and their positions in the field of view are reported in Fig.\,\ref{fig1}.
For further details about reduction and calibration procedures, see Cracco et al. (\cite{Cr11}).

\begin{figure*}
\centering
\includegraphics[width=0.26\textwidth]{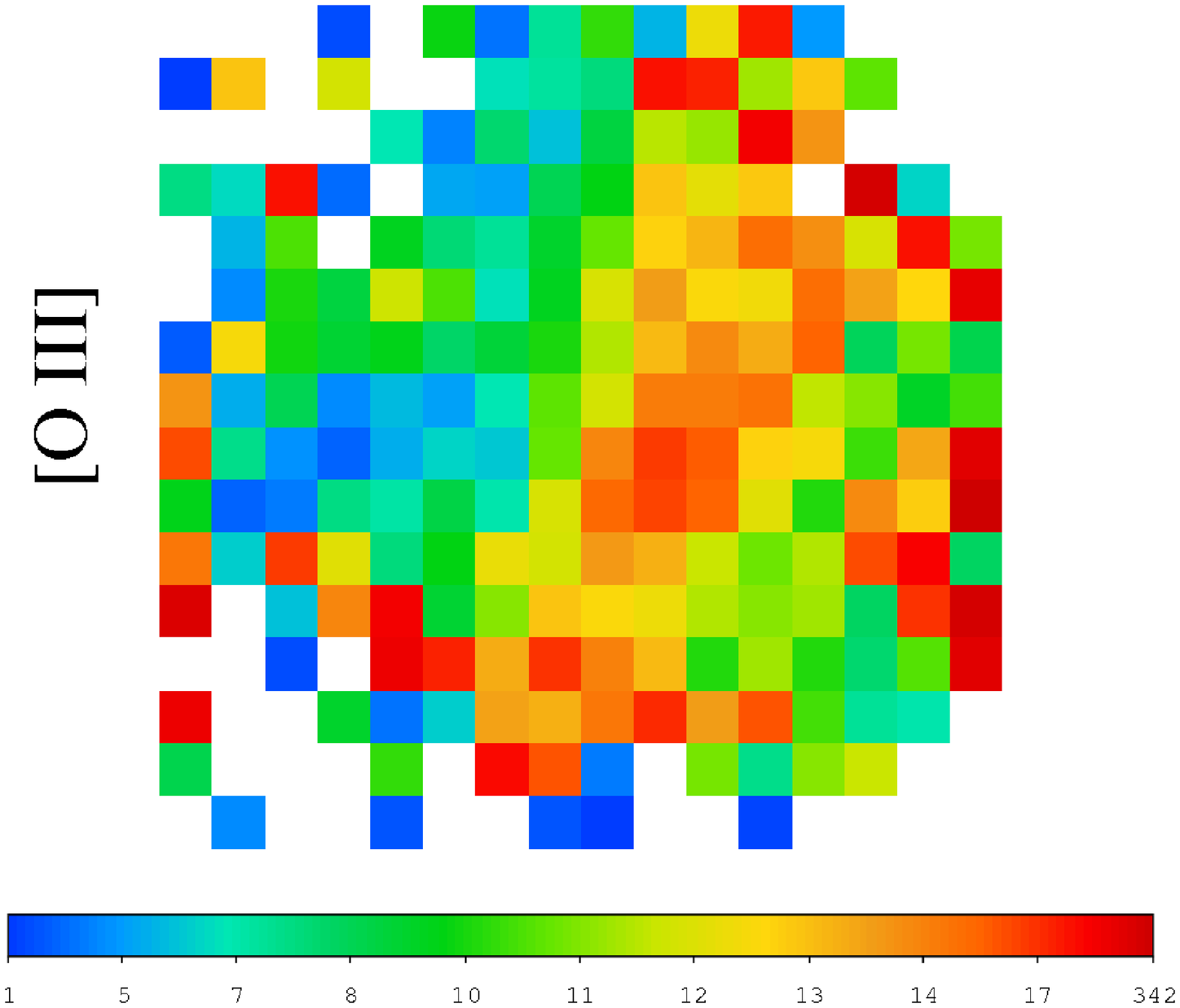}
\includegraphics[width=0.26\textwidth]{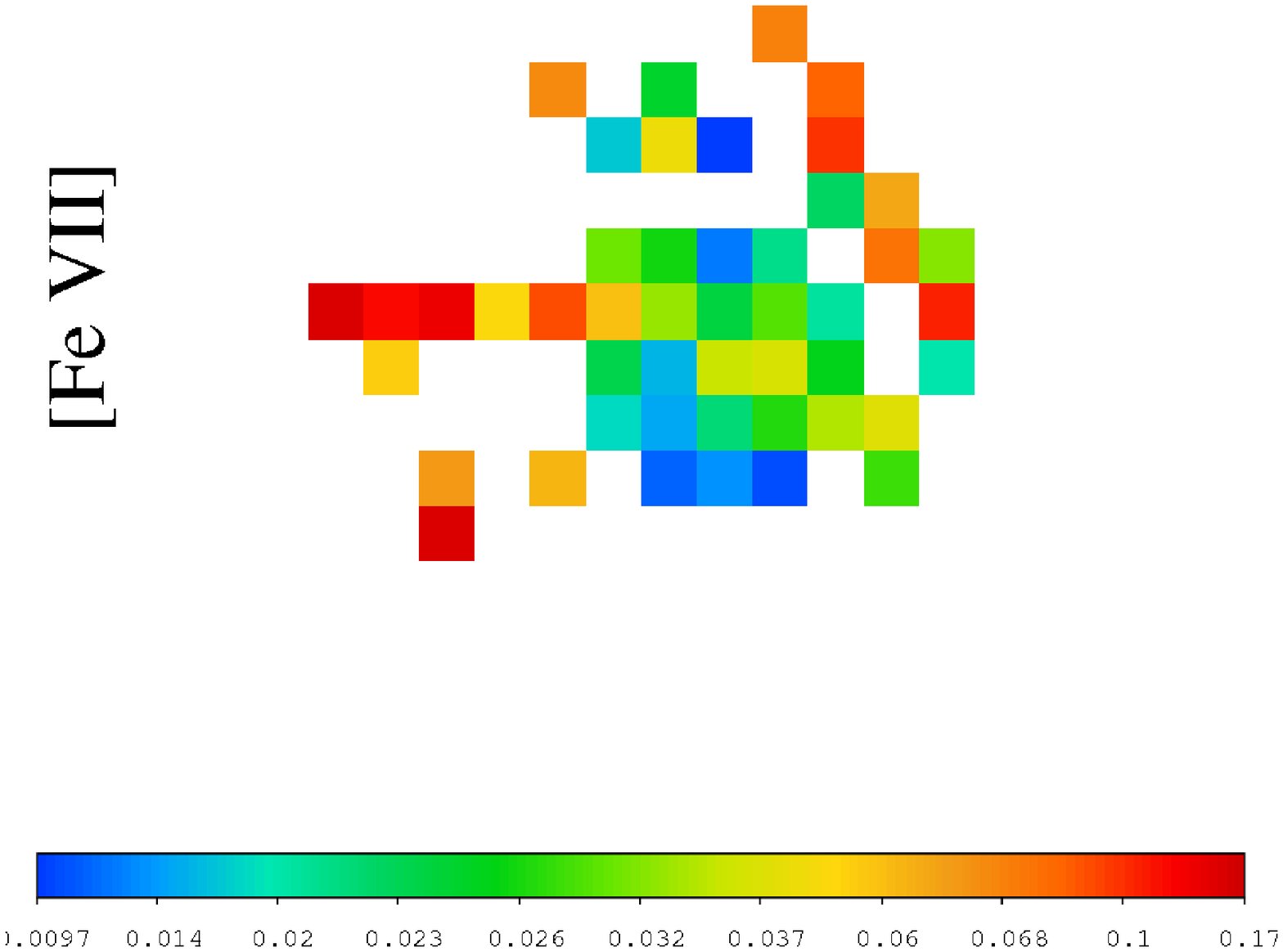}
\includegraphics[width=0.26\textwidth]{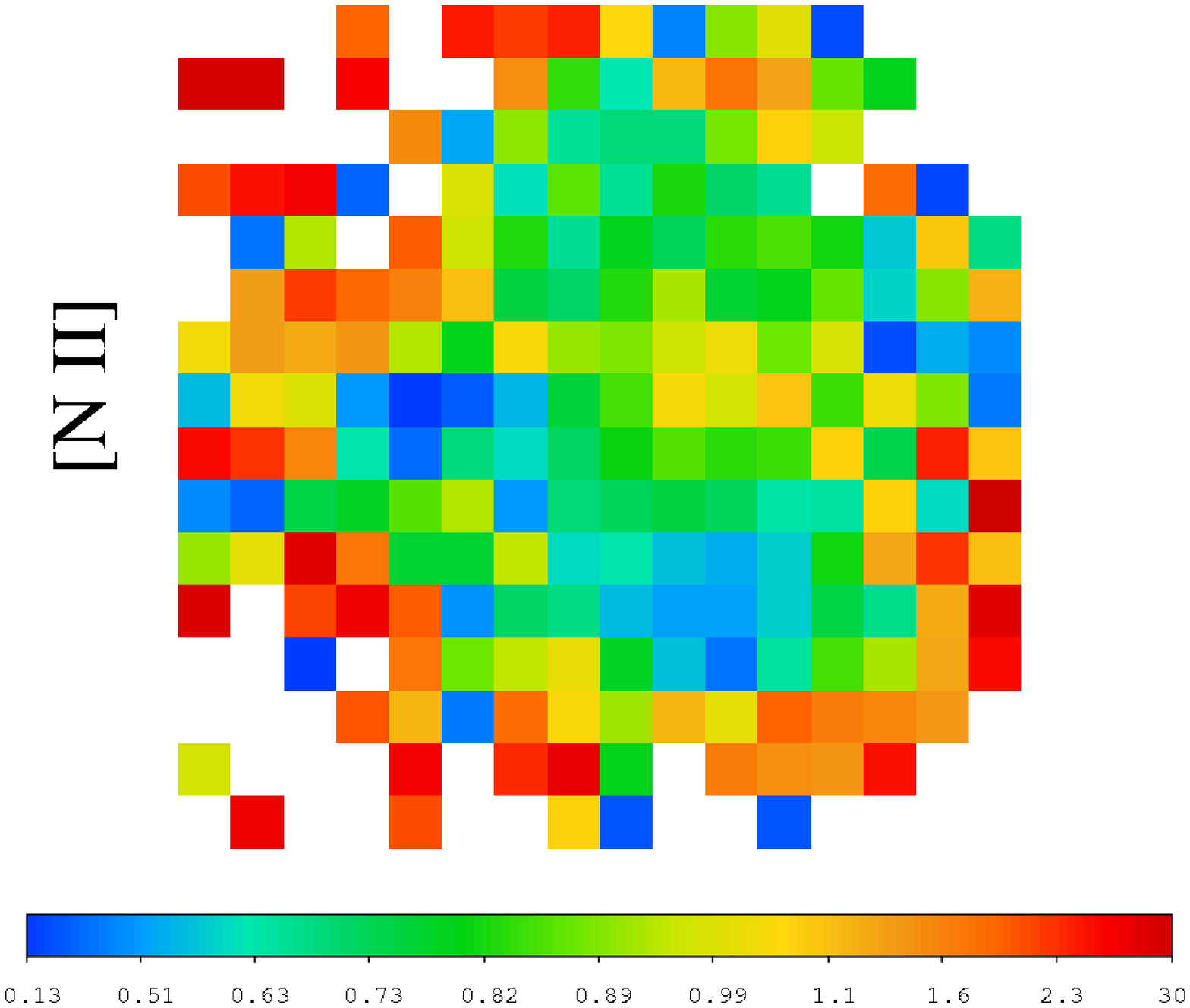}\\
\includegraphics[width=0.26\textwidth]{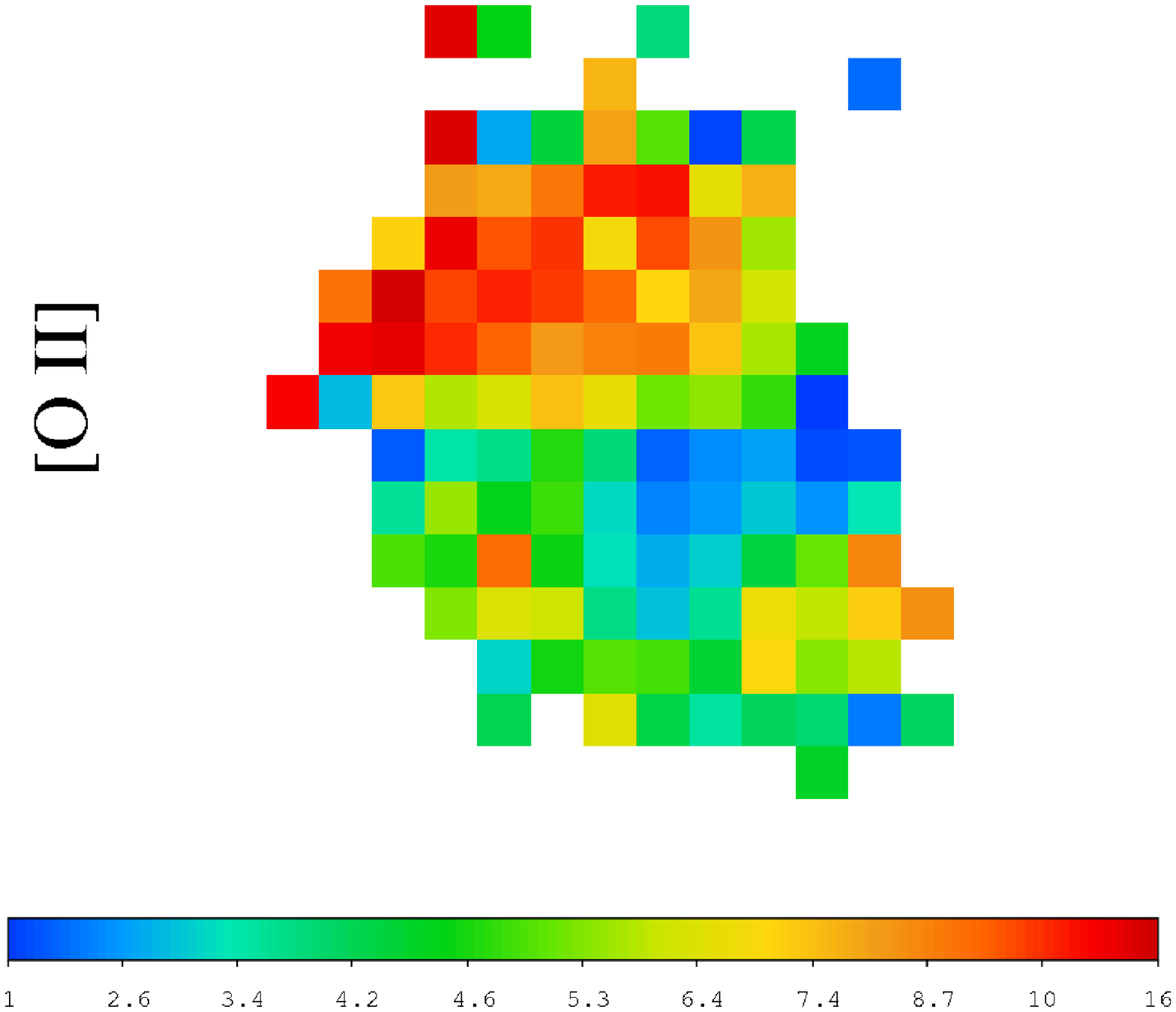}
\includegraphics[width=0.26\textwidth]{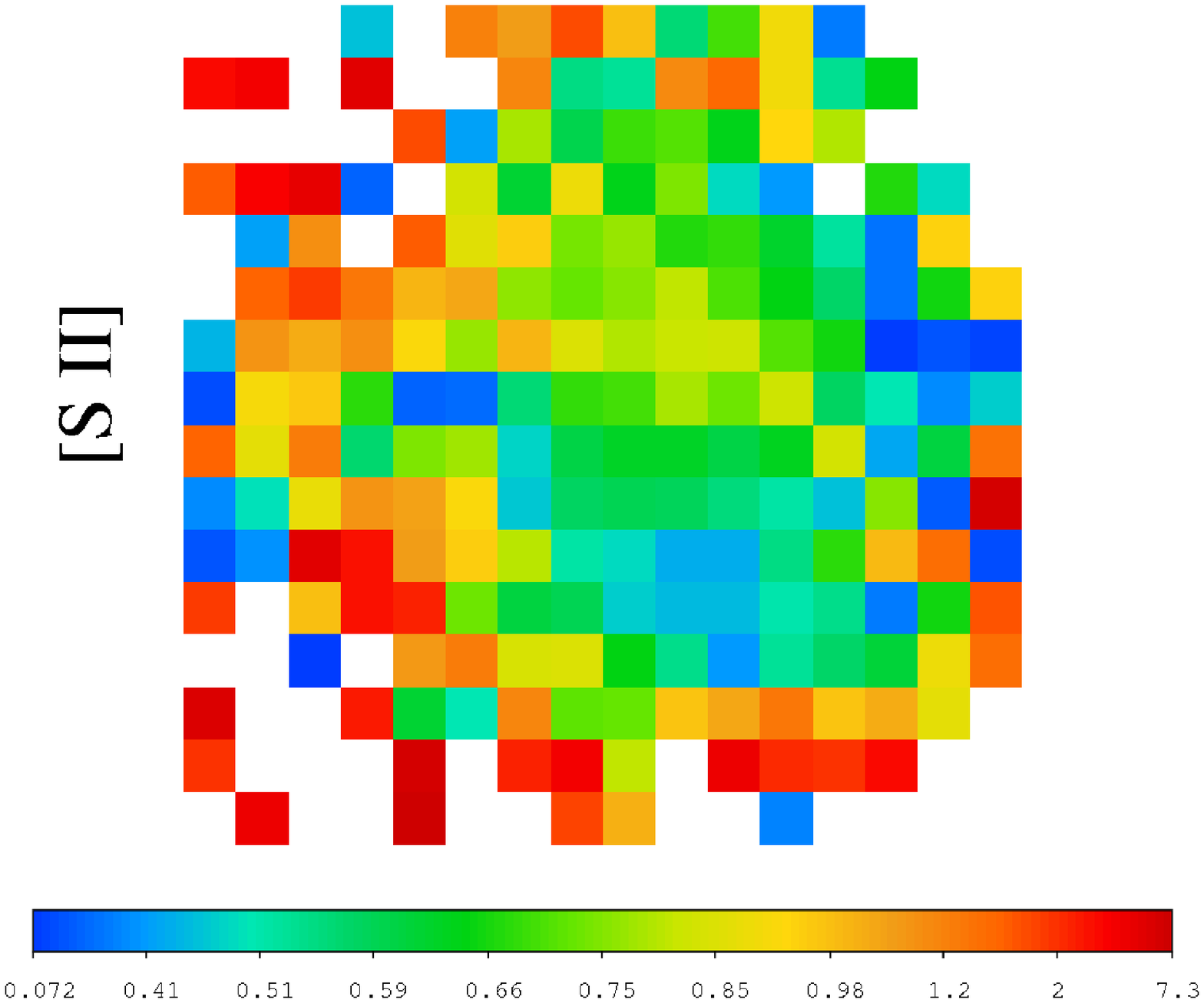}
\includegraphics[width=0.26\textwidth]{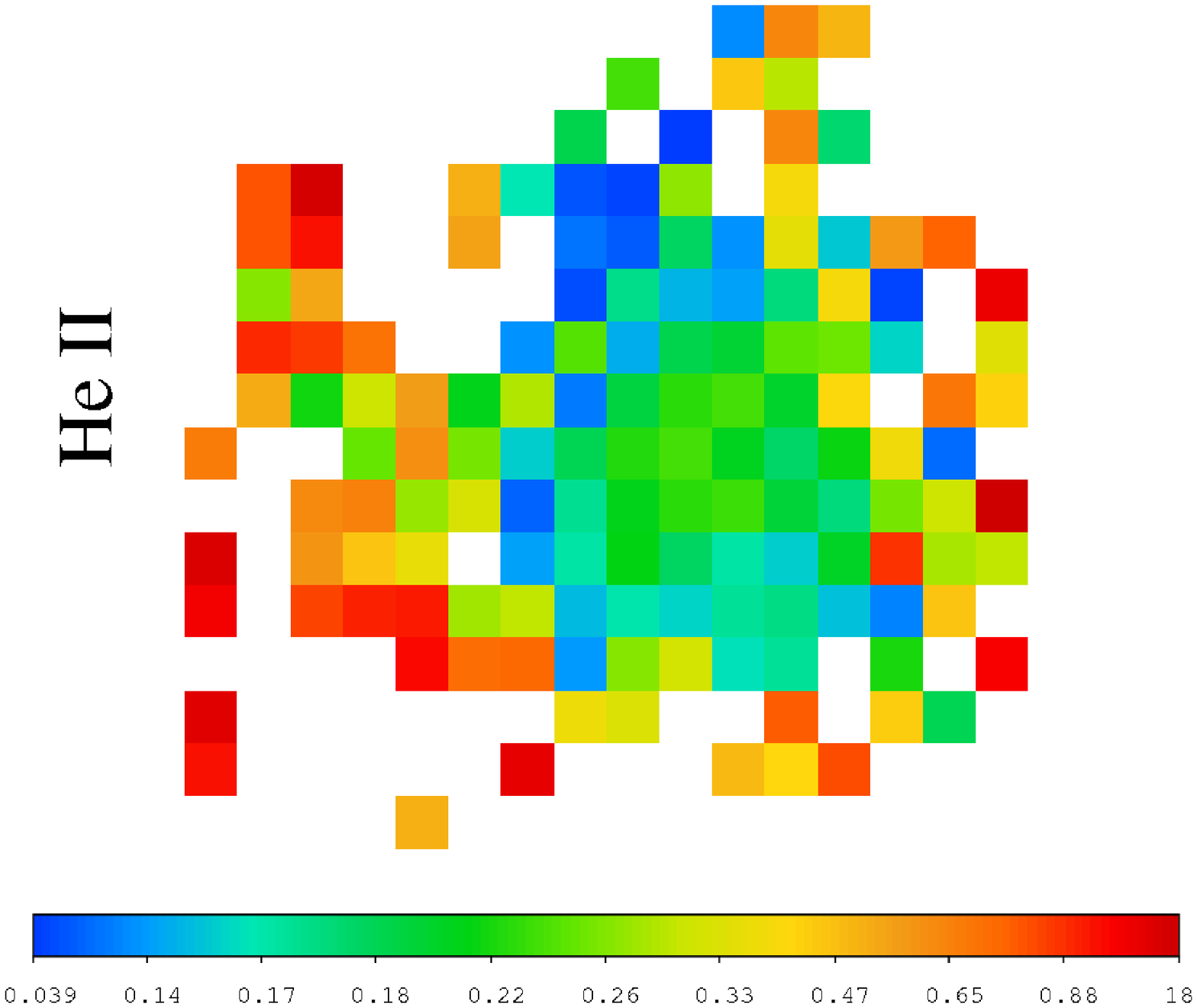}\\
\includegraphics[width=0.26\textwidth]{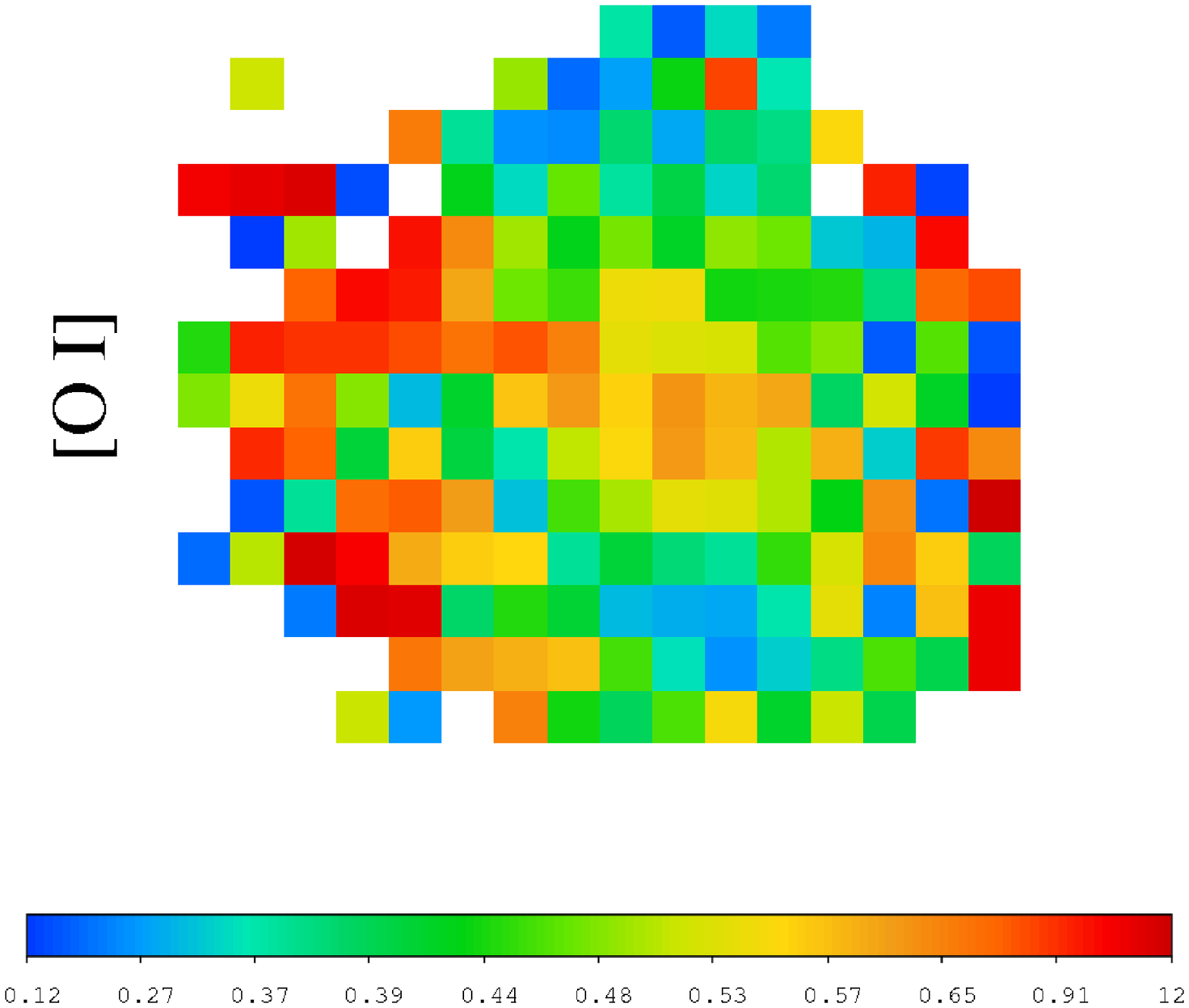}
\includegraphics[width=0.26\textwidth]{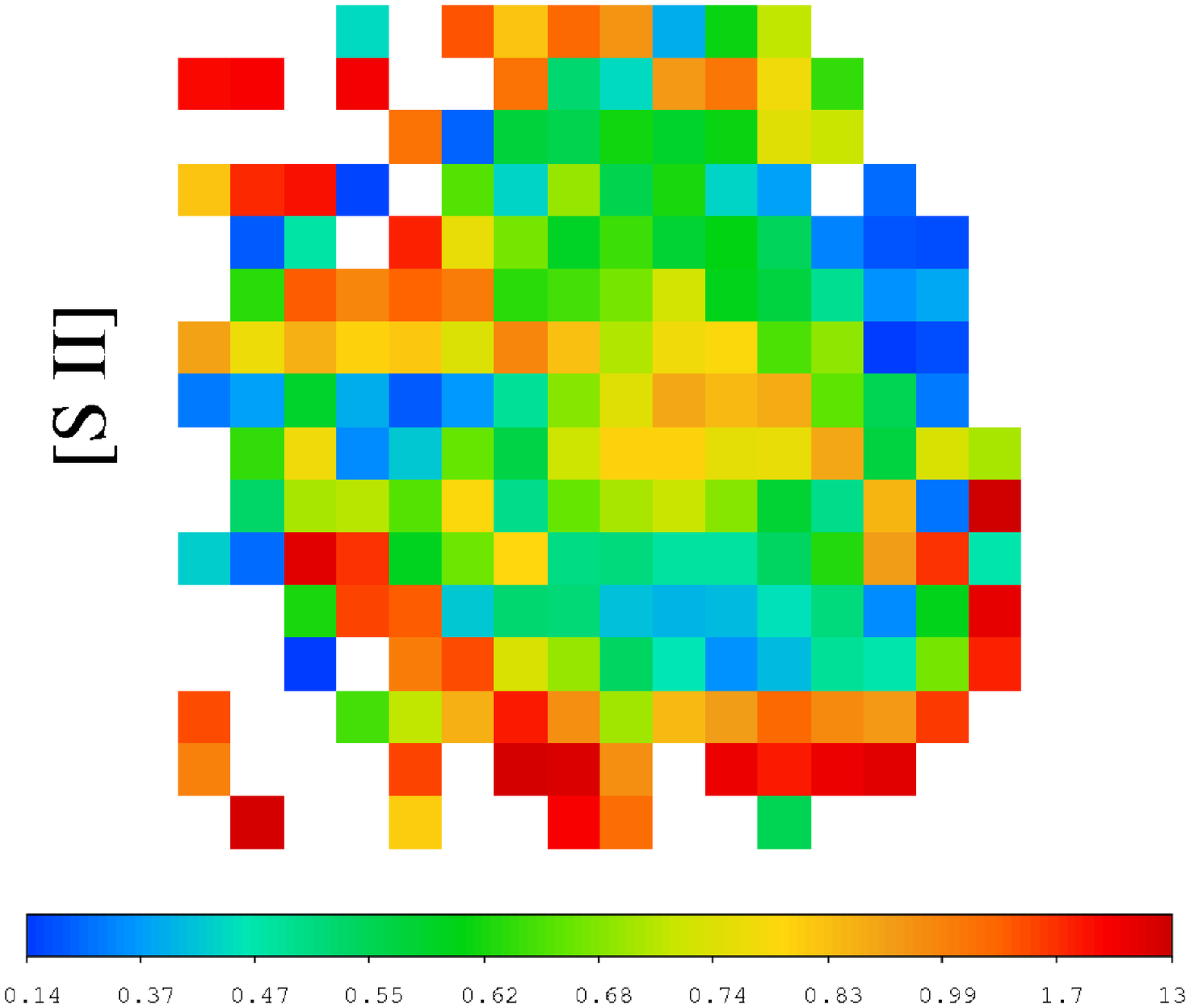}
\includegraphics[width=0.26\textwidth]{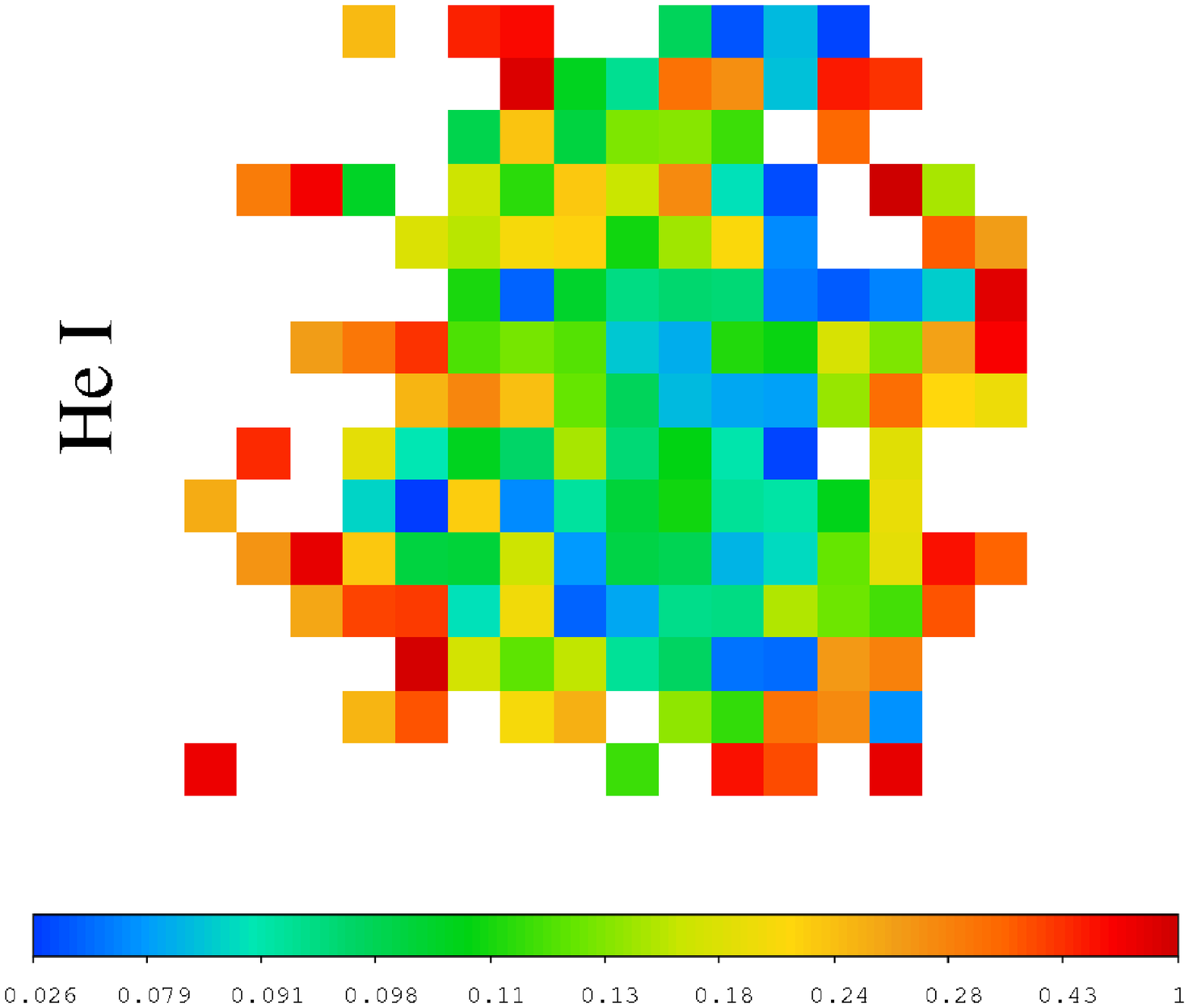}
\caption{Maps of the line ratios to H$\beta$. From left to right: [\ion{O}{III}], [\ion{Fe}{VII}] and [\ion{N}{II}] (top); [\ion{O}{II}], [\ion{S}{II}]6716 and \ion{He}{II} (middle); [\ion{O}{I}], [\ion{S}{II}]6731 and \ion{He}{I} (bottom). Spatial scale is 1 arcsec px$^{-1}$. North is up and East is to the left.}
\label{fig2}
\end{figure*}

In Fig.\,\ref{fig2} we  show  a few significant observed line ratios throughout the observed region in order to obtain a first hint about the ionization conditions of the emitting gas.  The diagrams of the left column describe the spatial distribution of [\ion{O}{III}], [\ion{O}{II}], and [\ion{O}{I}] line ratios to H$\beta$. It can be noticed that photoionization from the AGN dominates in the central regions where the [\ion{O}{III}]/H$\beta$  and the [\ion{O}{II}]/H$\beta$ show smooth opposite trends. However, throughout all the ENLR, [\ion{O}{I}]6300/H$\beta$ line ratios show moderate fluctuations (Fig.\,\ref{fig2}) ranging between
$\sim$ 0.3--0.6. These line ratios are relatively low, implying that the spectra correspond to matter bound clumps. Such a narrow range is unusual in matter fragmented by the underlying turbulence which accompanies relatively high velocity shocks. Eventually, [\ion{O}{I}]/H$\beta$ can be about 0, where the emitting cloud is so thin that there is not enough gas at temperatures  $\leq$ 10$^4$ K.  
 The [\ion{O}{I}]/H$\beta$ line ratios observed in NGC 7212 may suggest that the contribution from the galactic interstellar medium (ISM), that is ionized gas not perturbed by high 
velocity shocks, emerges at some locations in the ENLR. 

In this work, we have tried to fit the [\ion{O}{I}] lines consistently with [\ion{O}{III}] and [\ion{O}{II}].
Only in a few cases did the fit of the three oxygen ionization level lines to H$\beta$ require certain ISM contribution, even when accounting for the typical observational errors  of 0.12, 0.18 and 0.32, for [\ion{O}{II}]/H$\beta$, [\ion{O}{III}]/H$\beta$ and [\ion{O}{I}]/H$\beta$, respectively.

\section{The models}\label{model}
Complex line  profiles, as those observed in each position of NGC 7212, indicate a complex network of emitting clouds. The models help to disentangle the contributions of the different heating and ionizing sources (e.g. photoionization from the AGN, diffuse secondary radiation, shock waves, etc.) 
to  each  spectrum by reproducing a large number of observed line intensity ratios.
In this way we obtain the physical conditions prevailing in each position.

\subsection{Analysis of the spectra}
 
The clumpy and irregular morphology of NGC 7212 and the relative large FWHM of the spectral lines
indicate that in addition to the photoionization flux from the AGN, shocks are heating and ionizing the ENLR gas.
We choose the shock velocity on the basis of both a close approximation to the observed FWHM of the line profiles and the best fit to the observed line intensities. 
Actually, we  consider that the FWHM represents roughly the shock velocity.
So, when  different  profiles are intermingled in one spectrum and show a broad
component  compared  with the bulk of the narrow profiles  observed throughout the ENLR,
we present both the  spectra calculated by high and low velocity shocks.

The physical parameters are combined throughout the calculation of forbidden and permitted lines 
(see Osterbrock 1989) emitted from a shocked nebula. 

In pure photoionization models, the density $n$ is constant throughout the nebula, while in a shock wave 
regime, $n$ is calculated downstream by the compression equation in each single slab. 
Compression depends on $n$, the magnetic field $B$ and the shock velocity \Vs.
In models accounting for the shocks, both the electron temperature  $T_{\rm e}$  and the electron density \ne 
are far from constant throughout each cloud, 
showing the characteristic profiles in the downstream region (see Sect.\,\ref{fe2}).
So, even  sophisticated calculations reproduce approximately the highly inhomogeneous
conditions of the gas, giving rise to some discrepancies between the calculated and observed line ratios.

The ranges of the physical conditions in the nebula are inferred, as an initial guess, from the observed 
lines (e.g. the shock velocity from the FWHM) and from the 
characteristic line ratios (e.g. $n_{\rm e}$ and $T_{\rm e}$).
The physical parameters revealed by the observations indicate that a steady state situation can be applied (Cox 1972) in NGC 7212. In this case, 
the time $t$ required for a parcel of gas to cross the cooling region from the shock front to the recombination zone, for shock waves with \Vs = 100 \kms, is found to be approximately 1000/\ne yr (calculated by the recombination coefficients), such that for an electron density \ne= 100 \cm3, $t = 10$ yr. 
Shock velocities  are not likely to change appreciably in such a short time, and therefore the steady state calculation may be regarded as adequate.

\subsection{The code: input parameters}
In this paper, the line and continuum spectra emitted by the gas downstream of the shock front 
are calculated by {\sc suma} (see http://wise-obs.tau.ac.il/~marcel/suma/index.htm for a detailed description). 
The code simulates the physical conditions in an emitting gaseous cloud under the coupled effect of 
photoionization from an external radiation source and shocks. The line and continuum emission 
from the gas are calculated consistently with dust-reprocessed radiation in a plane-parallel geometry.

The input parameters which refer to the shock are: the 
shock velocity \Vs, the pre-shock density \n0, the pre-shock magnetic field \B0. 
The input parameters related to the radiation field are the power-law
flux from the active nucleus $F$  in number of photons cm$^{-2}$ s$^{-1}$ eV$^{-1}$ at the Lyman limit,
if the photoionization source is an active nucleus, and the spectral indices  $\alpha_{\rm UV}=-1.5$
and $\alpha_{\rm X}=-0.7$.
 $F$  is combined with the ionization parameter $U$ by
$U$ = ($F$/($n$ c ($\alpha -1$)) (($E_{\rm H})^{-\alpha +1}$ - ($E_{\rm C})^{-\alpha +1}$)
(Contini \& Aldrovandi 1983), where
$E_{\rm H}$ is H ionization potential  and $E_{\rm C}$ is the high energy cutoff,
$n$ the density, $\alpha$ the spectral index, and c the speed of light.
The secondary diffuse radiation emitted from the slabs of gas heated by the shocks is also calculated.
The flux from the active center and the secondary radiation are calculated by
radiation transfer throughout the slabs downstream.

The dust-to-gas ratio ($d/g$) and the  abundances of He, C, N, O, Ne, Mg, Si, S, A, Fe relative to H,
are also accounted for. They  affect the calculation of the cooling rate.
The dust grains are heated radiatively by photoionization and collisionally by the shock.
The distribution of the grain radii downstream results from sputtering.
The geometrical thickness of the emitting nebula ($D$), determines whether the model is radiation-bound or matter-bound.

The input parameters providing the best fit of the line ratios
establish the physical conditions in each of the observed positions.

\subsection{The code: calculation process}

The calculations initiate at the shock front where the gas is compressed and thermalized adiabatically, 
reaching a maximum temperature in the immediate post-shock region ($T\sim 1.5 \times 10^5$ (\Vs/100 \kms )$^2$). 
$T$ decreases downstream following recombination. The cooling rate is calculated in each slab. 
The downstream region is cut up into a maximum of 300 plane-parallel slabs with different geometrical 
widths calculated automatically, in order to account for the temperature gradient (Contini 1997 and references therein).

In each slab, compression is calculated by the Rankine--Hugoniot equations for the conservation of mass, 
momentum and energy throughout the shock front. Compression ($n/n_{\rm 0}$) downstream ranges between 4 
(the adiabatic jump) and $\geq$ 10, depending on \Vs and \B0. The stronger the magnetic field, the lower the 
compression downstream is, while a higher shock velocity corresponds to a higher compression.

The ionizing radiation from an external source is characterized by its spectrum  and by 
the flux intensity. The flux is calculated at 440 energies, from a few eV to keV. Due to radiative transfer, 
the spectrum changes throughout the downstream slabs, each of them contributing to the optical depth. 
In addition to the radiation from the primary source, the effect of the diffuse radiation created by 
the gas line and continuum emission is also taken into account, using 240 energies to calculate the spectrum.

For each slab of gas, the fractional abundance of the ions of each chemical element is obtained by solving 
the ionization equations. These equations account for the ionization mechanisms 
(photoionization by the primary and diffuse radiation, and collisional ionization) and recombination 
mechanisms (radiative, dielectronic recombinations), as well as charge transfer effects. 
The ionization equations are coupled to the energy equation if collision processes dominate, and 
to the thermal balance if radiative processes dominate. The latter balances the heating of the gas due 
to the primary and diffuse radiations reaching the slab with the cooling due to recombinations and 
collisional excitation of the ions followed by line emission, dust collisional ionization and thermal bremsstrahlung. 
The coupled equations are solved for each slab, providing the physical conditions necessary for calculating the slab optical depth, as well as its line and continuum emissions. The slab contributions are integrated throughout the cloud.

In particular, the absolute line fluxes corresponding to the ionization level $i$ of element K are calculated by the 
term $n_{\rm K}(i)$, which represents the density of the ion $i$. We consider that $n_{\rm K}(i)={\rm X}(i){\rm [K/H]}n_{\rm H}$, where X($i$) is 
the fractional abundance of the ion i calculated by the ionization equations, [K/H] is the relative 
abundance of the element K to H and $n_{\rm H}$ is the density of H (in number \cm3). In models including shock, 
$n_{\rm H}$ is calculated by the compression equation (Cox 1972) in each slab downstream. 
So the abundances of the elements are given relative to H as input parameters.

Dust grains are coupled to the gas across the shock front by the magnetic field (Viegas \& Contini 1994). 
They are heated by radiation from the AGN and collisionally by the gas to a maximum temperature which is a 
function of the shock velocity, of the chemical composition and of the radius of the grains, 
up to the evaporation temperature ($T_{\rm dust}$ $\geq$ 1500 K). The grain radius distribution downstream is determined 
by sputtering, which depends on the shock velocity and on the density. Throughout shock fronts and downstream, 
the grains might be destroyed by sputtering.

Summarizing, the code starts by adopting an initial $T_{\rm e}$ ($\sim$ 10$^4$ K) and the input parameters for the first slab (Sections 3.1 and 3.2). It then calculates the density from the compression equation, the fractional abundances of the ions from each level for each element, line emission, free-free emission and free-bound emission. It re-calculates $T_{\rm e}$ by thermal balancing or the enthalpy equation, and calculates the optical depth of the slab and the primary and secondary fluxes. Finally, it adopts the parameters found in slab $i$ as initial conditions for slab $i+1$. 
The line intensities and the bremsstrahlung are calculated from the gas which emits the spectrum. 
The data are observed at Earth, therefore they diverge by a factor which depends on the distance of the nebula 
from the active centre ($r$), and  on the distance ($d$) of the galaxy to Earth ($r^2/d^2$).
We then calculate the line ratios to a specific line (in the present case \Hb which is a strong line), and we compare them with the observed line ratios.

\section{Modelling the NGC 7212 spectra}\label{model_result}

The spectra are rich in number of lines, therefore the calculated spectra are strongly constrained by the observed line ratios.
The line ratios are different in each of the observed spectra, revealing  different  physical conditions from region to region. 
At each stage of the modelling process, if a satisfactory fit is not found for all the lines,
a new iteration is initiated with a different set of input parameters. When a ratio is not reproduced, we check how it depends on the physical parameters and decide consequently how to change them. Each ratio has a different weight, however we generally consider that the observed spectrum is satisfactorily fitted by a model when the strongest line ratios are reproduced by the calculation within a 20\%
discrepancy and weak line ratios within 50\%.

\begin{figure}
 \centering
\includegraphics[width=0.5\textwidth]{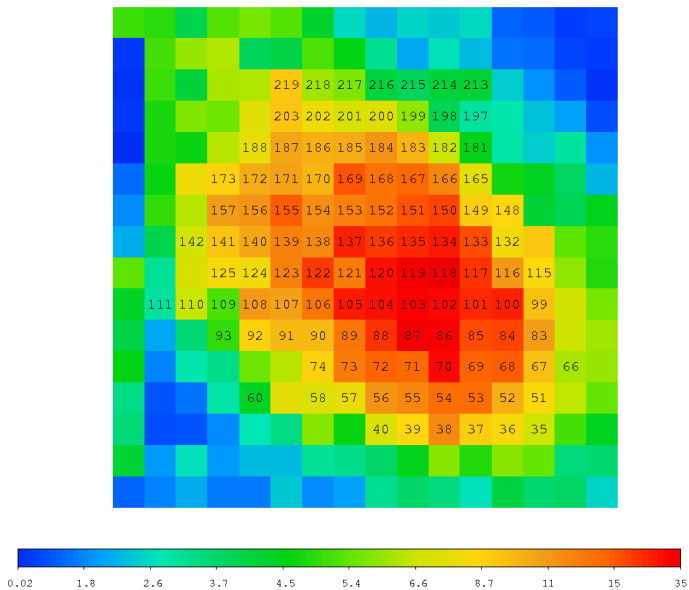}
\caption{2D map of the S/N ratio of the stellar continuum. The numbers indicate the spectra successfully modelled.}
\label{sn}
\end{figure}

We have run a grid of models ($\sim$ 100 for each position) in order to select the most appropriate one.
The fit was successful in 111 out of the 256 spectra, since they were the most complete in number of lines. 
Fig.\,\ref{sn} shows a map of the S/N ratio of the stellar continuum with the number of the modelled spectrum overlaid. Most of these spectra have S/N larger than 6.   
The final gap between  observed and  calculated line ratios is due to both random and systematic observational errors, as well as to the uncertainties of the atomic parameters  adopted by the code, such as recombination coefficients, collision strengths etc., which are continuously updated, and to the choice of the model. 
Models are generally allowed to reproduce the data within a factor of 2. This leads to input parameter
ranges of a few per cent.

\subsection{First step:  choice of physical parameters}

\begin{figure}
\includegraphics[width=0.50\textwidth]{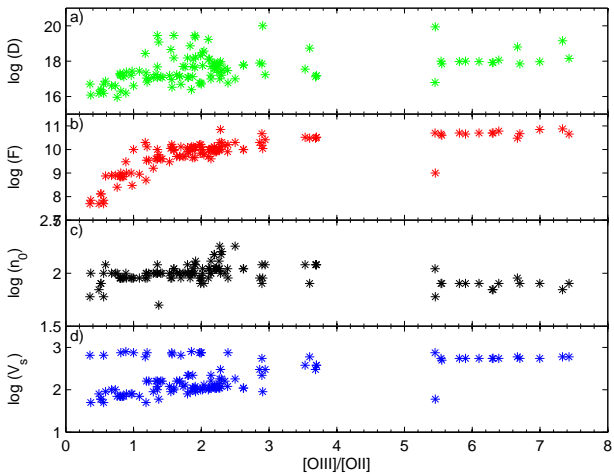}
\caption{The correlation of the input parameters with 
calculated [\ion{O}{III}]5007+/[\ion{O}{II}]3727+ line ratios.}
\label{fig3}
\end{figure}

In the first iterations, 
we try to reproduce the [\ion{O}{III}]5007+4959/H$\beta$ line ratio (5007+4959 will
be referred to in the following as 5007+; the + indicates that  
the doublet 5007, 4959 is summed up), the highest ratio in general, by readjusting $F$ and \Vs (\Vs however, is constrained  throughout a small range by the observed FWHM). 
The  higher $F$, the higher the [\ion{O}{III}]/H$\beta$ and the [\ion{O}{III}]/[\ion{O}{II}] line ratios are (see Fig.\,\ref{fig3}),
as well as  \ion{He}{II}/H$\beta$. Moreover,
a high $F$ maintains the gas ionized far from the source, yielding enhanced  [\ion{O}{I}]
and [\ion{S}{II}] lines.  Recall that these lines behave similarly because S first ionization
potential (10.36 eV) is lower than that of O (13.61 eV).

Then, the [\ion{O}{II}]3727+ doublet is considered. 
If the flux from the active center  is low ($F$ $\leq 10^9$ ph cm$^{-2}$ s$^{-1}$ eV$^{-1}$), 
a shock dominated  regime is found, which is characterised by  relatively 
high [\ion{O}{II}]/[\ion{O}{III}] ( $\geq$ 1, as can be seen in Fig.\,\ref{fig3}).
These lines can be drastically reduced by collisional deexcitation at high electron densities
(\ne $>$ 3000 \cm3). 

The gas density is a crucial parameter. 
In each cloud, it reaches its upper limit  downstream and  remains nearly constant, while the electron density decreases following recombination. 
A high density, increasing the cooling rate, speeds up the
recombination process of the gas,  enhancing the low ionization lines. 
In fact, each line is produced in a region  of gas  at a different $n_{\rm e}$ and $T_{\rm e}$,  
depending on the ionization level and the  atomic parameters characteristic of the ion.  
The density $n$, which can be roughly inferred from the [\ion{S}{II}]6716/6731 doublet ratio, 
is  related  with \n0 by 
compression downstream ($n$/\n0), which ranges between $\sim$  4 and $\sim$ 100, depending on \Vs and \B0.  
The  [\ion{S}{II}] lines are also characterised by a relatively low critical density for collisional deexcitation.
In NGC 7212 the [\ion{S}{II}]6716/6731 line ratios are $\sim$ 1 in all observed positions, constraining the choice of \n0. 
In some cases the [\ion{S}{II}]6716/6731 line ratio  varies from  $>$1 to  $<$1 
throughout a relatively small region,  since the [\ion{S}{II}] line ratios depend on both the temperature and
electron density of the emitting gas (Osterbrock \cite{O89}). The median error of the modelled ratios is about 10\%, and 90\% of this sample has an error lower than 30\%.

\begin{figure*}
 \centering
\includegraphics[width=0.30\textwidth]{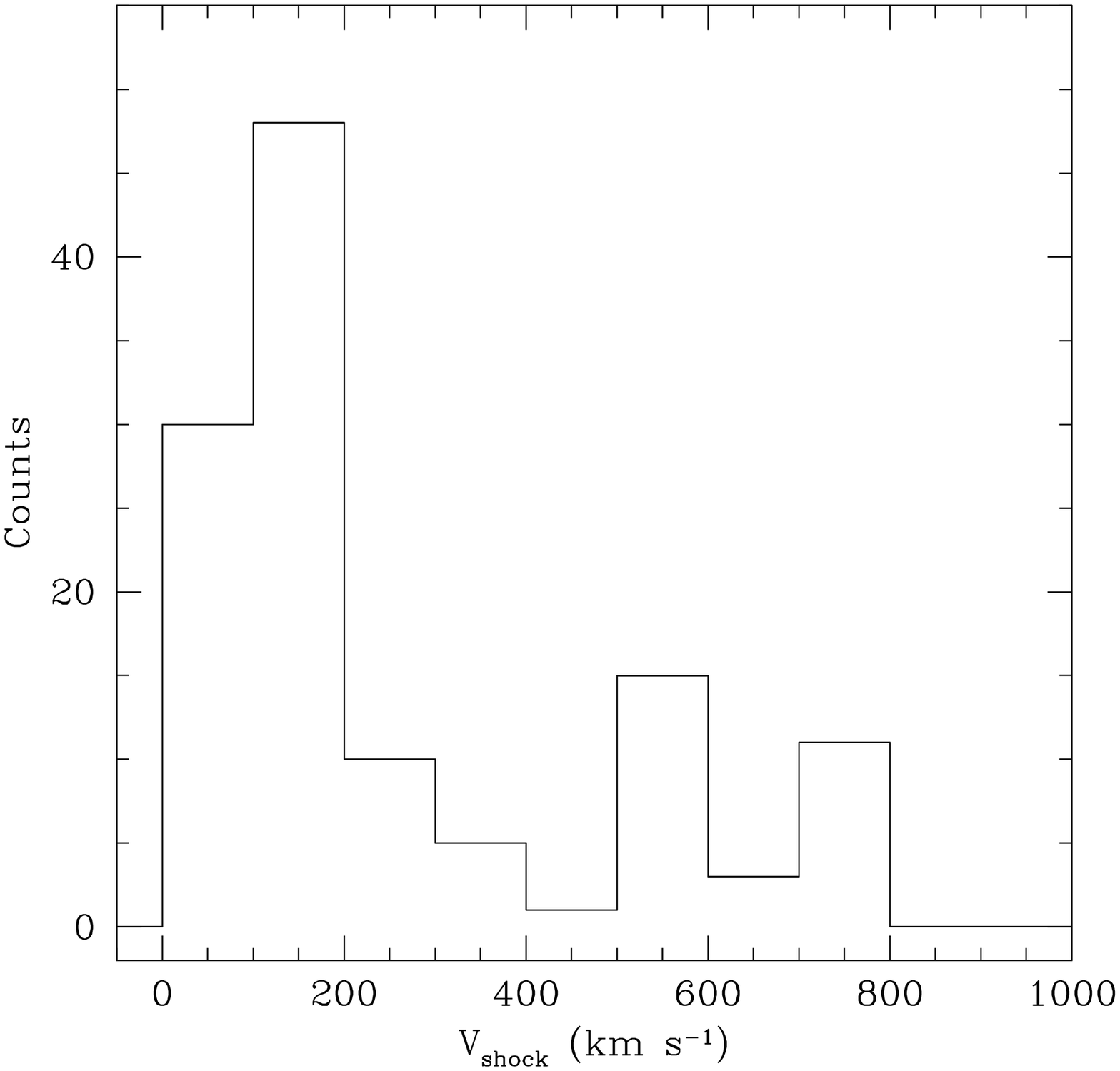}
\includegraphics[width=0.30\textwidth]{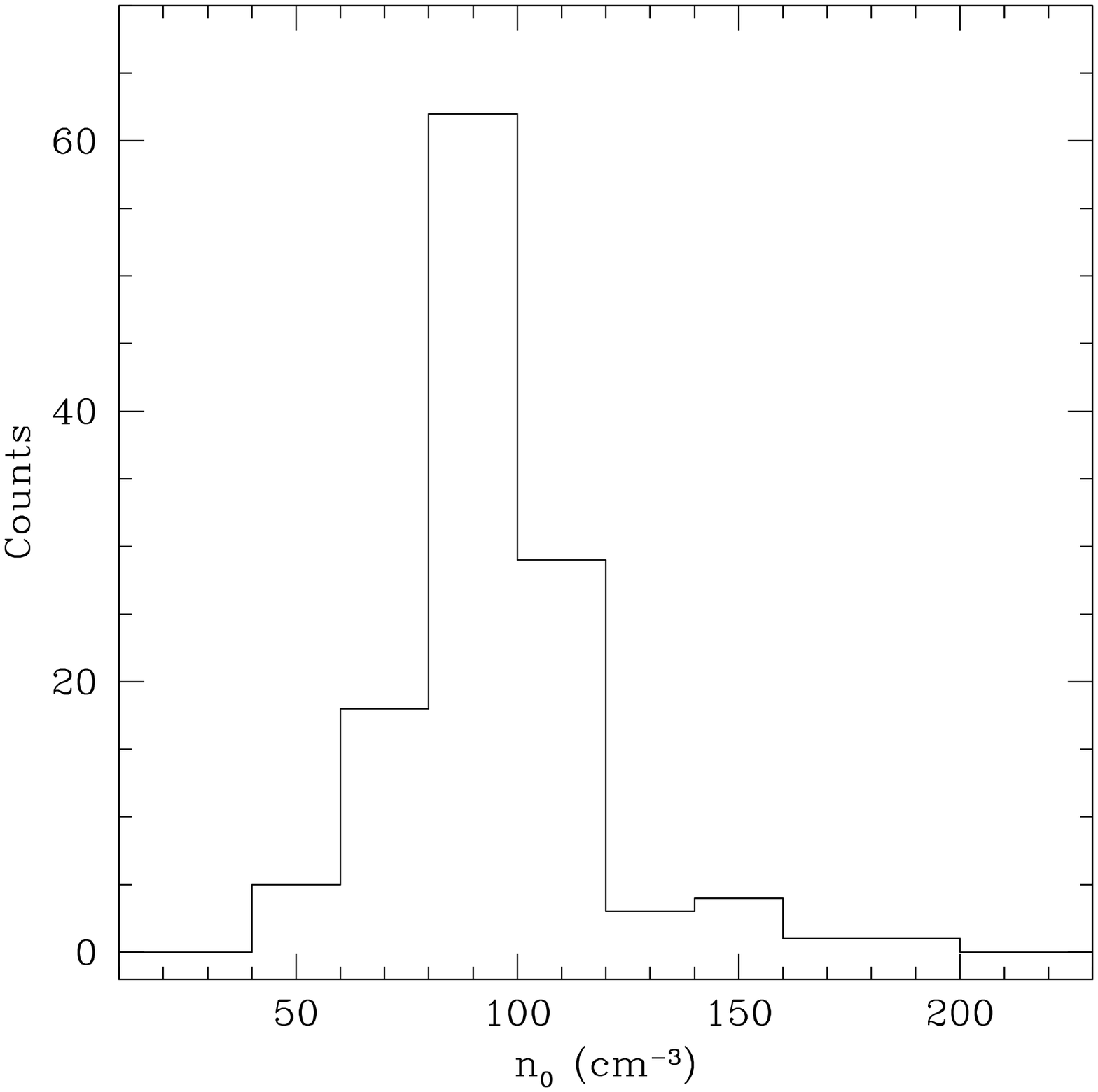}
\includegraphics[width=0.30\textwidth]{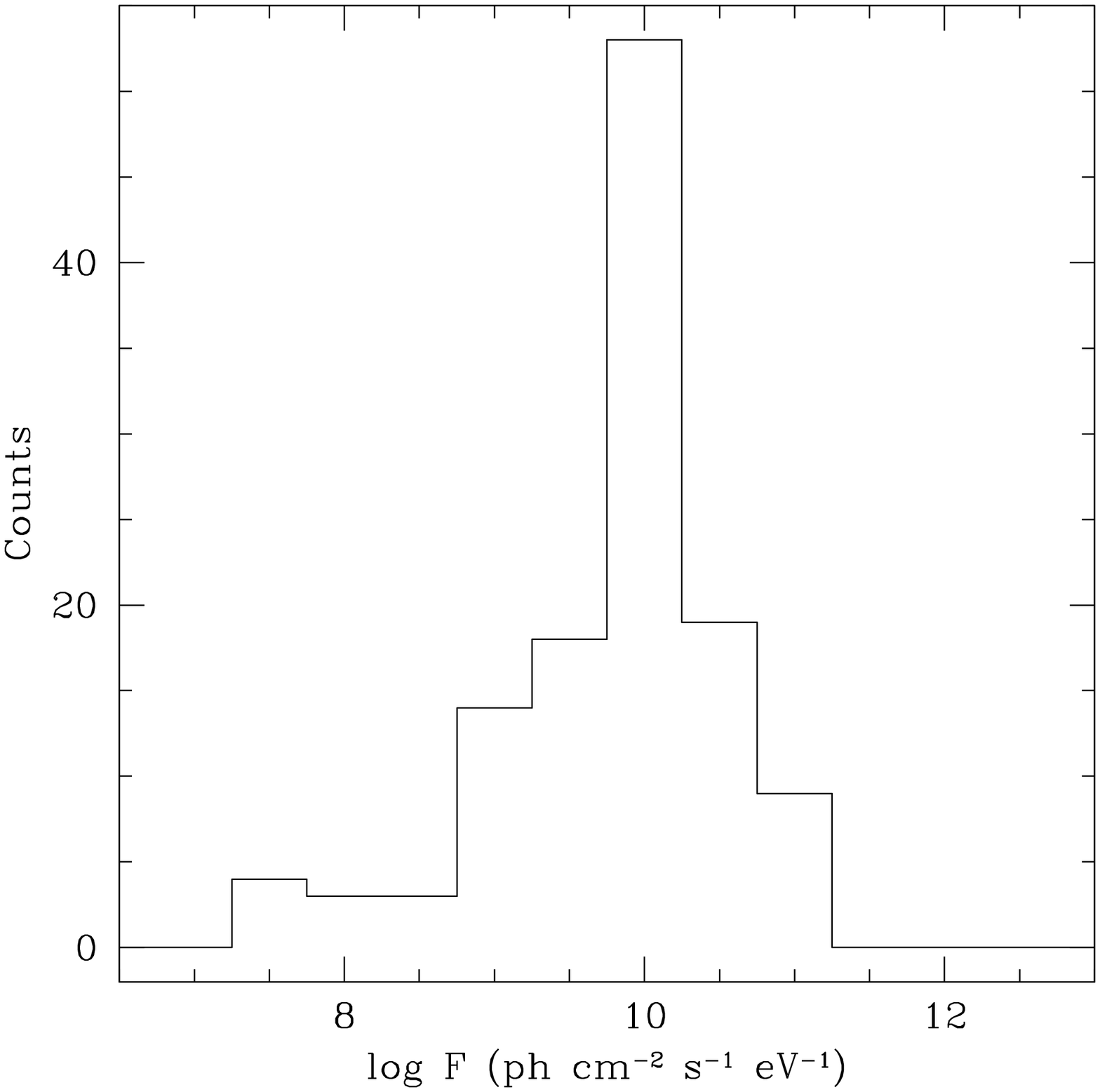}
\includegraphics[width=0.30\textwidth]{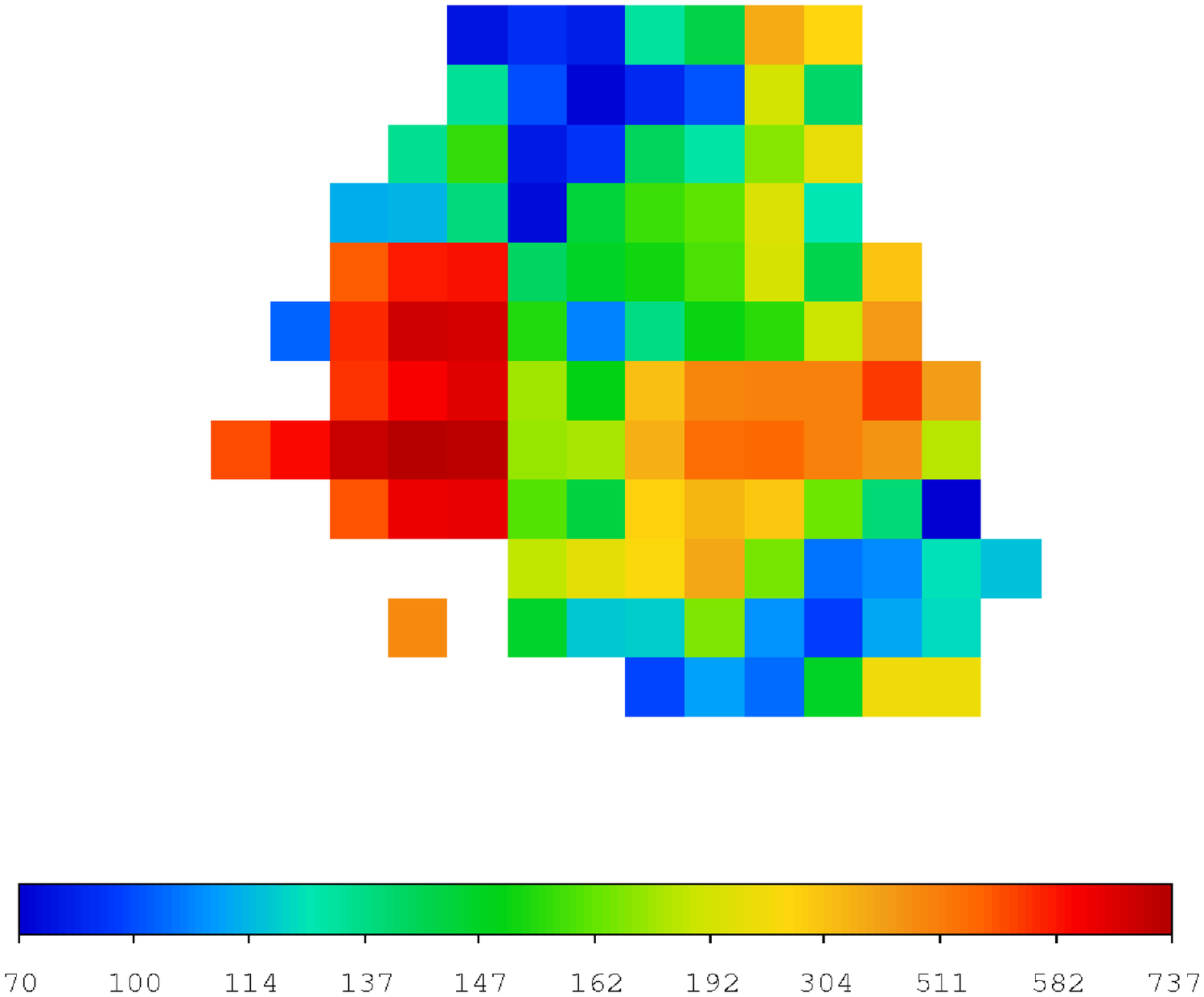}
\includegraphics[width=0.30\textwidth]{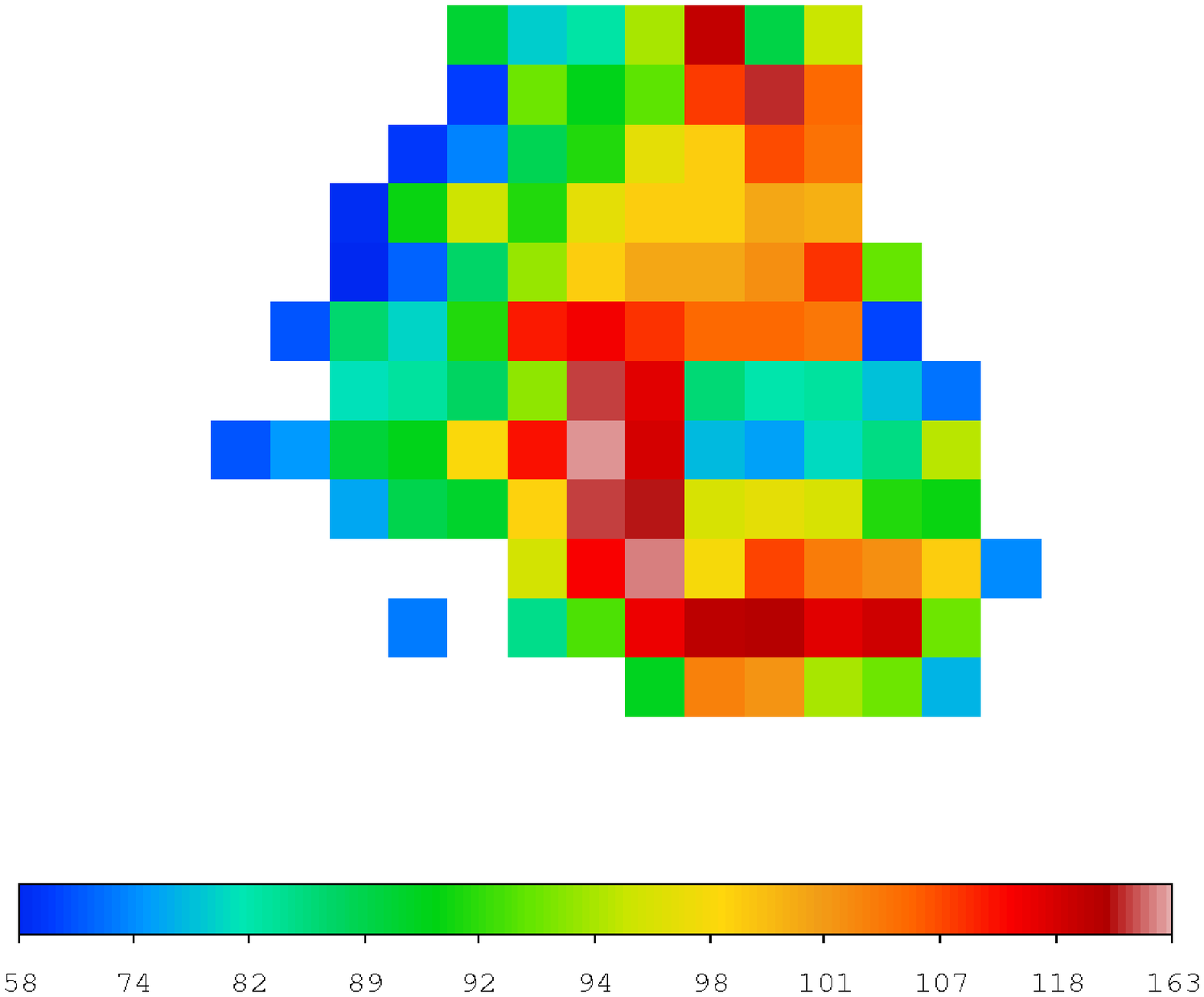}
\includegraphics[width=0.30\textwidth]{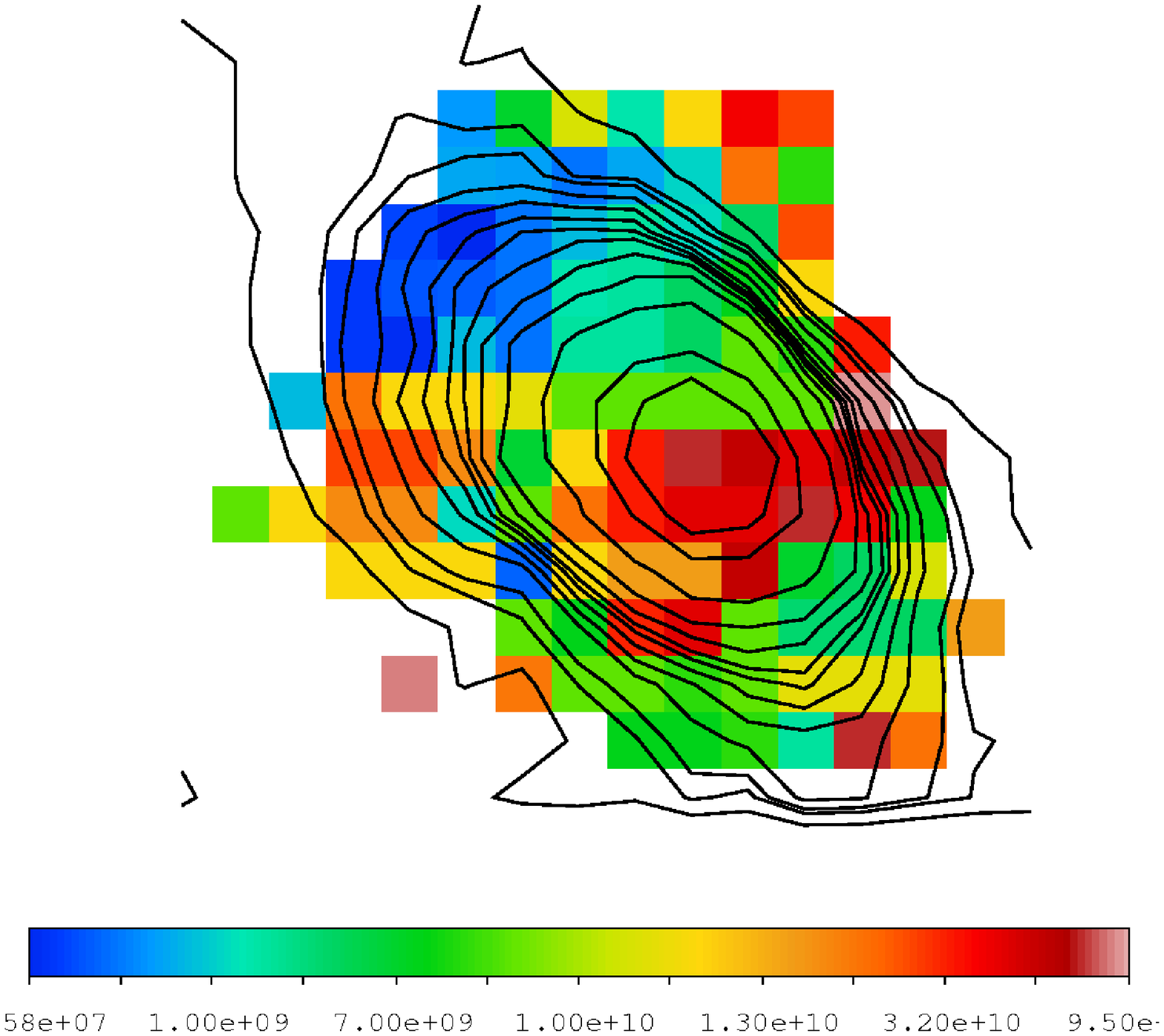}
\caption{ Top: histograms of the shock velocities (left), pre-shock density (middle) and the AGN ionizing flux (right). 
Bottom: 2D maps of the same parameters. The units are km s$^{-1}$ (left), cm$^{-3}$ (middle) and ph cm$^{-2}$ s$^{-1}$ eV$^{-1}$ (right)}. Spatial scale is 1 arcsec px$^{-1}$. Contours of the stellar continuum are overlapped onto the map of the ionizing flux. 
North is up and East is to the left.
\label{fig7}
\end{figure*}

We fit the lines of a single spectrum  adopting an \n0 which is combined with the
other input parameters, including the pre-shock magnetic field \B0.
The magnetic field  has an important role in models accounting for the shock.
The stronger \B0, the lower the compression is. So, lower densities are compensated 
by a  lower magnetic field. 
If \n0  in one spectrum is  inconsistent with \n0 calculated in the surrounding
regions, we can  reduce the \n0  divergence by changing the magnetic field, because
\n0 and \B0 are strongly (anti)correlated.
This procedure would lead to a smoother distribution of \n0 throughout the ENLR of
NGC 7212,  which  is however not expected because collisions and merging phenomena render the density field dishomogeneous.
So we have  adopted  a constant \B0  = 10$^{-4}$ G, which is suitable for the ENLR of AGN (Beck 2011).
This choice is slightly
unrealistic but it is justified by the fact that the preshock densities range between values within a factor of 2 ($\sim$ 80--150 \cm3, Table\,\ref{tab1}). 
Table 1  shows that  \B0 could be modified in regions 103-105, 121-122, 132-133, and 156-157.
These regions are too few as to yield a trend to \B0 throughout the ENLR.
Moreover,  \B0  fluctuations would not affect the ensemble of the  results.

The final choice of a model is dictated by  [\ion{O}{I}]6300+/\Hb,
which is calculated consistently with [\ion{O}{III}]/H$\beta$ and [\ion{O}{II}]/H$\beta$.
The  geometrical thickness ($D$) of the emitting clouds is mainly determined by [\ion{O}{I}]/\Hb
because [\ion{O}{I}] is emitted from gas at relatively low temperature,  
at the cloud edge opposite the shock front. 

A first analysis of the data shows that the models are mostly matter bound, 
namely, the integration of the calculated line fluxes downstream is 
interrupted at such a distance from the
shock front that all the line ratios of a  single spectrum are satisfactorily reproduced  (by 10 per cent
for the strongest ones). 
The low [\ion{O}{I}]6300/H$\beta$ ($\leq$  0.6, Fig.\,\ref{fig2}) observed by 
Cracco et al. are hardly reproduced  by radiation bound models, which  
are revealed in  a few cases (represented by the maximum $D$ in Fig.\,\ref{fig3}).  
The [\ion{S}{II}]6716, 6731 lines  are also emitted by relatively cool gas. 
Therefore, their ratios to H$\beta$ are determined by [\ion{O}{I}]/H$\beta$.

%\section{Modelling results}\label{model_result}
%\subsection{Line ratios and relative abundances}
So far, the results of modelling  correspond to the  set  of  input parameters  presented in Table\,\ref{tab1}. The relative maps are shown in Fig.\,\ref{fig7}.
 The uncertainties of the input parameters are approximately 10\%. In fact, a change of 10\% corresponds to variations of the resulting line ratios within 20\% for the strongest lines up to 50\% for the weakest lines.
For all the models, solar abundances were used. 
In Table\,\ref{tab2} we  compare the  spectra calculated  in each position  with the observed  corrected line ratios to H$\beta$. Each of the observed  line ratios  is followed  by the calculation results. 
The numbers in the first column refer to Fig.\,\ref{fig1}. Not all the lines used for modelling are included in Table\,\ref{tab2}. 
We have focused on  [\ion{O}{III}]5007+, [\ion{Ne}{III}]3869+, [\ion{O}{II}]3727+, [\ion{O}{I}]6300+, \ion{He}{II} 4686, \ion{He}{I} 5876, [\ion{N}{II}]6548+, [\ion{S}{II}]6716,6731.
For all the spectra, we assumed that the reddening corrected ratio is H$\alpha$/H$\beta$=2.8, model calculations give H$\alpha$/H$\beta$ $\sim 2.9$ and  H$\gamma$/H$\beta$ $\sim 0.46$.

\begin{figure*}
\centering
\includegraphics[width=0.32\textwidth]{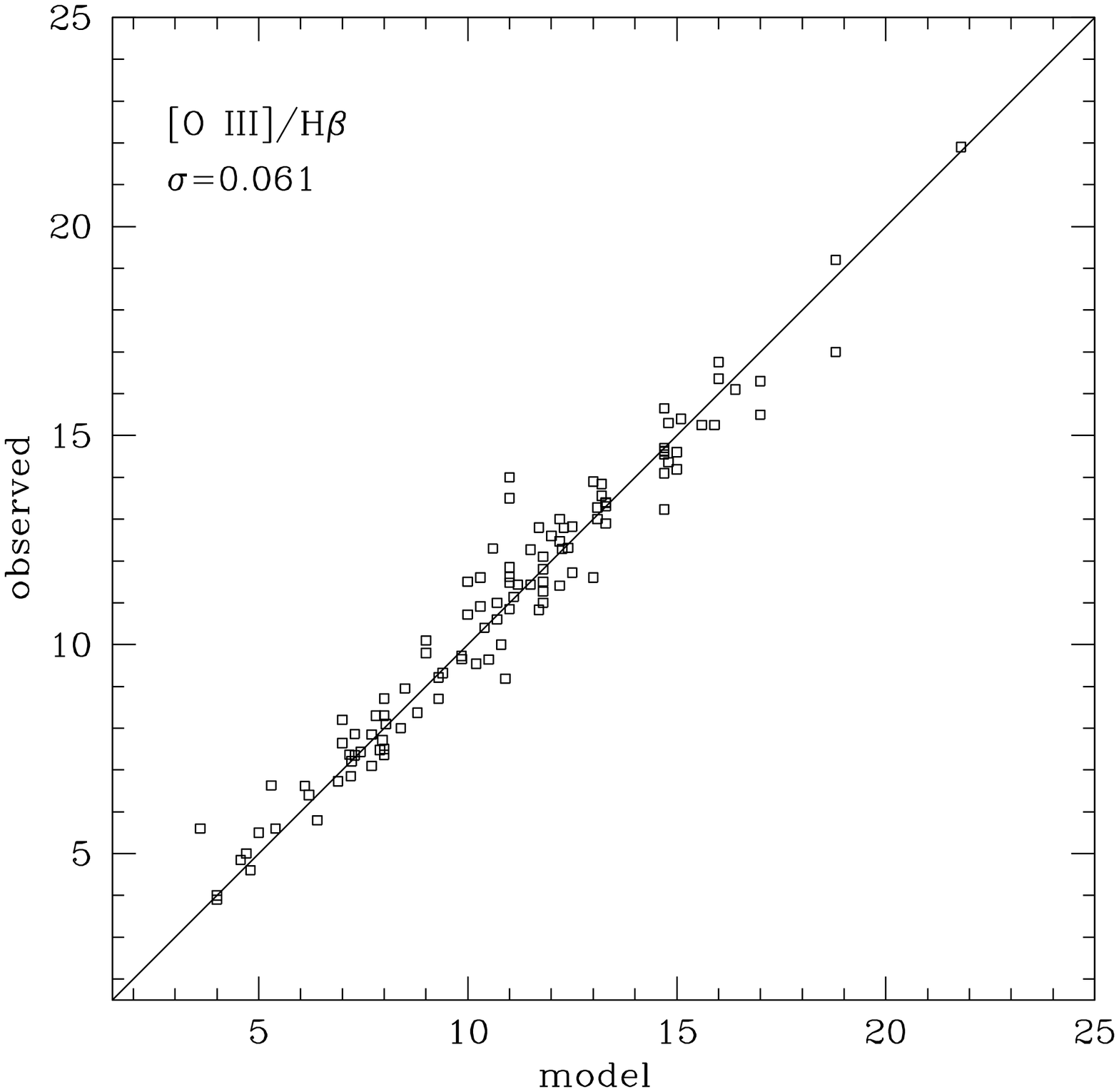}
\includegraphics[width=0.32\textwidth]{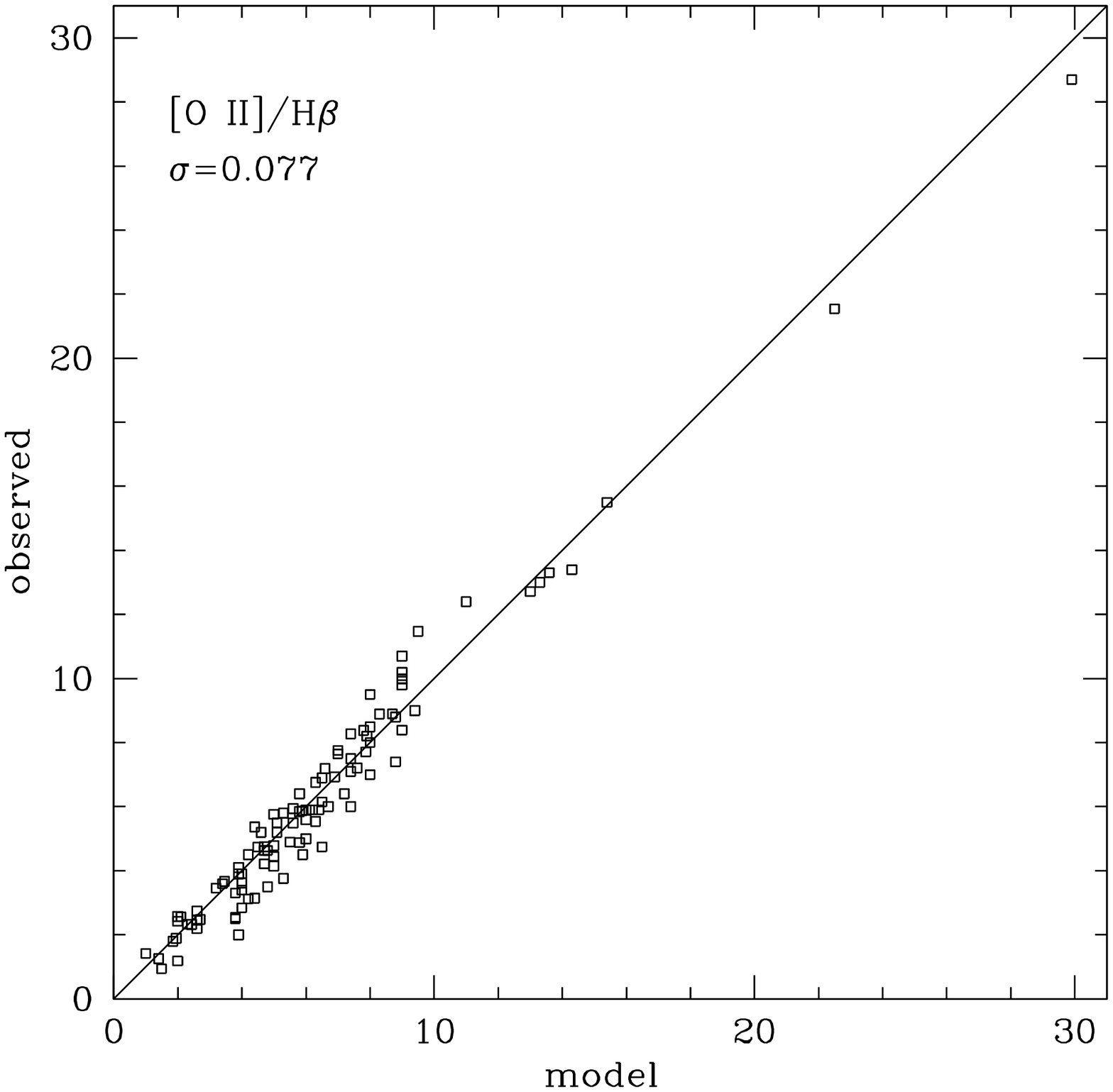}
\includegraphics[width=0.32\textwidth]{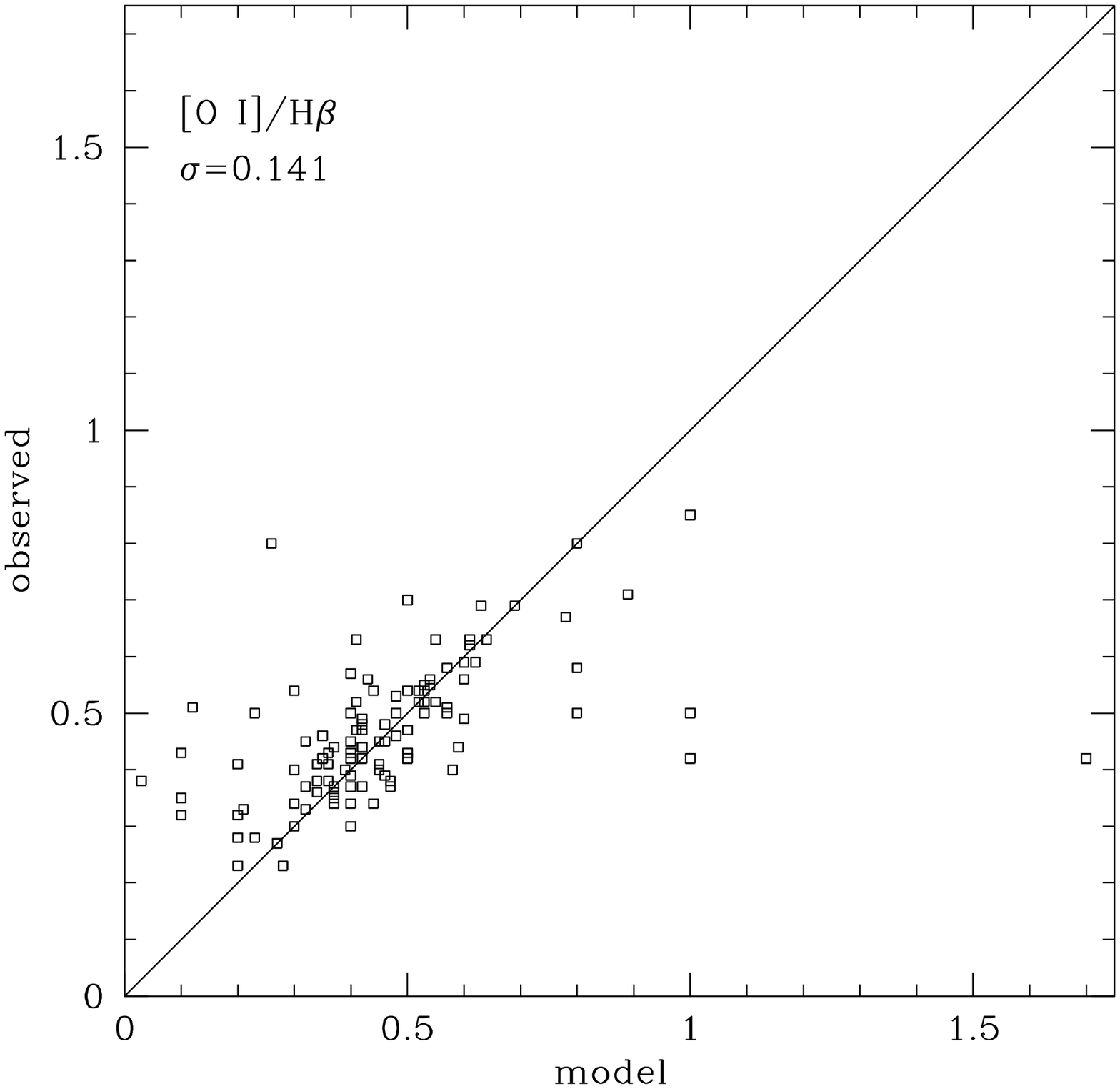}\\
\includegraphics[width=0.32\textwidth]{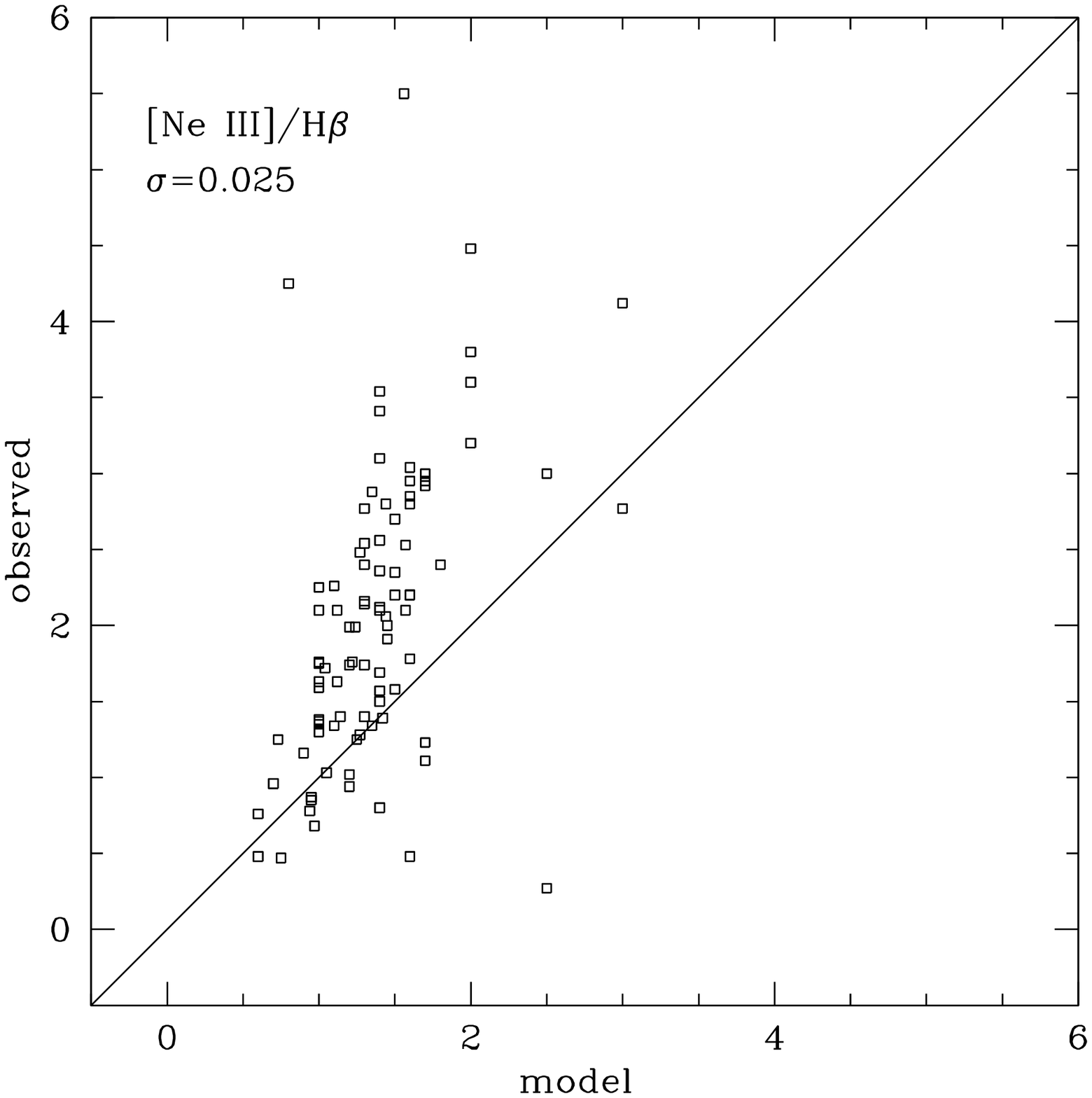}
\includegraphics[width=0.32\textwidth]{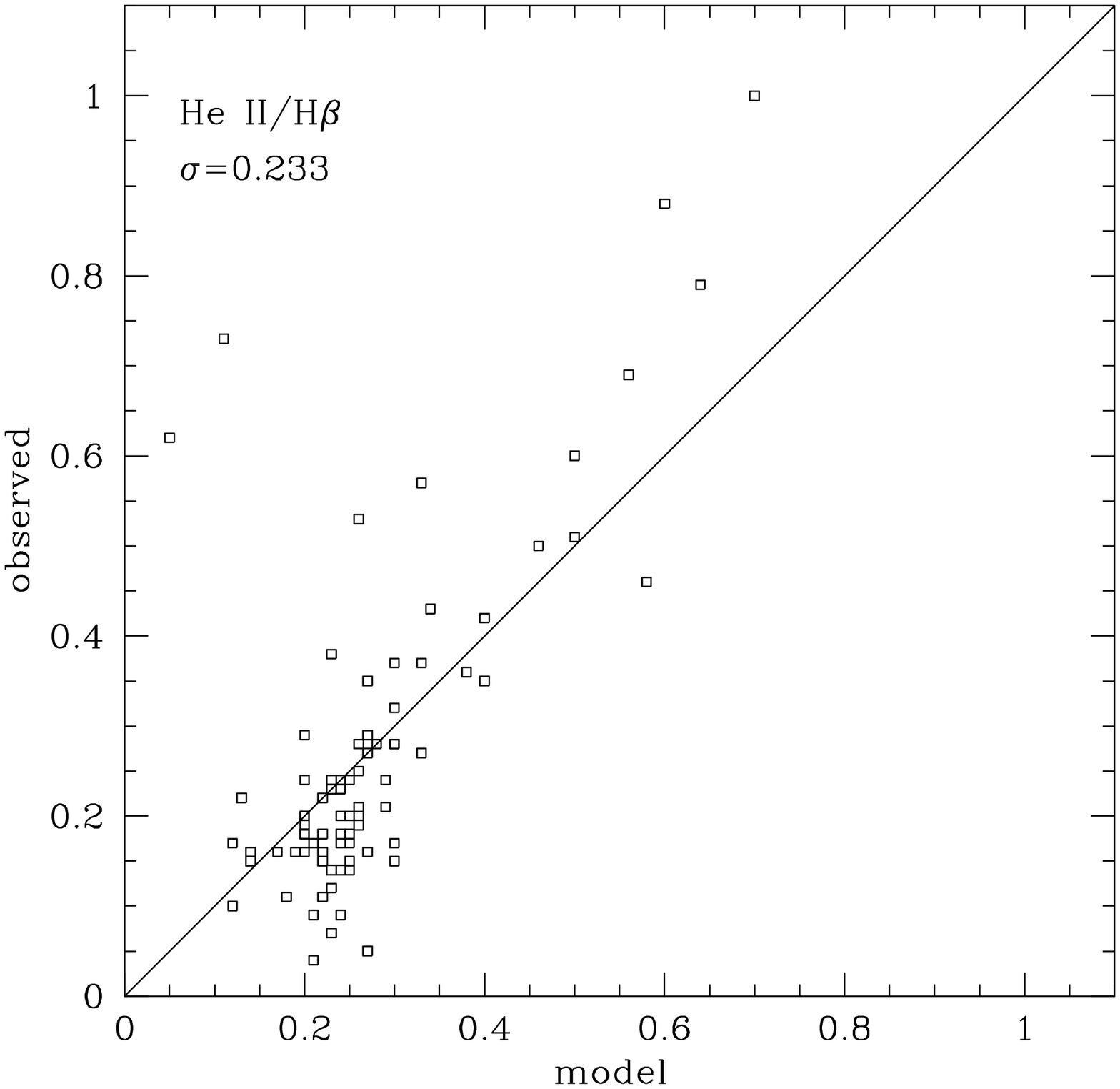}
\includegraphics[width=0.32\textwidth]{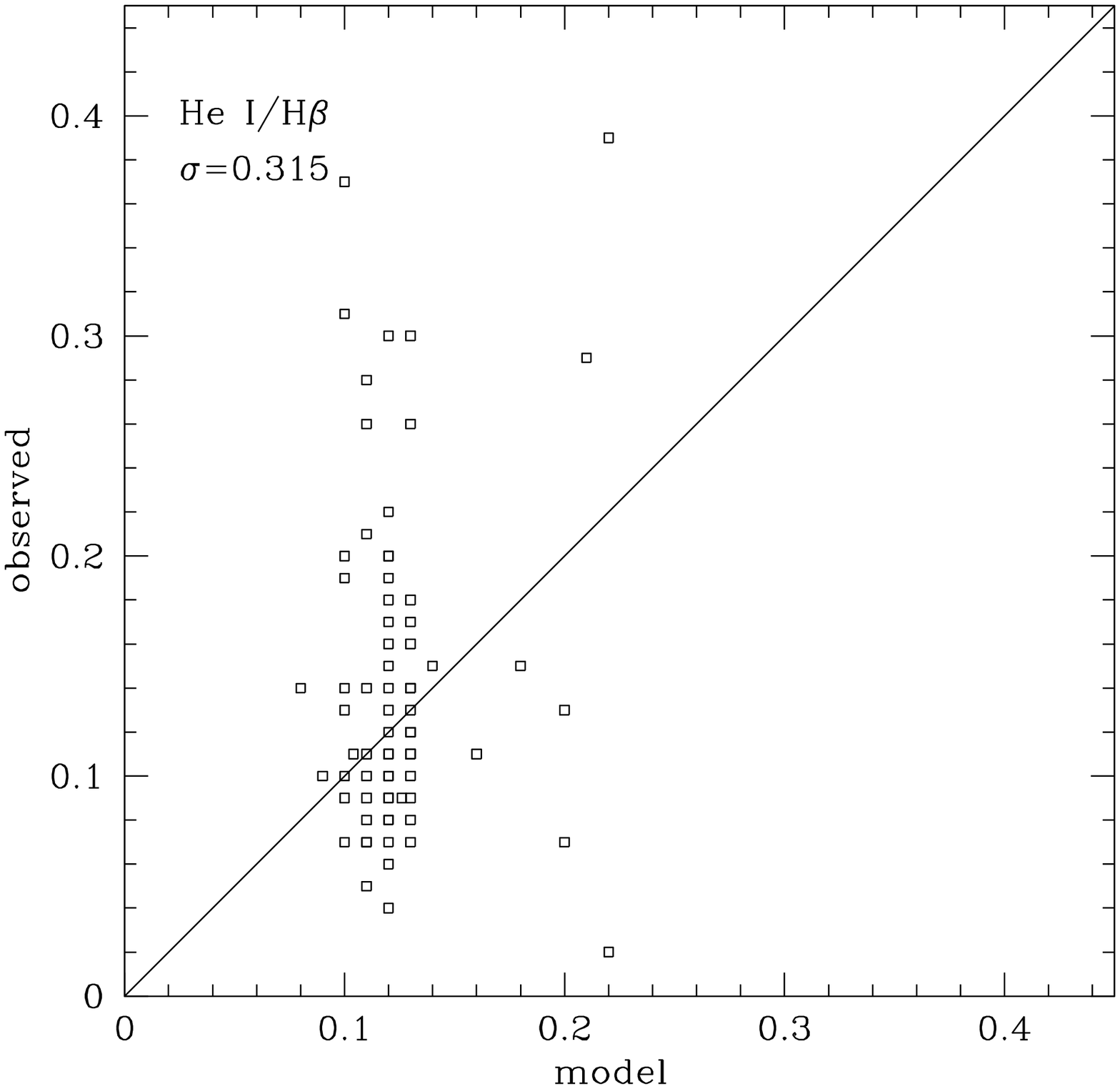}\\
\includegraphics[width=0.32\textwidth]{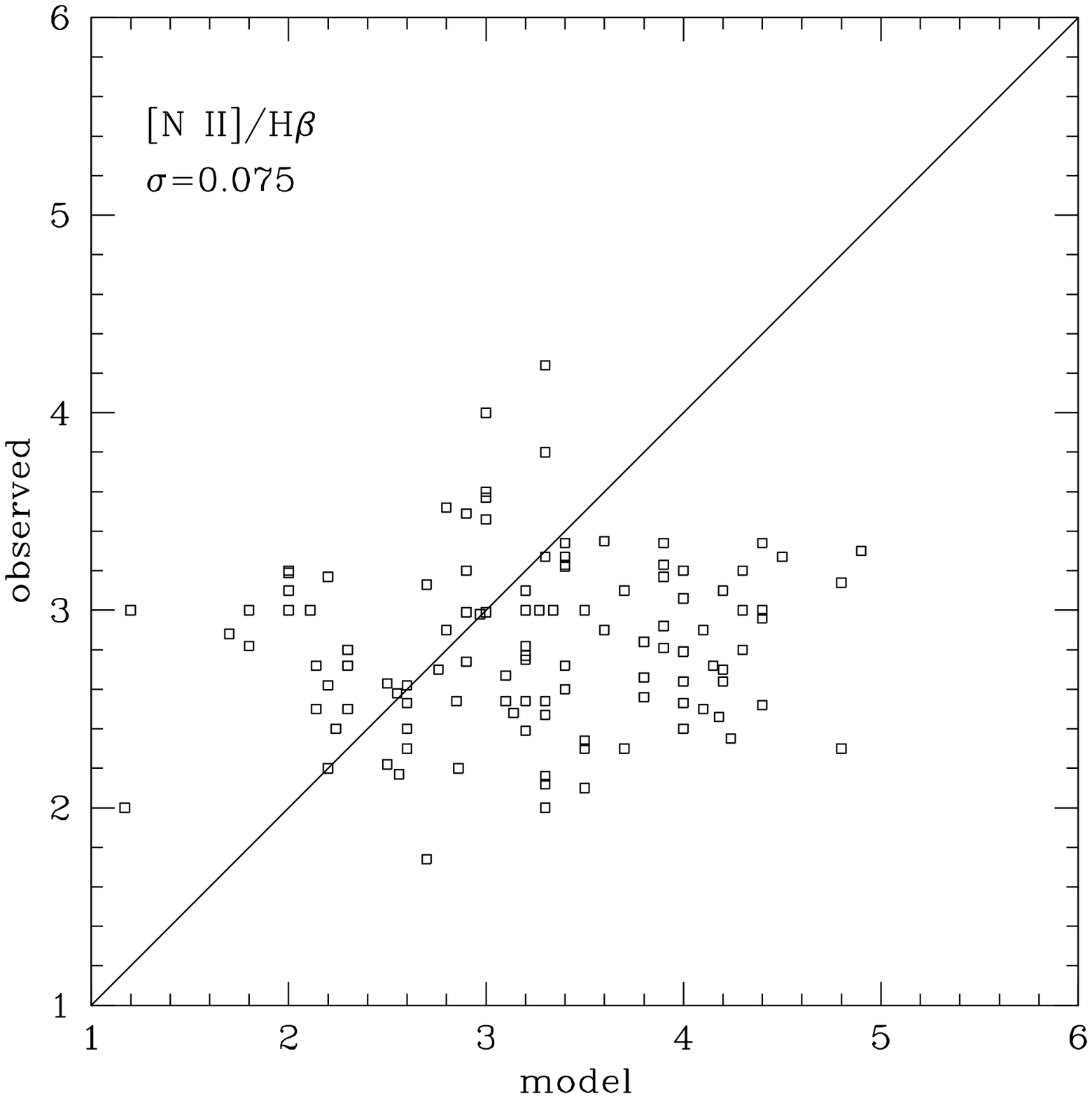}
\includegraphics[width=0.32\textwidth]{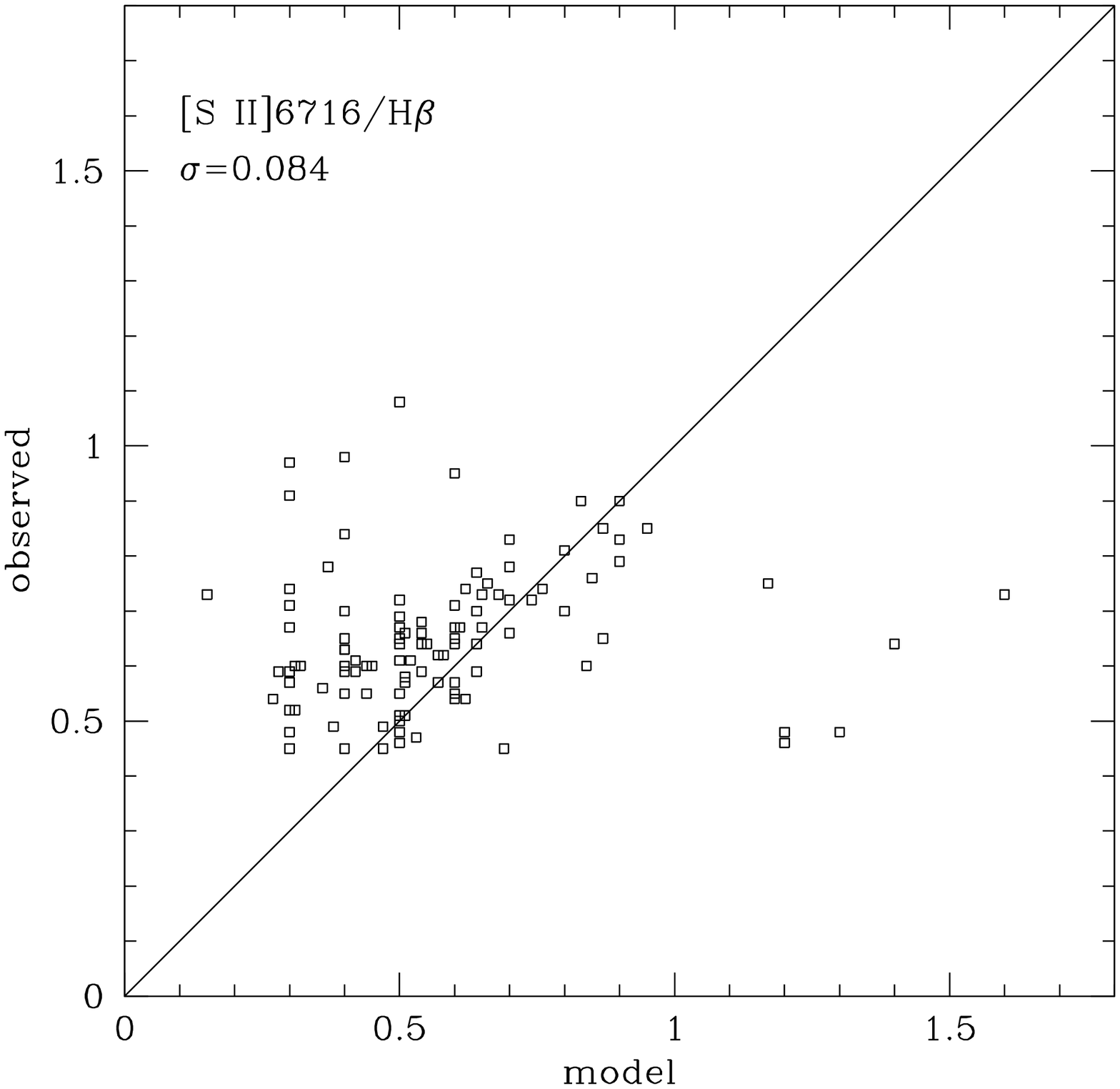}
\includegraphics[width=0.32\textwidth]{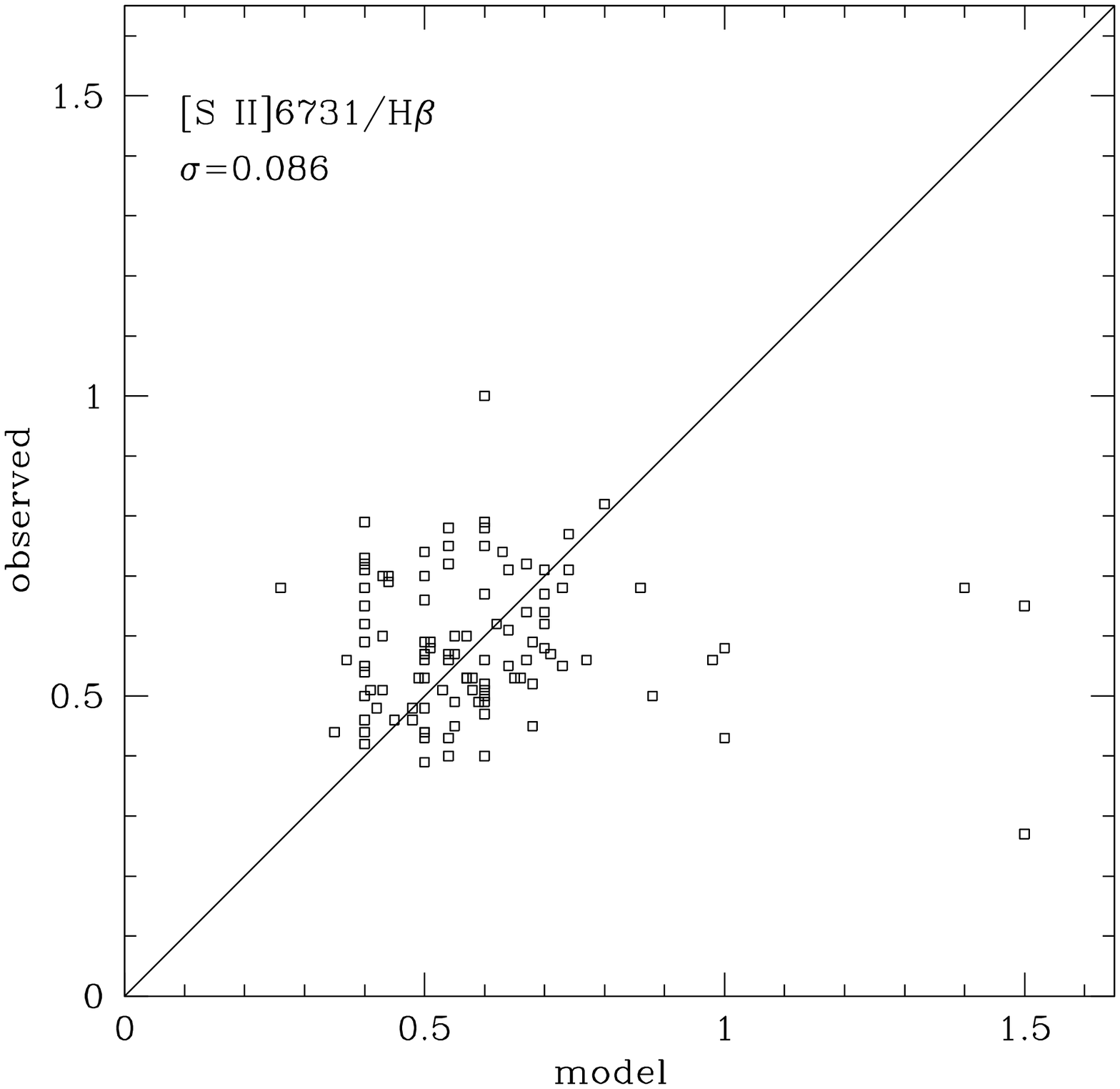}
\caption{Comparison between the observed and the calculated emission line ratios for the lines in Table 2. $\sigma$ is the median error for the observed ratios. The solid line is the 1:1 line.}
\label{fig4a}
\end{figure*}

Table\,\ref{tab2}  and Fig.\,\ref{fig4a} show that the observed  [\ion{O}{III}]/H$\beta$,  [\ion{O}{II}]/H$\beta$ and [\ion{O}{I}]/H$\beta$ are reproduced by the models with relatively high precision (the  discrepancy is $<20$ \%).

The Ne/H relative abundance used in the calculation of the spectra is 10$^{-4}$. 
Comparison with the data shows that Ne/H higher by a factor of 1.5--2 should be adopted. This would approach the $\rm Ne/H = 1.95\times10^{-4}$ solar value calculated by Bahcall et al. (\cite{B05}). 
Table\,\ref{tab2} shows that [\ion{Fe}{VII}]6087/H$\beta$ is not negligible in some regions, in the nucleus and in the regions to the East of the nucleus, even if the iron line is faint and the median error is about 0.3.

The modelled \ion{He}{I}/\Hb and \ion{He}{II}/\Hb ratios are in agreement with the observed values, within their large error bars. [\ion{N}{II}]/\Hb,  [\ion{S}{II}]/\Hb line ratios show  a large scatter (see Fig.\,\ref{fig4a}), which does not depend on the observational errors. We point out that at this stage of modelling, we considered solar abundances.
We will now determine the metallicity relative to N and S, by  constraining in the following
calculations  the relative abundances of N/H and S/H,  in order to obtain a perfect fit of calculated to observed line ratios.

\begin{figure*}
\centering
\includegraphics[width=0.33\textwidth]{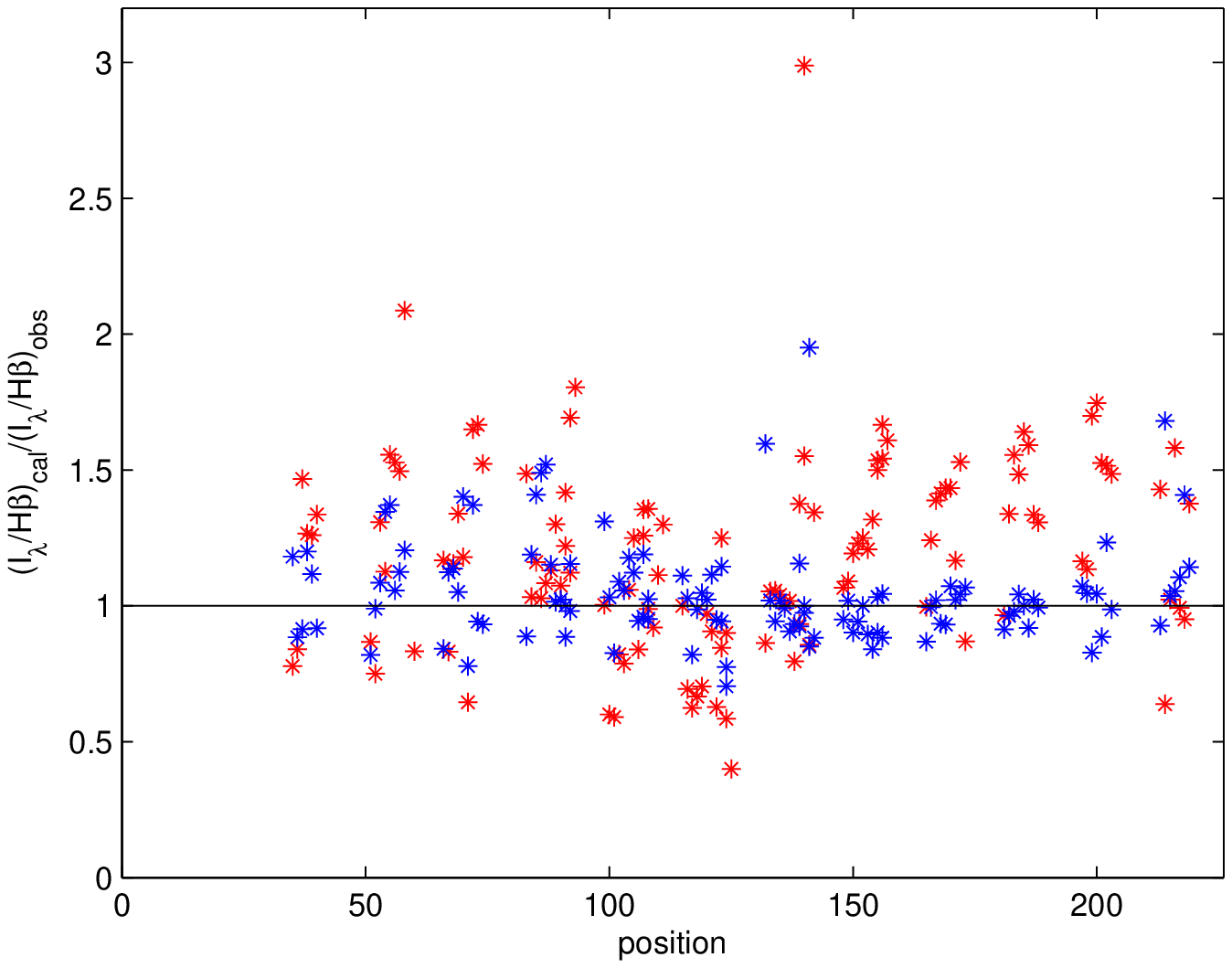}
\includegraphics[width=0.33\textwidth]{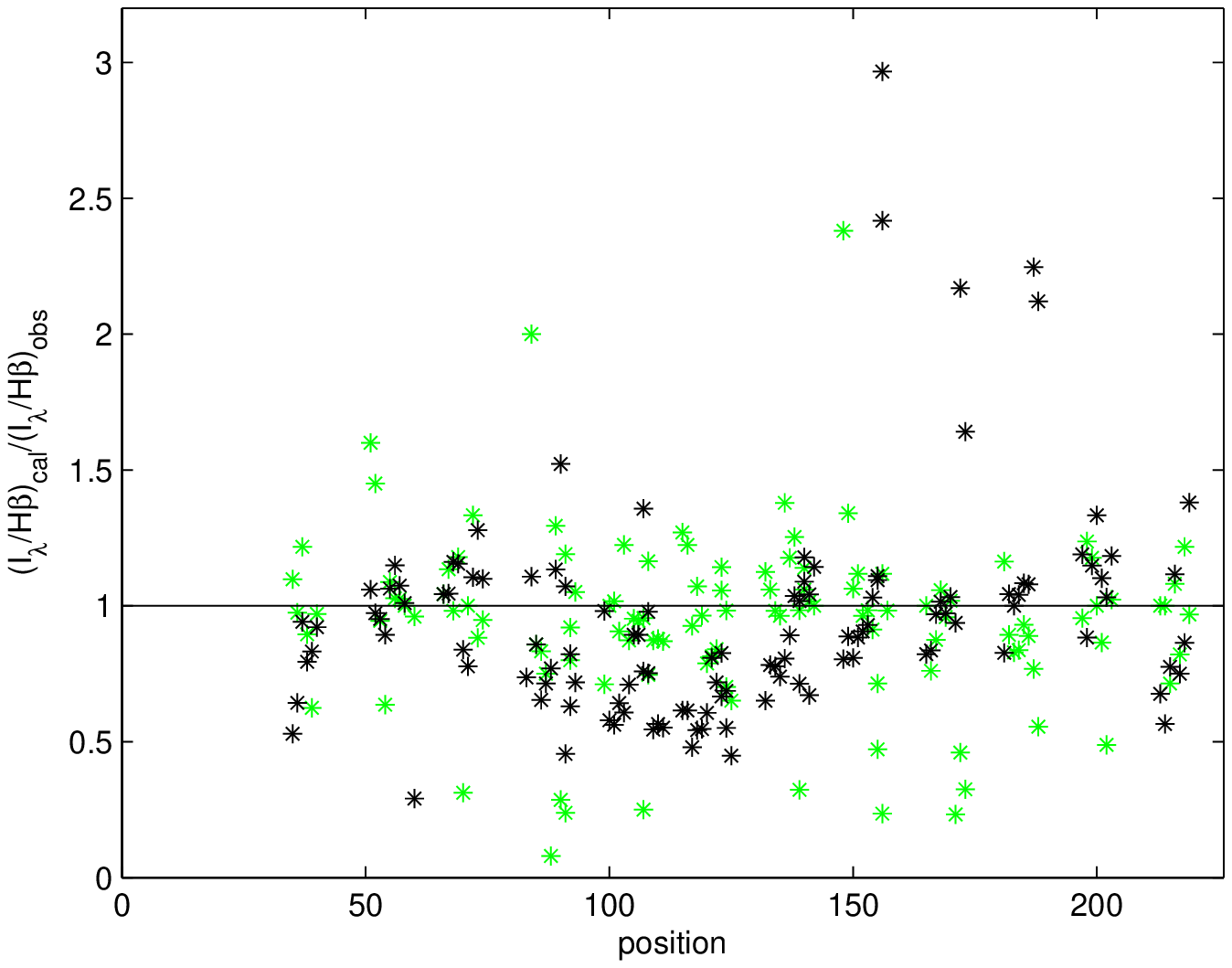}
\includegraphics[width=0.33\textwidth]{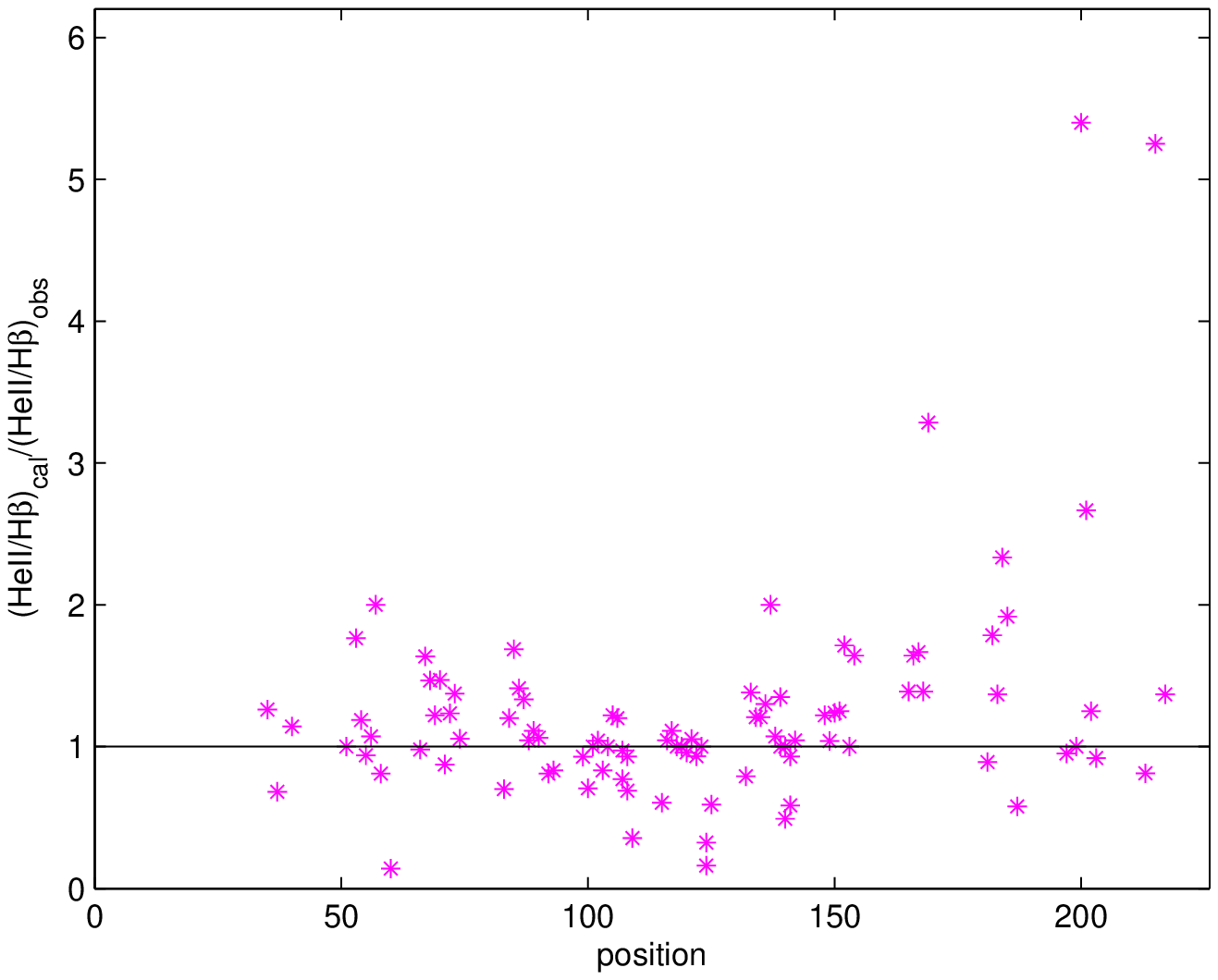}
\caption{Variations  of  calculated/observed line ratios as a function of the number of spectra (see Fig.\,\ref{fig1}). 
Left: [\ion{N}{II}]/H$\beta$ (red asterisks), [\ion{O}{II}]/H$\beta$ (blue asterisks); middle
: [\ion{O}{I}]/H$\beta$ (green asterisks), [\ion{S}{II}]6716+6731/H$\beta$ (black asterisks); right: \ion{He}{II}/H$\beta$ (magenta asterisks). }
\label{fig4b}
\end{figure*}

\begin{table*}
\caption{Physical parameters resulting from modelling the observed spectra.}
\label{tab1}
\begin{center}
\small{
\begin{tabular}{l l l l l l llllll} \\ \hline \hline
\ pos & \Vs\tablefootmark{1} & \n0\tablefootmark{2} &  $F$\tablefootmark{3}   &  $D$\tablefootmark{4}  &  H$\beta$ abs\tablefootmark{5} & pos & \Vs \tablefootmark{1} & \n0\tablefootmark{2} &  $F$\tablefootmark{3}   &  $D$\tablefootmark{4}  &  H$\beta$ abs\tablefootmark{5} \\ 
\hline
\ 35 &  300. & 90. &  20  &  7.8&  0.096       &         123&   750. &   90 &   16 &0.74&  0.04\\        
\ 36 &  300.  &100. &  70 & 9.5&  0.25       &         124&   60.  & 60. &  1 &  860   &0.0047\\       
\ 37 &  160.  &100. &  4  & 0.76&  0.0073     &         124&   600. &   80. &   30 & 54&  0.3\\        
\ 38 &  110. &  110.&   9 & 5.6&  0.02      &         125&   600. &   90. &   30 & 63& 0.39  \\      
\ 39 &  120.  &  110.&   8&  2.9 & 0.0165   &         132&  550. &  60. &  95 & 50   &0.5  \\        
\ 40 &  120. &   110.&   8&   3.5 & 0.018    &         133&  110. &  110. &  10 & 6   & 0.022\\       
\ 51 &  120. &   100.&   12& 230 & 0.04     &          134&  110. &  110. &  10 & 6   & 0.022\\       
\ 52 &  90.  &   130.&   12 &17 & 0.038    &          135&  110. &  110. &  10 & 6   & 0.022\\       
\ 53 &  80.  &   120.&   12& 15 & 0.029    &          136&  110. &  110. &  10 & 13  & 0.025\\       
\ 54 &  110.  &  130.&   10 & 5.  & 0.032    &         137&  100. &  110. &  10 & 30   & 0.026\\       
\ 55 &  140. &   130.&   9 & 2.3  & 0.026    &         138&  100. &  120. &  12 &280  & 0.04\\        
\ 56 &  110. &   110.&   10& 5.6  & 0.022    &         139&  100. &  120. &  10 & 280  & 0.042\\       
\ 57 &  110.  &  100.&   10& 6.8  & 0.02     &         139&   750. &  90.&    13& 0.5  &0.057\\        
\ 58 &  180.  &  100.&  18&3.6   & 0.025    &          140&  90.   & 90. &  11 &73   &0.046\\        
\ 60 &  750.  &   110.&   80&    0.1&  0.13   &         140&   750.  & 80. &   13&0.92  &0.037 \\       
\ 66 &  160. &   100. &  15&4.  &  0.026    &          141&  90.   & 90. &  11& 1000    &0.06\\         
\ 67 &  130. &   100. &  5 & 4.7 &  0.015    &         141&   600.  & 90.&    20&  27.2&0.156 \\       
\ 68 &  110. &   100. & 4.5 & 5.  &  0.01     &        142&  80.   &  90.  & 1  & 2.4  & 0.0023\\      
\ 69 &  110. &   100. & 4.5 & 4.4 &  0.01     &        148&  380.  & 120.  & 33& 3.5  & 0.1\\         
\ 70 &  110. &   110. & 10 &  6. &  0.027    &         149&  110.  & 110. &  9  & 7.6  & 0.021\\       
\ 71 &  550. &    80.&  50 & 9.5 &  0.3      &         150&  220.  & 100. &  10 & 1.6  & 0.02\\        
\ 72 &  220. &   190.& 32 & 1.  &  0.09     &          151&  160.  & 100. &  5  & 1.7  & 0.011\\       
\ 73 &  220.  &  100.&  8  & 1.23&  0.02     &         152&  160.  & 100. &  4  & 1.3  & 0.009\\       
\ 74 &  220. &   110.&  10 & 1.33&  0.025    &         153&  160.  & 100. &  4  & 1.75 & 0.01\\        
\ 83 &  60.  &   100.& 11 &  50 &  0.019    &          154&   70.  &  90. &  0.7  & 2   & 0.0016\\      
\ 84 &  100. &    90.&  5  & 170 &  0.013    &         155&   70.  &  90. &  0.7  & 1.6  & 0.0015\\      
\ 85 &  110. &   100.& 11 & 12 &  0.03     &          155&   750. &  90. &   1 &  0.2  &0.0013\\       
\ 85 &  110. &   100.&  7  & 5.5 &  0.02     &         156&   50.  & 100. &  0.07  & 0.15  & 0.00009\\       
\ 86 &  320. &   100.&   60& 7  &  0.23     &         156&   600. &  60.&   0.04 &  0.5  &0.0003 \\       
\ 87 &  320. &   100.&  15& 7  &  0.28     &          157&   650.  & 60.&    0.05 &  0.5  &0.0003 \\       
\ 88 &  280. &   120.&  15& 7  &  0.28     &          165&  110.  & 110. & 13 &13   &0.039\\        
\ 89 &  120. &   150.&  13& 7  &  0.04     &          166&  220.  & 100. &  8  & 1.2  & 0.019\\       
\ 90 &  80.  &    90.&   0.3& 2.7 &  0.0001    &        167&  160.  & 100. &  5  & 1.4  & 0.011\\       
\ 91 &  120. &   100.&  5  & 5  &  0.013    &         168&  160.  & 100. &  4  & 1.2  & 0.0087\\      
\ 91 &  800. &   100.& 13 & 0.24 & 0.028     &          169&  160.  & 100. &  3.5 & 1.1  & 0.0085\\      
\ 92 &  120. &   100.&  4  & 4.7 &  0.013    &         170&   70.  &  90. &  0.7  & 1.7  & 0.0016\\      
\ 92 &  700. &  100. &  13&  0.33&  0.039    &          171&   70.  & 100. &  0.25 & 0.09   & 0.000135\\     
\ 93 &  650. &   90. &   13& 0.49&  0.036    &          172&   50.  & 100. &  0.07  & 0.12  & 0.000067\\      
\ 99 &  120. &   110.&  8  & 3.8 &  0.018    &         173&   80.  &  70. &  0.05  & 0.15  & 0.0001\\      
\ 100&  550. &    80.&  45& 10  &  0.28     &          181&  300.  & 120. &  26& 1.7  & 0.07\\        
\ 101&  550. &    80.&  70& 22  &  0.45     &         182&  160.  & 100. &  5  & 1.4  & 0.011\\       
\ 102&  550. &   70. &  50 & 8.3 & 0.2       &         183&  120.  & 100. &  3  & 1.5  & 0.006\\       
\ 103&  550. &   70. &  50 & 12.4& 0.22      &         184&  160.  & 100. &  4  & 1.2  & 0.001\\       
\ 104&  380. &  120. &  33& 1.5 &  0.11     &          185&  80.   &  90. &  1  & 1.6  & 0.0022\\      
\ 105&  180. &  180. &  20 & 1.  &  0.05     &         186&  70.   &  90. &  0.7  & 1.5  & 0.0015\\      
\ 106&  110. &  110. &  10 & 13 &  0.04     &         187& 180.   &  70. &  0.03  & 0.2   & 0.00019 \\     
\ 107&  50.  &  100. &  0.5  & 2.2 &  0.001    &       188& 180.   &  70. &  0.06  & 0.14  & 0.00011 \\     
\ 107&   800.  &   100.&   3&   0.156& 0.002  &         197&  110.  & 110. &  9   & 6.1  & 0.021\\       
\ 108&  100. &  110. &  9 &  280 &  0.033    &         198&  180.  & 160. &  20  & 1.6  & 0.051\\       
\ 108&   750. &    90. &    16& 0.58&  0.07  &          199&  90.   & 100. &  1.6  & 1.2  & 0.003\\       
\ 109&   750.  &   90. &   16& 0.58&  0.07   &          200&  70.   &  90. &  0.9  & 2.7  & 0.0018\\      
\ 110&   750.  &   90. &   13& 0.65 & 0.06   &          201&  70.   &  90. &  0.7  & 1.3  & 0.0015\\      
\ 111&   750.  &   90. &   10 & 0.4  & 0.04   &         202&  100.  &  100.&  0.8 & 0.4   & 0.0017\\      
\ 115&  600.  &  80. &  65 &12  & 0.36     &          203&  160.  &  60. &  0.9  & 1   & 0.002 \\      
\ 116&  550.  &  80. &  60  &7   & 0.28     &         213&  300. &   120.&  30  &  1.8 & 0.05  \\      
\ 117&  550.  &  80. &  50  &9.5  & 0.3      &         214&  500.  &   80.&  40  & 7   & 0.225\\       
\ 118&  550.  &  80. &  60  &11.3 & 0.3      &         215&  120.   & 150.&  13  & 2.7  & 0.037\\       
\ 119&  550.  &  80. &  70  &9.2  & 0.32     &         216&  160.  &  100.&  3.5 & 1   & 0.0085\\      
\ 120&  380.  & 120. &  33 &1.5  & 0.11     &          217&  90.   &   90.&  11   & 45  & 0.0045\\      
\ 121&  120.  & 150. &  13& 4  & 0.04     &          218&  100.  &   80.&  7     & 20   & 0.023\\       
\ 122&  130.  &  80. &  6  & 37  & 0.028    &         219&  90.   & 120.  & 0.75 & 0.73  & 0.002\\ 
\ 123&  110.  & 110. &  10 & 15  & 0.04 & &&&&&\\ \hline 
\end{tabular}}
\tablefoot{
\tablefoottext{1}{in \kms;}
\tablefoottext{2}{in \cm3;}
\tablefoottext{3}{in $10^9$ photons cm$^{-2}$ s$^{-1}$ eV$^{-1}$ at the Lyman limit;}
\tablefoottext{4}{in $10^{17}$ cm;}   
\tablefoottext{5}{in \erg.}}
\end{center}
\end{table*}

\subsection{Second step: relative abundances}

The  ratios of calculated to observed [\ion{S}{II}]/H$\beta$ and [\ion{N}{II}]/H$\beta$ in each position are shown in Table\,\ref{tab2}.
Unfortunately there are no data for S lines from higher ionization levels, which could indicate whether the choice of the  model is misleading, or different relative abundances should be adopted.
Since sulphur can be easily depleted from the gaseous phase and trapped into dust grains and molecules, we considered that the [\ion{S}{II}]/H$\beta$ line ratios indicate the S/H relative abundance.

The same is valid for the [\ion{N}{II}] lines.
The observations do not contain any other  strong line of N, which could confirm  whether the  discrepancy between the calculated and the observed [\ion{N}{II}]/H$\beta$ line ratios can be resolved by changing the N/H relative abundance or the other input physical parameters.
Solar nitrogen is less abundant than other solar elements e.g. oxygen, neon, etc., therefore it is not an efficient coolant, namely, it does not affect strongly the cooling rate of the gas downstream.
Varying N/H will change only the [\ion{N}{II}]/H$\beta$ line ratio, but not the other line ratios, while changing O/H the whole input parameter set must be readjusted.

\begin{table*}
\caption{Comparison between calculated and observed line ratios.}
\label{tab2}
\begin{tabular}{p{0.3cm}p{0.35cm}p{0.35cm}p{0.35cm}p{0.35cm}p{0.35cm}p{0.35cm}p{0.35cm}p{0.35cm}p{0.35cm}p{0.35cm}p{0.35cm}p{0.35cm}p{0.35cm}p{0.35cm}p{0.35cm}p{0.35cm}p{0.35cm}p{0.35cm}p{0.35cm}p{0.35cm}p{0.35cm}p{0.35cm}}
\hline
  \multicolumn{1}{c}{\phantom{n}} &
  \multicolumn{2}{c}{[\ion{O}{II}]} &
  \multicolumn{2}{c}{[\ion{Ne}{III}]} &
  \multicolumn{2}{c}{\ion{He}{II}} &
  \multicolumn{2}{c}{[\ion{Ar}{IV}]} &
  \multicolumn{2}{c}{[\ion{O}{III}]} &
  \multicolumn{2}{c}{\ion{He}{I}} &
  \multicolumn{2}{c}{[\ion{Fe}{VII}]} &
  \multicolumn{2}{c}{[\ion{O}{I}]} &
  \multicolumn{2}{c}{[\ion{N}{II}]} &
  \multicolumn{2}{c}{[\ion{S}{II}]} &
  \multicolumn{2}{c}{[\ion{S}{II}]} \\
  \multicolumn{1}{c}{\phantom{n}}&
  \multicolumn{2}{c}{3727+} &
  \multicolumn{2}{c}{3869+} &
  \multicolumn{2}{c}{4686 } &
  \multicolumn{2}{c}{4713 } &
  \multicolumn{2}{c}{5007+} &
  \multicolumn{2}{c}{5876 } &
  \multicolumn{2}{c}{6087 } &
  \multicolumn{2}{c}{6300+} &
  \multicolumn{2}{c}{6583+} &
  \multicolumn{2}{c}{6718 } &
  \multicolumn{2}{c}{6731 } \\
  \multicolumn{1}{c}{n}&
  \multicolumn{2}{c}{Obs}{Mod} &
  \multicolumn{2}{c}{Obs}{Mod} &
  \multicolumn{2}{c}{Obs}{Mod} &
  \multicolumn{2}{c}{Obs}{Mod} &
  \multicolumn{2}{c}{Obs}{Mod} &
  \multicolumn{2}{c}{Obs}{Mod} &
  \multicolumn{2}{c}{Obs}{Mod} &
  \multicolumn{2}{c}{Obs}{Mod} &
  \multicolumn{2}{c}{Obs}{Mod} &
  \multicolumn{2}{c}{Obs}{Mod} &
  \multicolumn{2}{c}{Obs}{Mod} \\
\hline
  35 & 2.2 & 2.6 & 1.23 & 1.7  & 0.46 & 0.58 & 0.0 & 0.07 & 7.43 & 7.44& 0.08 & 0.11 & 0.0 & 0.08 & 0.41 & 0.45 & 4.24 & 3.3 & 1.08 & 0.5 & 1.0 & 0.6 \\
  36 & 5.2 & 4.6 & 2.26 & 1.1 & 0.0 & 0.3 & 0.0 & 0.04 & 9.64 & 10.5 & 0.2 & 0.1 & 0.0 & 0.01 & 0.4 & 0.39 & 3.57 & 3.0 & 0.7 & 0.4 & 0.7 & 0.5\\
  37 & 7.65 & 7.0 & 4.12 & 3.   & 0.88 & 0.6  & 0.0 & 0.03 & 14.6 & 15.  & 0.19 & 0.1  & 0.0 & 0.01 & 0.23 & 0.28 & 3.0 & 4.4 & 0.62 & 0.57 & 0.57 & 0.55\\
  38 & 5.0 & 6.0 & 0.0 & 1.38 & 0.0 & 0.27 & 0.0 & 0.06 & 13.4 & 13.3 & 0.09 & 0.11 & 0.0 & 0.05 & 0.38 & 0.34 & 2.48 & 3.14 & 0.67 & 0.5 & 0.59 & 0.5\\
  39 & 6.0 & 6.7 & 1.78 & 1.6 & 0.0 & 0.28 & 0.0 & 0.07 & 16.76 & 16.0 & 0.11 & 0.11 & 0.0 & 0.04 & 0.32 & 0.2 & 2.54 & 3.2 & 0.6 & 0.45 & 0.53 & 0.49\\
  40 & 7.19 & 6.6 & 0.0  & 1.5 & 0.35 & 0.40 & 0.03 & 0.07 & 14.62 & 14.7 & 0.0 & 0.12 & 0.0 & 0.03 & 0.33 & 0.32 & 2.47 & 3.3 & 0.57 & 0.51 & 0.57 & 0.54\\
  51 & 5.37 & 4.4 & 2.25 & 1.0 & 0.23 & 0.23 & 0.0 & 0.03 & 8.0 & 8.4 & 0.3 & 0.12 & 0.0 & 0.04 & 0.5 & 0.8 & 3.46 & 3.0 & 0.67 & 0.65 & 0.51 & 0.6\\
  52 & 4.75 & 4.7 & 2.1 & 1.0 & 0.0 & 0.24 & 0.0 & 0.04 & 9.8 & 9.0 & 0.28 & 0.11 & 0.0 & 0.05 & 0.4 & 0.58 & 4.0 & 3.0 & 0.59 & 0.54 & 0.51 & 0.53\\
  53 & 5.9 & 6.4 & 2.4 & 1.3 & 0.17 & 0.3 & 0.0 & 0.04 & 11.43 & 11.5 & 0.07 & 0.1 & 0.0 & 0.09 & 0.36 & 0.34 & 2.6 & 3.4 & 0.58 & 0.51 & 0.46 & 0.48\\
  54 & 3.12 & 4.2 & 1.37 & 1.0 & 0.16 & 0.19 & 0.06 & 0.04 & 10.1 & 9.0 & 0.0 & 0.12 & 0.0 & 0.03 & 0.33 & 0.21 & 2.22 & 2.5 & 0.55 & 0.44 & 0.48 & 0.48\\
  55 & 4.74 & 6.5 & 1.58 & 1.5 & 0.32 & 0.3  & 0.05 & 0.08 & 13.23 & 14.7 & 0.1 & 0.11 & 0.0 & 0.06 & 0.34 & 0.37 & 2.12 & 3.3  & 0.51 & 0.5  & 0.43 & 0.5\\
  56 & 6.15 & 6.5 & 2.7 & 1.5 & 0.28 & 0.3 & 0.0 & 0.08 & 14.1 & 14.7 & 0.0 & 0.11 & 0.0 & 0.06 & 0.36 & 0.37 & 2.16 & 3.3 & 0.48 & 0.5 & 0.39 & 0.5\\
  57 & 6.4 & 7.2 & 3.0 & 1.7 & 0.15 & 0.3 & 0.09 & 0.09 & 16.36 & 16.0 & 0.13 & 0.1 & 0.0 & 0.09 & 0.39 & 0.4 & 2.34 & 3.5 & 0.51 & 0.51 & 0.43 & 0.5\\
  58 & 4.15 & 5.0 & 2.77 & 3.  & 0.79 & 0.64& 0.0 & 0.16 & 13.0 & 13.1 & 0.1 & 0.09 & 0.0 & 0.18 & 0.42 & 0.42& 2.3 & 4.8 & 0.55 & 0.5  & 0.49 & 0.55\\
  60 & 0.0 & 4.  & 0.0 & 2.  & 1.7 & 0.24 & 0.0 & 0.0 & 21.9 & 21.8 & 0.8 & 0.12 & 0.0 & 0.1   & 0.5 & 0.48 & 3.6 & 3.  & 0.73 & 0.15 & 0.68 & 0.26\\
  66 & 9.5 & 8.0 & 3.6 & 2.0 & 0.51 & 0.5 & 0.0 & 0.1 & 17.0 & 18.8 & 0.31 & 0.1 & 0.0 & 0.1 & 0.46 & 0.48 & 3.34 & 3.9 & 0.49 & 0.47 & 0.44 & 0.5\\
  67 & 4.45 & 5.0 & 1.59 & 1.0 & 0.11 & 0.18 & 0.03 & 0.03 & 8.37 & 8.8 & 0.16 & 0.13 & 0.0 & 0.008 & 0.37 & 0.42 & 3.49 & 2.9 & 0.57 & 0.57 & 0.53 & 0.58\\
  68 & 5.54 & 6.3 & 3.1 & 1.4 & 0.15 & 0.22& 0.0 & 0.04 & 11.5 & 11.8 & 0.14 & 0.13 & 0.0 & 0.014 & 0.55 & 0.54 & 2.77 & 3.2 & 0.54 & 0.62 & 0.51 & 0.6\\
  69 & 5.9 & 6.2 & 2.54 & 1.3 & 0.18 & 0.22 & 0.0 & 0.04 & 11.27 & 11.8 & 0.16 & 0.12 & 0.0 & 0.014 & 0.39 & 0.46 & 2.39 & 3.2 & 0.54 & 0.6 & 0.49 & 0.59\\
  70 & 3.14 & 4.4 & 1.63 & 1.0 & 0.17 & 0.25 & 0.04 & 0.05 & 11.5 & 10.0 & 0.1 & 0.11 & 0.0 & 0.04 & 0.32 & 0.1 & 2.17 & 2.56 & 0.49 & 0.38 & 0.44 & 0.4\\
  71 & 2.57 & 2.0 & 1.35 & 1.  & 0.16 & 0.14 & 0.05 & 0.05 & 12.47 & 12.2 & 0.1 & 0.12 & 0.0 & 0.04 & 0.3 & 0.3 & 3.1 & 2.0 & 0.48 & 0.3 & 0.42 & 0.4\\
  72 & 3.5 & 4.8 & 2.8 & 1.44 & 0.17 & 0.21 & 0.05 & 0.04 & 12.6 & 12.  & 0.08 & 0.12 & 0.0 & 0.02 & 0.3 & 0.4 & 2.0 & 3.3 & 0.45 & 0.4  & 0.4 & 0.54\\
  73 & 6.9 & 6.5 & 3.54 & 1.4 & 0.16 & 0.22 & 0.06 & 0.034 & 12.8 & 11.7 & 0.06 & 0.12 & 0.0 & 0.015 & 0.34 & 0.3 & 2.1 & 3.5 & 0.46 & 0.5 & 0.4 & 0.6\\
  74 & 6.76 & 6.3  & 3.41 & 1.4 & 0.36 & 0.38& 0.08 & 0.04 & 11.0 & 11.8 & 0.18 & 0.12 & 0.0 & 0.02 & 0.38 & 0.36& 2.3 & 3.5 & 0.5 & 0.5  & 0.5 & 0.6 \\
  83 & 12.4 &11.  & 3.8 & 2.  & 1.0 & 0.7  & 0.16 & 0.11 & 15.4 & 15.1 & 0.14 & 0.08& 0.0 & 0.17 & 0.42 & 1.7 & 2.96 & 4.4 & 0.6 & 0.44 & 0.54 & 0.4\\
  84 & 4.88 & 5.8 & 1.69 & 1.4 & 0.2 & 0.24& 0.11 & 0.06 & 11.41 & 12.2 & 0.13 & 0.2 & 0.03 & 0.03 & 0.5 &   1.& 3.1 & 3.2 & 0.76 & 0.85& 0.64 & 0.7\\
  85 & 3.76 & 5.3 & 1.4 & 1.14 & 0.16 & 0.27 & 0.0 & 0.04 & 11.0 & 10.7 & 0.1 & 0.1  & 0.0 & 0.06 & 0.49 & 0.42 & 2.67 & 3.1 & 0.61 & 0.52 & 0.59 & 0.51\\
  86 & 2.55 & 3.8 & 1.38 & 1.   & 0.17 & 0.24 & 0.02 & 0.05 & 11.85 & 11.0 & 0.09 & 0.1  & 0.01 & 0.09 & 0.42 & 0.35 & 2.53 & 2.6 & 0.52 & 0.3  & 0.55 & 0.4 \\
  87 & 2.5 & 3.8 & 1.3 & 1.   & 0.18 & 0.24 & 0.04 & 0.05 & 13.5 & 11.  & 0.10& 0.1  & 0.02 & 0.09 & 0.4 & 0.3  & 2.4 & 2.6 & 0.48 & 0.3 & 0.5 & 0.4\\
  88 & 3.3 & 3.8 & 1.76 & 1.   & 0.23 & 0.24& 0.05 & 0.05 & 14.0 & 11.  & 0.1 & 0.1  & 0.01 & 0.01 & 0.38 & 0.03 & 2.3 & 2.6  & 0.45 & 0.3  & 0.46 & 0.4 \\
  89 & 4.63 & 4.7  & 2.1 & 1.12& 0.18 & 0.2  & 0.02 & 0.05 & 11.6 & 10.3 & 0.08 & 0.12 & 0.0 & 0.02 & 0.34 & 0.44& 2.2 & 2.86& 0.45 & 0.47 & 0.45 & 0.55\\
  90 & 13.0 & 13.3 & 4.48 & 2.0 & 0.16 & 0.17& 0.09 & 0.07 & 11.6 & 13.0 & 0.13 & 0.12 & 0.04 & 0.01 & 0.35 & 0.1 & 3.35 & 3.6 & 0.45 & 0.69& 0.45 & 0.68\\
  91 & 5.87 & 5.2 & 1.02 & 1.26 & 0.0 & 0.2 & 0.0 & 0.01 & 9.18 & 9.0 & 0.09 & 0.12 & 0.0 & 0.03 & 0.42 & 0.1 & 2.54 & 3.6 & 0.65 & 0.18 & 0.47 & 0.33\\
  91 & 5.87 & 5.88& 1.02 & 1.2 & 0.0 & 0.2  & 0.0 & 0.03 & 9.18 &10.9 & 0.09 & 0.126& 0.0 & 0.01  & 0.42 & 0.5 & 2.54 & 3.1  & 0.65 & 0.6 & 0.47 & 0.6\\
  92 & 5.2 & 6.0 & 1.75 & 1.33 & 0.37 & 0.3  & 0.0 & 0.012 & 7.5 & 8.1 & 0.1 & 0.13 & 0.06 & 0.02 & 0.5 & 0.46 & 2.54 & 4.3 & 0.95 & 0.34 & 0.51 & 0.58\\
  92 & 5.2 & 5.1 & 1.75 & 1.  & 0.37 & 0.3 & 0.0 & 0.02 & 7.5 & 8.   & 0.1 & 0.13 & 0.06 & 0.01  & 0.5 & 0.4  & 2.54 & 2.85 & 0.95 & 0.6  & 0.51 & 0.6 \\
  93 & 0.0 & 6.3 & 0.0 & 1.5 & 0.6 & 0.5  & 0.0 & 0.02 & 10.6 & 10.7 & 0.12 & 0.13 & 0.0 & 0.018& 0.59 & 0.62& 2.35 & 4.24& 0.78 & 0.37 & 0.53 & 0.57\\
  99 & 4.5 & 5.9  & 1.34 & 1.35 & 0.28 & 0.26 & 0.08 & 0.06 & 13.9 & 13.  & 0.15 & 0.12  & 0.0 & 0.03 & 0.45 & 0.32 & 2.99 & 3.  & 0.51 & 0.51& 0.56 & 0.54\\
  100 & 1.89 & 1.95& 0.87 & 0.95& 0.17 & 0.12 & 0.05 & 0.05 & 10.0 & 10.8 & 0.13 & 0.12 & 0.05 & 0.03 & 0.52 & 0.52& 3.0 & 1.8 & 0.59 & 0.3 & 0.62 & 0.4\\
  101 & 2.42 & 2.  & 1.34 & 1.1 & 0.2 & 0.2 & 0.03 & 0.05 & 12.1 & 11.8 & 0.1 & 0.12 & 0.04 & 0.05 & 0.59 & 0.6  & 2.88 & 1.7  & 0.6 & 0.32 & 0.68 & 0.4 \\
  102 & 2.48 & 2.7 & 1.5 & 1.4 & 0.24 & 0.25& 0.0 & 0.08 & 15.5 & 17.0 & 0.09 & 0.12 & 0.03 & 0.07 & 0.53 & 0.48 & 2.8 & 2.3  & 0.54 & 0.27 & 0.66 & 0.5 \\
  103 & 2.46 & 2.6 & 1.57 & 1.4  & 0.24 & 0.2  & 0.05 & 0.08 & 16.1 & 16.4 & 0.11 & 0.12 & 0.02 & 0.07 & 0.49 & 0.6  & 2.72 & 2.14& 0.52 & 0.31& 0.65 & 0.4 \\
  104 & 3.4 & 4.  & 1.91 & 1.45 & 0.23 & 0.23 & 0.04 & 0.05 & 15.3 & 14.8 & 0.1 & 0.12 & 0.01 & 0.03 & 0.47 & 0.41& 2.74 & 2.9 & 0.63 & 0.4  & 0.51 & 0.41\\
  105 & 4.9 & 5.5 & 2.06 & 1.44 & 0.18 & 0.22 & 0.02 & 0.04  & 11.72 &12.5  & 0.09 & 0.12 & 0.02 & 0.02 & 0.42 & 0.4  & 2.72 & 3.4 & 0.61 & 0.42& 0.51 &0.58\\
  106 & 3.6 & 3.4 & 1.25 & 0.73& 0.1 & 0.12 & 0.05 & 0.033& 7.37 & 7.17& 0.08 & 0.11 & 0.0 & 0.03  & 0.36 & 0.34 & 2.62 & 2.2 & 0.55 & 0.5  & 0.57 & 0.5 \\
  107 & 7.4 & 7.1 & 0.94 & 1.5 & 0.35 & 0.34 & 0.09 & 0.06 & 8.3 & 8.4 & 0.17 & 0.125 & 0.0 & 0.4 & 0.44 & 0.11 & 3.1 & 3.9 & 0.6 & 0.3 & 0.52 & 0.55\\
  107 & 7.4 & 8.8 & 0.94 & 1.2 & 0.35 & 0.27& 0.09 & 0.02 & 8.3 & 7.8 & 0.17 & 0.12 & 0.0 & 0.003& 0.44 & 0.42 & 3.1 & 4.2  & 0.6 & 0.84& 0.52 & 0.68\\
  108 & 4.1 & 4.2 & 1.16 & 0.97 & 0.29 & 0.27 & 0.0 & 0.01 & 7.1 & 6.7 & 0.17 & 0.14 & 0.0 & 0.01 & 0.67 & 0.5  & 2.58 & 3.5  & 0.83 & 0.4 & 0.5 & 0.6\\
  108 & 4.1 & 3.9 & 1.16 & 0.9  & 0.29 & 0.2  & 0.0 & 0.03 & 7.1 & 7.7 & 0.02 & 0.22 & 0.0 & 0.03 & 0.67 & 0.78 & 2.58 & 2.55 & 0.83 & 0.7 & 0.5 & 0.6\\
  109 & 0.0 & 3.3 & 0.0 & 0.1  & 0.73 & 0.11& 0.0 & 0.01 & 7.48 & 7.9 & 0.08 & 0.13 & 0.0 & 0.01  & 0.63 & 0.55 & 2.5 & 2.3  & 0.91 & 0.3 & 0.56 & 0.5\\
  110 & 0.0 & 4.0 & 0.0 & 0.75 & 0.0 & 0.10 & 0.14 & 0.01 & 4.6 & 4.8 & 0.0 & 0.14 & 0.0 & 0.018 & 0.41 & 0.36& 3.0 & 3.34& 0.98 & 0.4 & 0.79 & 0.6 \\
  111 & 0.0 & 4.  & 0.0 & 0.75& 0.0 & 0.11& 0.12 & 0.01 & 4.0 & 4.   & 0.0 & 0.14 & 0.0 & 0.02 & 0.23 & 0.2 & 2.54 & 3.3 & 0.71 & 0.3 & 0.74 & 0.5\\
  115 & 1.26 & 1.4 & 0.0 & 1.7  & 0.38 & 0.23 & 0.07 & 0.04 & 10.4 &10.4  & 0.2 & 0.12 & 0.02 & 0.07 & 0.37 & 0.47 & 2.2 & 2.2 & 0.84 & 0.4  & 0.51 & 0.43\\
  116 & 1.8 & 1.85& 0.48 & 1.6  & 0.23 & 0.24 & 0.0 & 0.05 & 12.31 & 12.4 & 0.0 & 0.12 & 0.0 & 0.07 & 0.49 & 0.6 & 3.17 & 2.2 & 0.65 & 0.4 & 0.7 & 0.43\\
  117 & 2.56 & 2.1 & 1.03 & 1.05 & 0.18 & 0.2  & 0.03 & 0.05 & 13.0 & 12.2 & 0.04 & 0.12 & 0.03 & 0.04 & 0.54 & 0.5 & 3.2 & 2.0 & 0.67 & 0.3 & 0.79 & 0.4\\
  118 & 2.33 & 2.3  & 1.28 & 1.27 & 0.22 & 0.22 & 0.04 & 0.06 & 15.65 & 14.7 & 0.09 & 0.12 & 0.04 & 0.06 & 0.56 & 0.6  & 3.0 & 2.   & 0.57 & 0.3  & 0.72 & 0.4 \\
  119 & 2.32 & 2.43& 1.39 & 1.42 & 0.24 & 0.24 & 0.05 & 0.07 & 16.3 & 17.  & 0.1 & 0.11 & 0.03 & 0.08 & 0.56 & 0.54 & 3.0 & 2.11& 0.57 & 0.3 & 0.71 & 0.4\\
  120 & 3.91 & 4.  & 2.0 & 1.45 & 0.24 & 0.23& 0.05 & 0.05 & 14.36 & 14.8 & 0.09 & 0.12 & 0.01 & 0.03 & 0.52 & 0.41 & 2.99 & 2.9 & 0.59 & 0.4 & 0.73 & 0.4\\
  121 & 4.22 & 4.7 & 1.63 & 1.12& 0.19 & 0.2  & 0.04 & 0.05 & 10.91 &10.3 & 0.15 & 0.12 & 0.02 & 0.02  & 0.54 & 0.44 & 3.2 & 2.9 & 0.61 & 0.5  & 0.75 & 0.6 \\
\hline
\end{tabular}
\tablefoot{
The + indicates that the doublet is summed up. \ion{He}{II}\,4686 is blended with [\ion{Fe}{III}]4658,4702.}

\end{table*}

\begin{table*}
\begin{tabular}{p{0.3cm}p{0.35cm}p{0.35cm}p{0.35cm}p{0.35cm}p{0.35cm}p{0.35cm}p{0.35cm}p{0.35cm}p{0.35cm}p{0.35cm}p{0.35cm}p{0.35cm}p{0.35cm}p{0.35cm}p{0.35cm}p{0.35cm}p{0.35cm}p{0.35cm}p{0.35cm}p{0.35cm}p{0.35cm}p{0.35cm}}
\hline
  \multicolumn{1}{c}{\phantom{n}} &
  \multicolumn{2}{c}{[\ion{O}{II}]} &
  \multicolumn{2}{c}{[\ion{Ne}{III}]} &
  \multicolumn{2}{c}{\ion{He}{II}} &
  \multicolumn{2}{c}{[\ion{Ar}{IV}]} &
  \multicolumn{2}{c}{[\ion{O}{III}]} &
  \multicolumn{2}{c}{\ion{He}{I}} &
  \multicolumn{2}{c}{[\ion{Fe}{VII}]} &
  \multicolumn{2}{c}{[\ion{O}{I}]} &
  \multicolumn{2}{c}{[\ion{N}{II}]} &
  \multicolumn{2}{c}{[\ion{S}{II}]} &
  \multicolumn{2}{c}{[\ion{S}{II}]} \\
  \multicolumn{1}{c}{\phantom{n}}&
  \multicolumn{2}{c}{3727+} &
  \multicolumn{2}{c}{3869+} &
  \multicolumn{2}{c}{4686 } &
  \multicolumn{2}{c}{4713 } &
  \multicolumn{2}{c}{5007+} &
  \multicolumn{2}{c}{5876 } &
  \multicolumn{2}{c}{6087 } &
  \multicolumn{2}{c}{6300+} &
  \multicolumn{2}{c}{6583+} &
  \multicolumn{2}{c}{6718 } &
  \multicolumn{2}{c}{6731 } \\
  \multicolumn{1}{c}{n}&
  \multicolumn{2}{c}{Obs}{Mod} &
  \multicolumn{2}{c}{Obs}{Mod} &
  \multicolumn{2}{c}{Obs}{Mod} &
  \multicolumn{2}{c}{Obs}{Mod} &
  \multicolumn{2}{c}{Obs}{Mod} &
  \multicolumn{2}{c}{Obs}{Mod} &
  \multicolumn{2}{c}{Obs}{Mod} &
  \multicolumn{2}{c}{Obs}{Mod} &
  \multicolumn{2}{c}{Obs}{Mod} &
  \multicolumn{2}{c}{Obs}{Mod} &
  \multicolumn{2}{c}{Obs}{Mod} \\
\hline
  122 & 2.74 & 2.6  & 0.76 & 0.6  & 0.15 & 0.14& 0.06 & 0.03 & 6.63 & 5.3 & 0.11 & 0.13 & 0.0 & 0.02 & 0.44 & 0.37 & 3.19 & 2.  & 0.61 & 0.5  & 0.7 & 0.44\\
  123 & 3.67 & 4.2 & 0.47 & 0.94 & 0.27 & 0.27 & 0.17 & 0.01 & 6.73 & 6.7 & 0.1 & 0.138 & 0.0 & 0.01 & 0.35 & 0.4  & 2.72 & 3.4 & 0.65 & 0.36 & 0.76 & 0.58\\
  123 & 3.67 & 3.46& 0.47 & 0.75 & 0.27 & 0.27 & 0.17 & 0.03 & 6.73 & 6.9 & 0.1 & 0.12& 0.0 & 0.08 & 0.35 & 0.37 & 2.72 & 2.3 & 0.65 & 0.5 & 0.56 & 0.5 \\
  124 & 1.42 & 1.1 & 0.0 & 0.55 & 0.62 & 0.2  & 0.32 & 0.02 & 5.6 & 6.0 & 0.1 & 0.13 & 0.0 & 0.01 & 0.57 & 0.56 & 2.0 & 1.8 & 0.74 & 0.4 & 0.44 & 0.41\\
  124 & 1.42 & 1.0 & 0.0 & 0.37 & 0.62 & 0.05 & 0.32 & 0.01 & 5.6 & 3.6 & 0.1 & 0.13 & 0.0 & 0.01 & 0.57 & 0.4 & 2.0 & 1.17 & 0.74 & 0.3 & 0.44 & 0.35\\
  125 & 0.0 & 0.6 & 0.0 & 0.7 & 0.22 & 0.13 & 0.0 & 0.01 & 3.9 & 4.   & 0.26 & 0.13 & 0.09 & 0.08 & 0.63 & 0.41 & 3.0 & 1.2 & 0.97 & 0.3  & 0.59 & 0.4 \\
  132 & 0.94 & 1.5 & 0.27 & 2.5  & 0.43 & 0.34 & 0.12 & 0.1  & 11.62 & 11.  & 0.14 & 0.10 & 0.0 & 0.26 & 0.4 & 0.45 & 3.13 & 2.7 & 0.6 & 0.4  & 0.69 & 0.44\\
  133 & 5.49 & 5.6 & 2.1 & 1.57 & 0.21 & 0.29 & 0.04 & 0.07 & 14.63 & 14.7 & 0.07 & 0.11 & 0.02 & 0.06 & 0.5 & 0.53 & 3.23 & 3.4 & 0.66 & 0.54 & 0.72 & 0.54\\
  134 & 5.94 & 5.6 & 2.53 & 1.57 & 0.24 & 0.29 & 0.03 & 0.07 & 14.55 & 14.7 & 0.08 & 0.11 & 0.03 & 0.06 & 0.54 & 0.53 & 3.22 & 3.4 & 0.64 & 0.54 & 0.75 & 0.54\\
  135 & 5.49 & 5.6 & 2.56 & 1.4 & 0.24 & 0.29 & 0.04 & 0.06 & 14.7 & 14.7 & 0.08 & 0.11 & 0.02 & 0.06 & 0.55 & 0.53& 3.27 & 3.4 & 0.68 & 0.54& 0.78 & 0.54\\
  136 & 5.85 & 5.8 & 2.36 & 1.4 & 0.2 & 0.26 & 0.04 & 0.06 & 12.79 & 12.3 & 0.1 & 0.11 & 0.03 & 0.05 & 0.58 & 0.8 & 3.27 & 3.3 & 0.71 & 0.6  & 0.78 & 0.6 \\
  137 & 6.4 & 5.8 & 2.1 & 1.4  & 0.11 & 0.22 & 0.12 & 0.06 & 10.83 &11.7 & 0.13 & 0.12 & 0.06 & 0.06 & 0.85 & 1.   & 3.34& 3.4 & 0.73 & 0.68 & 0.74 & 0.63\\
  138 & 4.5 & 4.2 & 0.68 & 0.97& 0.28 & 0.3  & 0.08 & 0.05 & 7.36 & 8.  & 0.29 & 0.21 & 0.10& 0.04 & 0.71 & 0.89 & 3.52& 2.8  & 0.72 & 0.74& 0.64 & 0.67\\
  139 & 3.46 & 4.0 & 0.0 & 0.84 & 0.2 & 0.27 & 0.08 & 0.01 & 5.8 & 5.4 & 0.39 & 0.14 & 0.07 & 0.01 & 0.62 & 0.2 & 2.4 & 3.3 & 0.55 & 0.3 & 0.6 & 0.52\\
  139 & 3.46 & 3.2 & 0.0 & 0.7 & 0.2 & 0.2  & 0.08 & 0.05 & 5.8 & 6.4 & 0.39 & 0.22 & 0.07 & 0.03 & 0.62 & 0.61 & 2.4 & 2.24& 0.55 & 0.6 & 0.6 & 0.57\\
  140 & 3.9 & 3.8 & 0.0 & 2.2 & 0.53 & 0.53 & 0.28 & 0.03 & 6.62 & 6.97& 0.37 & 0.14 & 0.23 & 0.02 & 0.5 & 0.53 & 1.74 & 5.2 & 0.57 & 0.5 & 0.44 & 0.69\\
  140 & 3.9 & 3.9 & 0.0 & 0.7  & 0.53 & 0.26 & 0.28 & 0.05 & 6.62 & 6.1  & 0.37 & 0.1 & 0.23 & 0.05 & 0.5 & 0.57 & 1.74 & 2.7  & 0.57 & 0.6  & 0.44 & 0.5 \\
  141 & 2.0 & 1.7  & 0.48 & 0.7 & 0.29 & 0.17 & 0.0 & 0.01 & 4.85 & 4.94 & 0.0  & 0.14 & 0.17 & 0.01  & 0.63 & 0.64 & 2.5 & 2.14& 0.9 & 0.43 & 0.53 & 0.53\\
  141 & 2.0 & 3.9 & 0.48 & 0.6 & 0.29 & 0.27 & 0.0 & 0.02 & 4.85 & 4.57& 0.0  & 0.21 & 0.17 & 0.04 & 0.63 & 0.64 & 2.5 & 2.14& 0.9 & 0.83& 0.53 & 0.66\\
  142 & 10.2 & 9.  & 0.0 & 1.3 & 0.23 & 0.24& 0.08 & 0.02 & 8.1 & 8.04 & 0.0  & 0.1  & 0.17& 0.13 & 0.69 & 0.69& 3.2 & 4.3 & 0.9  & 0.9 & 0.56 & 0.77\\
  148 & 4.74 & 4.5 & 1.11 & 1.7 & 0.27 & 0.33 & 0.08 & 0.06 & 15.25 & 15.9 & 0.16 & 0.13 & 0.06 & 0.03 & 0.42 & 1.   & 3.  & 3.2  & 0.55 & 0.4  & 0.57 & 0.5 \\
  149 & 5.89 & 6.0 & 2.12 & 1.4 & 0.25 & 0.26 & 0.06 & 0.07 & 13.56 & 13.2 & 0.11 & 0.11 & 0.0 & 0.05 & 0.44 & 0.59& 3.0 & 3.27& 0.64 & 0.55& 0.6 & 0.55\\
  150 & 7.76 & 7.0 & 2.95 & 1.7 & 0.21 & 0.26 & 0.01 & 0.05 & 14.19 & 15.  & 0.11 & 0.12 & 0.02 & 0.02 & 0.47 & 0.5  & 3.1 & 3.7 & 0.69 & 0.5  & 0.67 & 0.6 \\
  151 & 8.49 & 8.0 & 2.95 & 1.6 & 0.2 & 0.25 & 0.01 & 0.04 & 13.28 & 13.1 & 0.08 & 0.12 & 0.01 & 0.01 & 0.51 & 0.57 & 3.17 & 3.9 & 0.74 & 0.62 & 0.72 & 0.67\\
  152 & 8.0 & 8.0 & 2.7 & 1.5  & 0.14 & 0.24 & 0.01 & 0.03 & 11.43 & 11.2 & 0.09 & 0.13 & 0.03 & 0.01 & 0.54 & 0.52 & 3.2 & 4.  & 0.77 & 0.64& 0.71 & 0.7 \\
  153 & 8.27 & 7.4 & 2.88 & 1.35 & 0.24 & 0.24 & 0.07 & 0.04 & 9.65 & 9.86& 0.12 & 0.13 & 0.03 & 0.01  & 0.62 & 0.61 & 3.23 & 3.9 & 0.78 & 0.7 & 0.77 & 0.74\\
  154 & 10.7 & 9.0 & 1.25 & 1.25 & 0.14 & 0.23 & 0.0 & 0.01  & 6.85 & 7.2 & 0.12 & 0.13 & 0.0 & 0.003 & 0.69 & 0.63& 3.34 & 4.4 & 0.83 & 0.9 & 0.82 & 0.8\\
  155 & 9.98 & 10.3 & 1.76 & 1.72 & 0.0  & 0.12 & 0.12& 0.08 & 7.86 & 8.6 & 0.12 & 0.14 & 0.0 & 0.55 & 0.7 & 0.33 & 2.8 & 4.2 & 0.7 & 0.56 & 0.67 & 0.96\\
  155 & 9.98 & 9.   & 1.76 & 1.22& 0.0  & 0.23 & 0.12& 0.013& 7.86 & 7.3 & 0.12 & 0.12 & 0.0 & 0.0 & 0.7 & 0.5  & 2.8 & 4.3 & 0.7 & 0.8 & 0.67 & 0.7 \\
  156 & 21.54 & 19.0 & 0.0 & 2.0 & 0.0 & 0.09 & 0.15& 0.12 & 8.71 & 10.6 & 0.3 & 0.25 & 0.0 & 0.77 & 0.51 & 0.57 & 2.4 & 3.7 & 0.48 & 1.2 & 0.43 & 1.5\\
  156 & 21.54 & 22.5 & 0.0 & 1.8 & 0.0 & 0.07 & 0.15& 0.02 & 8.71 & 8.   & 0.3 & 0.13 & 0.0 & 0.0  & 0.51 & 0.12 & 2.4 & 4.  & 0.48 & 1.2 & 0.43 & 1. \\
  157 & 0.0 & 19. & 0.0 & 2.   & 0.0 & 0.09& 0.0 & 0.1  & 12.3 &10.6 & 0.07 & 0.2  & 0.0 & 0.77 & 0.58 & 0.57 & 2.3 & 3.7  & 0.46 & 1.2  & 0.27 & 1.5 \\
  165 & 5.76 & 5.  & 1.72 & 1.04& 0.18 & 0.25& 0.08 & 0.05 & 9.32 & 9.4  & 0.07 & 0.11 & 0.02 & 0.06  & 0.45 & 0.45 & 2.98 & 2.97& 0.66 & 0.51& 0.58 & 0.51\\
  166 & 6.93 & 6.9  & 2.35 & 1.5 & 0.14 & 0.23 & 0.03 & 0.03 & 12.82 & 12.5 & 0.10& 0.12 & 0.0   & 0.02 & 0.46 & 0.35 & 2.9 & 3.6 & 0.72 & 0.5 & 0.62 & 0.62\\
  167 & 7.71 & 7.87& 2.8 & 1.6 & 0.15 & 0.25& 0.03 & 0.04 & 13.84 & 13.2 & 0.09 & 0.12 & 0.0   & 0.01  & 0.48 & 0.42 & 2.81 & 3.9  & 0.67 & 0.6  & 0.61 & 0.64\\
  168 & 8.89 & 8.3 & 3.04 & 1.6 & 0.18 & 0.25& 0.03 & 0.03 & 11.8 &11.8 & 0.1 & 0.12 & 0.0 & 0.01  & 0.52 & 0.55 & 2.9 & 4.1 & 0.7 & 0.64 & 0.62 & 0.7 \\
  169 & 8.38 & 7.8 & 1.74 & 1.3  & 0.07 & 0.23 & 0.05 & 0.02 & 9.21 & 9.3  & 0.11 & 0.13 & 0.0 & 0.01& 0.48 & 0.46 & 2.79 & 4.  & 0.75 & 0.66 & 0.68 & 0.73\\
  170 & 8.39 & 9.   & 1.99 & 1.24 & 0.0 & 0.23 & 0.0 & 0.01 & 7.21 & 7.22 & 0.07 & 0.13 & 0.0 & 0.0   & 0.52 & 0.53& 3.0 & 4.3 & 0.85 & 0.87& 0.71 & 0.74\\
  171 & 12.72 & 13.  & 5.5 & 1.56& 0.0 & 0.15 & 0.0 & 0.04 & 9.73 & 9.85 & 0.09& 0.12 & 0.0 & 0.01& 0.43 & 0.1  & 3.0 & 3.5 & 0.72 & 0.7 & 0.71 & 0.64\\
  172 & 28.7 & 29.9 & 3.0 & 2.5 & 0.0 & 0.07 & 0.32 & 0.03 & 10.85 & 11. & 0.0 & 0.13 & 0.0 & 0.0 & 0.5 & 0.23 & 3.14 & 4.8 & 0.48 & 1.3  & 0.58 & 1.  \\
  173 & 13.4 &14.3 & 0.0 & 1.3 & 0.0 & 0.06 & 0.29 & 0.05 & 8.2 &  7.  & 0.0 & 0.15 & 0.0 & 0.09 & 0.8 & 0.26& 3.8 & 3.3 & 0.75 & 1.17 & 0.56 & 0.98\\
  181 & 5.8 & 5.3 & 2.2 & 1.6 & 0.37 & 0.33& 0.0 & 0.07 & 15.25 & 15.6 & 0.07 & 0.12 & 0.09 & 0.03  & 0.43 & 0.5  & 2.9 & 2.8 & 0.56 & 0.36 & 0.48 & 0.5 \\
  182 & 8.2 & 7.9 & 2.85 & 1.6 & 0.14 & 0.25 & 0.0 & 0.04 & 13.31& 13.3 & 0.2 & 0.12 & 0.0 & 0.011& 0.47 & 0.42 & 2.84 & 3.8 & 0.62 & 0.58& 0.55 &0.64\\
  183 & 8.9 & 8.7 & 2.92 & 1.7 & 0.19 & 0.26 & 0.05 & 0.04 & 12.9 & 13.3 & 0.14 & 0.12 & 0.01 & 0.01 & 0.41 & 0.34 & 2.7 & 4.2 & 0.64 & 0.60 & 0.56 & 0.6 \\
  184 & 7.1 & 7.4 & 2.77 & 1.3  & 0.09 & 0.21 & 0.04 & 0.03 & 10.72 &10.  & 0.11 & 0.13 & 0.04 & 0.01& 0.43 & 0.36& 2.56 & 3.8 & 0.67 & 0.61 & 0.56 & 0.67\\
  185 & 8.8 & 8.8 & 2.48 & 1.27& 0.12 & 0.23 & 0.04 & 0.02 & 8.95 & 8.5  & 0.22 & 0.12 & 0.02 & 0.0 & 0.43 & 0.4 & 2.5 & 4.1 & 0.74 & 0.76& 0.59 &0.68\\
  186 & 9.8 & 9.   & 1.99 & 1.2 & 0.0 & 0.23 & 0.1 & 0.01 & 7.35 & 7.3 & 0.19 & 0.12 & 0.0 & 0.0 & 0.45 & 0.4 & 2.64 & 4.2 & 0.81 & 0.8  & 0.58 & 0.7 \\
  187 & 13.3 & 13.6 & 2.4 & 1.8 & 0.57 & 0.33 & 0.2 & 0.24 & 7.64 & 7.0 & 0.15 & 0.18 & 0.0 & 0.0 & 0.56 & 0.43& 2.92 & 3.9 & 0.73 & 1.6  & 0.65 & 1.5 \\
  188 & 15.5 &15.4 & 3.2 & 2.  & 0.0 & 0.07 & 0.0 & 0.04 & 8.3 & 8.   & 0.11 & 0.16 & 0.0 & 0.15 & 0.54 & 0.3  & 3.06 & 4.  & 0.64 & 1.4  & 0.68 & 1.4 \\
  197 & 5.6 & 6.0 & 1.4  & 1.3 & 0.42 & 0.4  & 0.0 & 0.06 & 13.39 & 13.3 & 0.05 & 0.11 & 0.08 & 0.05 & 0.44 & 0.42 & 2.75 & 3.2 & 0.47 & 0.53 & 0.43 & 0.54\\
  198 & 4.78 & 5.   & 2.16 & 1.3 & 0.0 & 0.21 & 0.0 & 0.04 & 12.27 & 11.5  & 0.11 & 0.12 & 0.0 & 0.02  & 0.38 & 0.47& 2.82 & 3.2  & 0.59 & 0.42 & 0.53 & 0.57\\
  199 & 11.47 & 9.5 & 2.2 & 1.6 & 0.28 & 0.28 & 0.0 & 0.03 & 12.29 & 12.26& 0.26 & 0.11 & 0.0 & 0.01  & 0.34 & 0.4 & 2.46 & 4.18& 0.59 & 0.64 & 0.49 & 0.6 \\
  200 & 9.0 & 9.4 & 2.2 & 1.5 & 0.05 & 0.27 & 0.0 & 0.02  & 9.54 &10.2 & 0.17 & 0.13 & 0.03 & 0.004 & 0.8 & 0.8  & 2.52 & 4.4  & 0.65 & 0.87& 0.55 &0.73\\
  201 & 10. & 9.  & 1.74 & 1.2  & 0.09 & 0.24& 0.1 & 0.01 & 7.85 & 7.7 & 0.19 & 0.12 & 0.0 & 0.0 & 0.37 & 0.32& 2.72 & 4.15& 0.66 & 0.7  & 0.52 & 0.6 \\
  202 & 6.  & 7.4 & 0.78 & 0.94 & 0.16 & 0.2 & 0.0 & 0.02 & 5.5 & 5.0 & 0.13 & 0.13 & 0.08 & 0.0 & 0.41 & 0.2 & 2.64 & 4.0 & 0.73 & 0.65 & 0.53 & 0.65\\
  203 & 7.5 & 7.4 & 0.8 & 1.4 & 0.5 & 0.46 & 0.0 & 0.02 & 5.6 & 5.4  & 0.18 & 0.13 & 0.0 & 0.02 & 0.45 & 0.46 & 3.3 & 4.9  & 0.85 & 0.95& 0.68 &0.86\\
  213 & 5.5 & 5.1 & 0.0 & 3.   & 0.69 & 0.56 & 0.2 & 0.16 & 19.2 & 18.8 & 0.0 & 0.1  & 0.0 & 0.16 & 0.27 & 0.27 & 2.66 & 3.8 & 0.6 & 0.31 & 0.48 & 0.42\\
  214 & 1.19 & 2.0 & 0.85 & 0.95 & 0.0 & 0.13 & 0.0 & 0.05 & 11.14 & 11.1 & 0.12 & 0.12 & 0.07 & 0.03 & 0.37 & 0.37 & 2.82 & 1.8 & 0.59 & 0.28 & 0.56 & 0.37\\
  215 & 4.63 & 4.8 & 0.0 & 1.12 & 0.04 & 0.21 & 0.17 & 0.05 & 11.48 & 11.0 & 0.14 & 0.11 & 0.0 & 0.03 & 0.28 & 0.2 & 2.7 & 2.76 & 0.63 & 0.4 & 0.53 & 0.5\\
  216 & 7.21 & 7.6 & 2.14 & 1.3 & 0.0 & 0.21 & 0.0 & 0.02 & 8.7 & 9.3 & 0.14 & 0.13 & 0.0 & 0.0   & 0.37 & 0.4 & 2.53 & 4.0 & 0.64 & 0.64 & 0.57 & 0.71\\
  217 & 3.62 & 4.0 & 0.96 & 0.7 & 0.19 & 0.26 & 0.0 & 0.03 & 6.4 & 6.2 & 0.11 & 0.104 & 0.0 & 0.05 & 0.28 & 0.23 & 2.62 & 2.6 & 0.64 & 0.5 & 0.6 & 0.43\\
  218 & 2.84 & 4.0 & 4.25 & 0.8 & 0.0 & 0.23 & 0.0 & 0.04 & 7.71 & 7.96 & 0.21 & 0.11 & 0.0 & 0.05 & 0.23 & 0.28 & 2.63 & 2.5 & 0.64 & 0.5 & 0.46 & 0.45\\
  219 & 7.0 & 8.0 & 0.0 & 1.1 & 0.0 & 0.2 & 0.0 & 0.01 & 5.0 & 4.7 & 0.15 & 0.14 & 0.0 & 0.002 & 0.63 & 0.61 & 3.27 & 4.5 & 0.79 & 0.9 & 0.5 & 0.88\\
\hline
\end{tabular}
\end{table*}

\subsubsection{[\ion{N}{II}] and [\ion{O}{II}]}

\begin{figure*}
\includegraphics[width=0.32\textwidth]{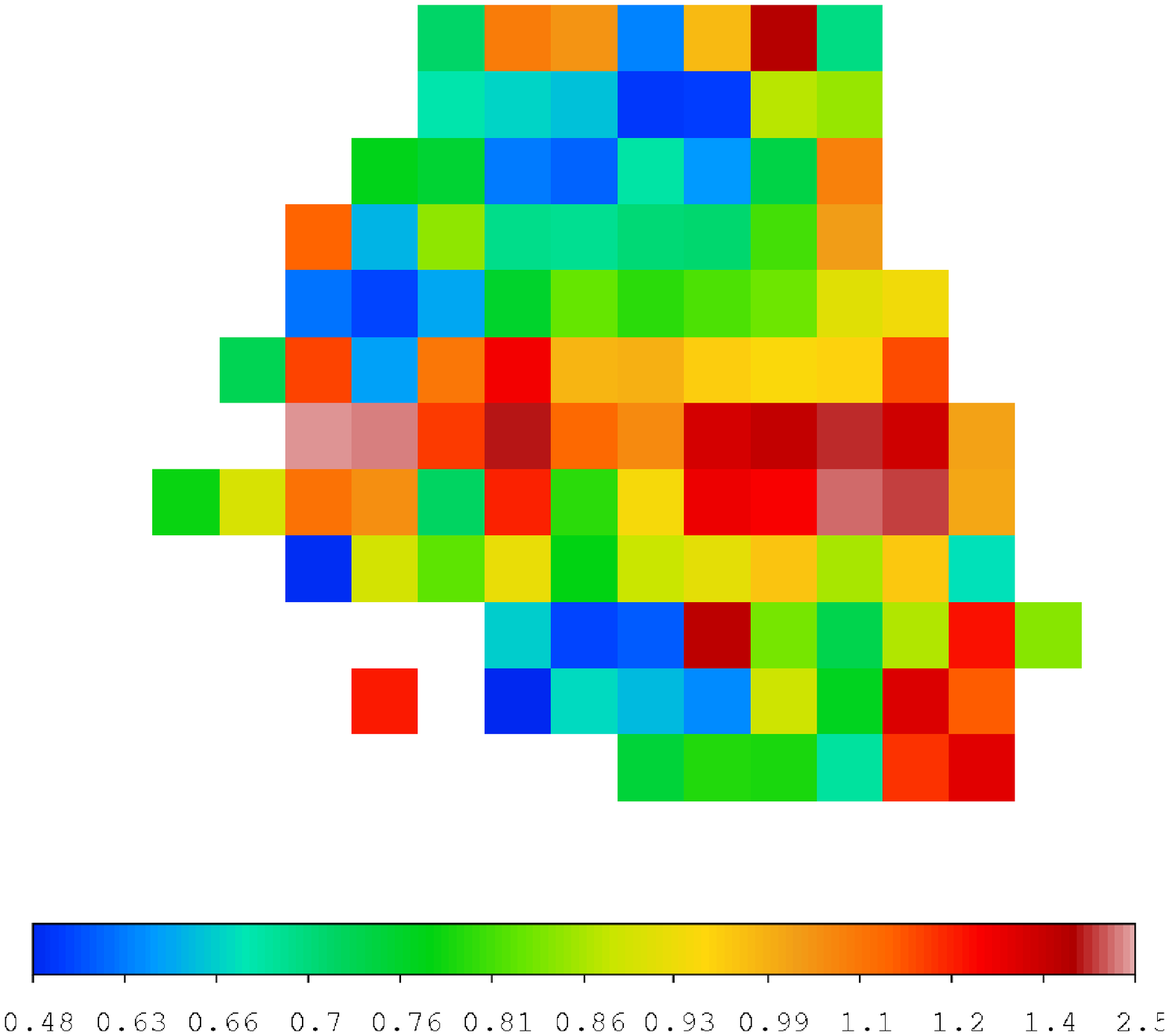}
\includegraphics[width=0.32\textwidth]{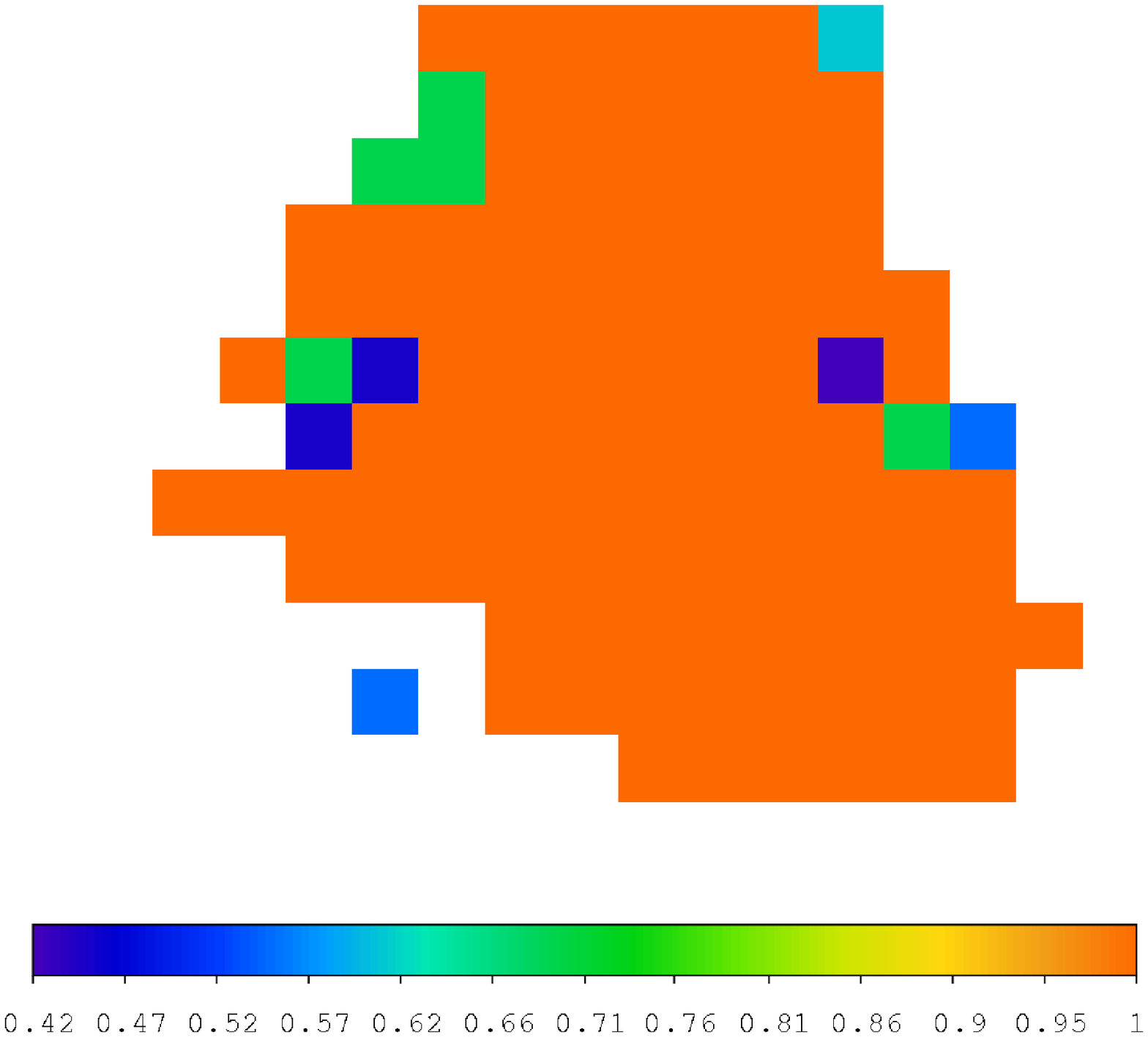}
\includegraphics[width=0.32\textwidth]{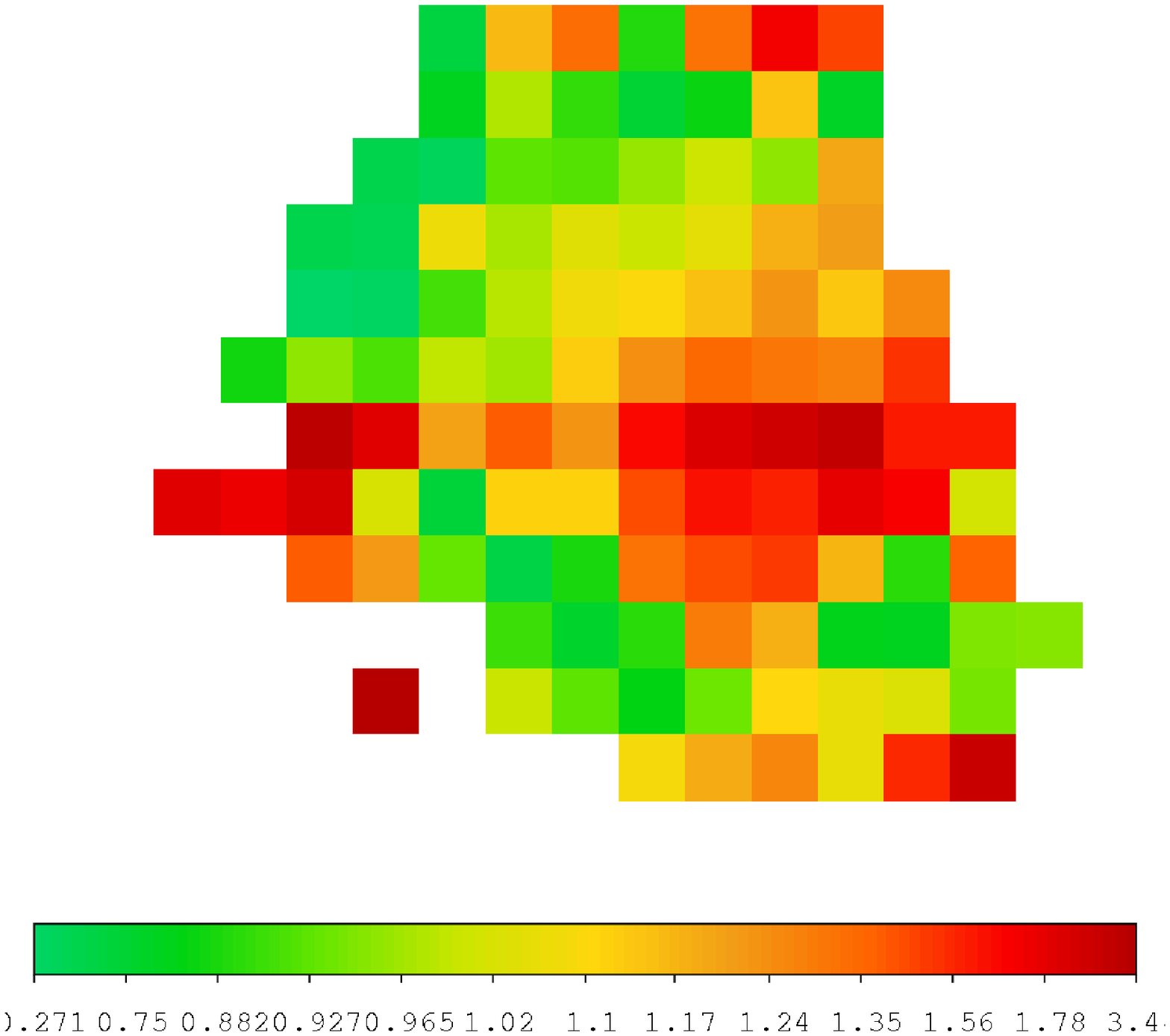}
\caption{The obs/calc maps of nitrogen (left), oxygen (middle) and sulphur (right). Values lower than 1 indicate depletion. Spatial scale is 1 arcsec px$^{-1}$. North is up and East is to the left.}
\label{fig5}
\end{figure*}

Table\,\ref{tab2} and Fig.\,\ref{fig4a} show that the  lines which systematically disagree with the observations are the [\ion{N}{II}] doublet at 6548+6584 \AA. The doublet is summed up in the table.
In the  calculations, solar N/H value $9.1\times10^{-5}$ (Allen \cite{Allen}) was adopted.
Table\,\ref{tab2} shows that the  observed [\ion{N}{II}]/H$\beta$ line ratios  are  mostly  overpredicted by the models.
The discrepancies between observed and calculated values are shown in Fig.\,\ref{fig4b}.  
 [\ion{N}{II}] lines come from the same region of the gas that emits the [\ion{O}{II}] lines.
[\ion{O}{II}]/H$\beta$ line ratios depend on the physical parameters and are 
constrained by [\ion{O}{III}]/H$\beta$ and [\ion{O}{I}]/H$\beta$,
so the discrepancy in the [\ion{N}{II}]/H$\beta$ line ratios can be explained by abnormal N/H relative abundances.
In the  left diagram of Fig.\,\ref{fig4b} ([\ion{N}{II}]/H$\beta$)$_{\rm calc}$/([\ion{N}{II}]/H$\beta$)$_{\rm obs}$ 
and ([\ion{O}{II}]/H$\beta$)$_{\rm calc}$/([\ion{O}{II}]/H$\beta$)$_{\rm obs}$ are plotted in the different positions.
If discrepancies between calculation  and observation were due to the model, both [\ion{N}{II}]/H$\beta$ and [\ion{O}{II}]/H$\beta$ 
would show the same variations. Contrariwise, in positions  beyond 100 (regions to the East of the nucleus) the [\ion{N}{II}]/H$\beta$ calculated line ratios overpredict the data by a factor of at least 1.5, whereas the observed [\ion{O}{II}]/H$\beta$ are always well reproduced by the models.  
This indicates that N/H should be depleted by a factor of at least 1.5, while  O/H is almost solar. 
The map showing the N/H depletion and excess (Table\,\ref{tab33})  appears in Fig.\,\ref{fig5}.

\subsubsection{[\ion{S}{II}] and [\ion{O}{I}]}

From the last two columns of Table\,\ref{tab2} we can realize that the doublet ratios [\ion{S}{II}]6716/6731$\sim$1, indicative of the emitting gas density, are relatively low ($\leq10^3$ cm$^{-3}$).
The [\ion{S}{II}] and [\ion{O}{I}] lines  are emitted  from  relatively cool gas in similar physical conditions. 
In Fig.\,\ref{fig4b} (middle diagram) both  ([\ion{S}{II}]/H$\beta$)$_{\rm calc}$/([\ion{S}{II}]/H$\beta$)$_{\rm obs}$ 
and ([\ion{O}{I}]/H$\beta$)$_{\rm calc}$/([\ion{O}{I}]/H$\beta$)$_{\rm obs}$ ratios show up.
Table\,\ref{tab2} and Fig.\,\ref{fig4a} suggest that [\ion{S}{II}] line ratios to H$\beta$  calculated adopting solar abundances are acceptable, except in some positions. 
At the North-East of the nucleus the calculated [\ion{S}{II}]/H$\beta$ ratios overpredict the data and at other positions they  even underpredict them (see Fig.\,\ref{fig4b}).
The map of the S/H depletion and excess (Table\,\ref{tab33}) is shown in Fig.\,\ref{fig5}.
Eventually, the [\ion{S}{II}] observed lines include a large contribution from the ISM cool gas.
The [\ion{S}{II}]4070+ lines are not included in Table\,\ref{tab2} because they do not appear in all the spectra and since they are blended with other lines at nearby wavelengths. However, model results indicate that the [\ion{S}{II}]4070+/H$\beta$ ratios are relatively high for high velocities and reproduce  the data in a few positions.

%%%%%%%%%%%%%%%%%%%%%%%%%%%%%%%%%%5

\subsubsection{\ion{He}{II} and the oxygen lines}

The first cycle of calculations yields \ion{He}{II}/H$\beta$ line ratios 
generally lower than those observed. Indeed, higher \ion{He}{II}/H$\beta$ can be obtained by increasing the photoionization flux, which however spoils the goodness of fit of the [\ion{O}{III}]/H$\beta$ and [\ion{O}{II}]/H$\beta$ line ratios.
Another option is to  reduce the geometrical thickness of some of the clouds, excluding the cooling region.
Accounting only for the high temperature zone close to the shock front downstream, only the \ion{He}{II} lines would be strong.
Summing up the spectra of these thinner clouds with the other ones would give high \ion{He}{II}/H$\beta$ ratios.
The three most important line ratios [\ion{O}{III}]/H$\beta$, [\ion{O}{II}]/H$\beta$ and [\ion{O}{I}]/H$\beta$ are all well fitted adopting the solar O/H relative abundance, $6.6\times10^{-4}$ (Allen \cite{Allen}).
However, Cracco et al. (\cite{Cr11}), compared the line ratios with models calculated by pure photoionization and found a low metallicity in NGC 7212.
Oxygen is a strong coolant. Decreasing  O/H in the models will induce a lower cooling rate downstream, yielding a larger zone of hot gas capable of emitting higher \ion{He}{II}/H$\beta$. 
Moreover, the \ion{He}{II} line sits on top of the [\ion{Fe}{III}] multiplet,
whose most prominent lines are at 4658 \AA\ and 4702 \AA. The lines are blended with \ion{He}{II} in high velocity regimes. 
Consequently, we adopted in the next cycle of calculations a lower O/H relative abundance in positions where the observed \ion{He}{II}/H$\beta$ lines ratios were underpredicted. This required an adjustment of the input parameters also regarding the physical conditions.
Finally, we  added to \ion{He}{II}/H$\beta$ the [\ion{Fe}{III}]/H$\beta$ line ratios, which can be as high as \ion{He}{II}/H$\beta$ in some positions.
The comparison of calculated with observed \ion{He}{II}/H$\beta$ improved (Fig.\,\ref{fig4b}, right diagram) except in some positions, e.g. 109, 124, 200, where observational errors are high (larger than 60\%).
In particular, we obtained $\rm O/H= 3.6\times 10^{-4}$ in positions 60 and 115, $4.6\times10^{-4}$ in positions 116, 141, 187, 188, 203 and $4\times10^{-4}$ in 213. Moreover, $\rm O/H =3\times10^{-4}$ was used in positions 125, 140, and $2.6\times10^{-4}$ in position 133. 
Interestingly, most of these positions refer to relatively high shock velocities (see Table\,\ref{tab1}).
For all other positions, the solar O/H value was adopted. 
The map of the oxygen depletion is shown in Fig.\,\ref{fig5}.

\begin{table}
\begin{center}
\caption{Calculated relative abundances.}
\label{tab33}
\begin{tabular}{ l c c l c c } \\ \hline \hline
\   pos.  & N/H  & S/H  & pos.&  N/H  & S/H\\ \hline
\     35. &   1.17$\times10^{-4}$  &  6.99$\times10^{-6}$   &   123. &    1.08$\times10^{-4}$  &   1.32$\times10^{-5}$\\
\     36. &   1.08$\times10^{-4}$  &  1.27$\times10^{-5}$   &   124. &    1.01$\times10^{-4}$  &   1.67$\times10^{-5}$\\
\     37. &   6.20$\times10^{-5}$  &  1.20$\times10^{-5}$    &  124. &    1.56$\times10^{-4}$  &   2.09$\times10^{-5}$\\
\     38. &   7.19$\times10^{-5}$  &  1.27$\times10^{-5}$    &  125. &    2.28$\times10^{-4}$  &   1.47$\times10^{-5}$\\
\     39. &   7.22$\times10^{-5}$  &  1.51$\times10^{-5}$   &   132. &    1.05$\times10^{-4}$  &   1.48$\times10^{-5}$ \\
\     40. &   6.81$\times10^{-5}$   & 1.34$\times10^{-5}$  &    133. &    8.64$\times10^{-5}$  &   1.07$\times10^{-5}$\\
\     51. &   1.05$\times10^{-4}$   & 1.08$\times10^{-5}$ &     134. &    8.62$\times10^{-5}$  &   1.07$\times10^{-5}$\\
\     52. &   1.21$\times10^{-4}$   & 1.36$\times10^{-5}$  &    135. &    8.75$\times10^{-5}$ &    1.01$\times10^{-5}$\\
\     53. &   6.96$\times10^{-5}$   & 1.55$\times10^{-5}$   &   136. &    9.02$\times10^{-5}$ &    8.95$\times10^{-6}$\\
\     54. &   8.08$\times10^{-5}$   & 1.69$\times10^{-5}$   &   137. &    8.94$\times10^{-5}$ &    8.31$\times10^{-6}$\\
\     55. &   5.85$\times10^{-5}$   & 1.70$\times10^{-5}$   &   138. &    1.14$\times10^{-4}$ &    8.34$\times10^{-6}$\\
\     56. &   5.96$\times10^{-5}$   & 1.84$\times10^{-5}$   &   139. &    6.62$\times10^{-5}$ &    1.70$\times10^{-5}$\\
\     57. &   6.08$\times10^{-5}$   & 1.69$\times10^{-5}$   &   139. &    9.75$\times10^{-5}$  &   1.19$\times10^{-5}$\\
\     58. &   4.36$\times10^{-5}$   & 1.47$\times10^{-5}$   &   140. &    3.05$\times10^{-5}$  &   1.33$\times10^{-5}$\\
\     60. &   1.09$\times10^{-4}$   & 2.77$\times10^{-5}$   &   140. &    5.86$\times10^{-5}$  &   1.44$\times10^{-5}$\\
\     66. &   7.79$\times10^{-5}$   & 1.77$\times10^{-5}$   &   141. &    1.06$\times10^{-4}$  &   1.17$\times10^{-5}$\\
\     67. &   1.10$\times10^{-4}$   & 1.26$\times10^{-5}$   &   141. &    1.06$\times10^{-4}$  &   7.51$\times10^{-6}$\\
\     68. &   7.88$\times10^{-5}$   & 1.25$\times10^{-5}$   &   142. &    6.77$\times10^{-5}$  &   6.56$\times10^{-6}$\\
\     69. &   6.80$\times10^{-5}$   & 1.31$\times10^{-5}$   &   148. &    8.53$\times10^{-5}$  &   1.59$\times10^{-5}$\\
\     70. &   7.71$\times10^{-5}$   & 2.21$\times10^{-5}$   &   149. &    8.35$\times10^{-5}$  &   1.17$\times10^{-5}$\\
\     71. &   1.41$\times10^{-4}$   & 2.54$\times10^{-5}$   &   150. &    7.62$\times10^{-5}$  &   1.07$\times10^{-5}$\\
\     72. &   5.52$\times10^{-5}$   & 2.00$\times10^{-5}$   &   151. &    7.40$\times10^{-5}$  &   8.50$\times10^{-6}$\\
\     73. &   5.46$\times10^{-5}$   & 1.69$\times10^{-5}$   &   152. &    7.28$\times10^{-5}$  &   8.07$\times10^{-6}$\\
\     74. &   5.98$\times10^{-5}$   & 1.45$\times10^{-5}$   &   153. &    7.54$\times10^{-5}$  &   7.17$\times10^{-6}$\\
\     83. &   6.12$\times10^{-5}$   & 1.67$\times10^{-5}$   &   154. &    6.91$\times10^{-5}$  &   5.70$\times10^{-6}$\\
\     84. &   8.82$\times10^{-5}$   & 7.37$\times10^{-6}$   &   155. &    6.07$\times10^{-5}$  &   7.68$\times10^{-6}$\\
\     85. &   7.84$\times10^{-5}$   & 1.29$\times10^{-5}$   &   155. &    5.93$\times10^{-5}$  &   7.79$\times10^{-6}$\\
\     86. &   8.85$\times10^{-5}$   & 2.14$\times10^{-5}$   &   156. &    5.90$\times10^{-5}$  &   6.51$\times10^{-6}$\\
\     87. &   8.40$\times10^{-5}$   & 2.33$\times10^{-5}$   &   156. &    5.46$\times10^{-5}$  &   7.99$\times10^{-6}$\\
\     88. &   8.05$\times10^{-5}$   & 2.51$\times10^{-5}$   &   157. &    5.66$\times10^{-5}$  &   8.12$\times10^{-6}$\\
\     89. &   7.00$\times10^{-5}$   & 1.74$\times10^{-5}$   &   165. &    9.13$\times10^{-5}$  &   1.27$\times10^{-5}$\\
\     90. &   8.47$\times10^{-5}$   & 1.30$\times10^{-5}$   &   166. &    7.33$\times10^{-5}$  &   1.07$\times10^{-5}$\\
\     91. &   6.42$\times10^{-5}$   & 2.80$\times10^{-5}$   &   167. &    6.56$\times10^{-5}$  &   1.01$\times10^{-5}$\\
\     91. &   7.46$\times10^{-5}$   & 1.19$\times10^{-5}$   &   168. &    6.44$\times10^{-5}$  &   9.05$\times10^{-6}$\\
\     92. &   5.38$\times10^{-5}$   & 1.19$\times10^{-5}$   &   169. &    6.35$\times10^{-5}$  &   8.05$\times10^{-6}$\\
\     92. &   8.11$\times10^{-5}$   & 9.13$\times10^{-6}$   &   170. &    6.35$\times10^{-5}$  &   6.37$\times10^{-6}$\\
\     93. &   5.04$\times10^{-5}$   & 1.30$\times10^{-5}$   &   171. &    7.80$\times10^{-5}$  &   8.35$\times10^{-6}$\\
\     99. &   9.07$\times10^{-5}$   & 1.42$\times10^{-5}$   &   172. &    5.95$\times10^{-5}$  &   6.56$\times10^{-6}$\\
\    100. &   1.52$\times10^{-4}$   & 1.89$\times10^{-5}$   &   173. &    1.05$\times10^{-4}$  &   5.68$\times10^{-6}$\\
\    101. &   1.54$\times10^{-4}$   & 1.74$\times10^{-5}$   &   181. &    9.43$\times10^{-5}$  &   1.79$\times10^{-5}$\\
\    102. &   1.11$\times10^{-4}$   & 1.73$\times10^{-5}$   &   182. &    6.80$\times10^{-5}$  &   1.12$\times10^{-5}$\\
\    103. &   1.16$\times10^{-4}$   & 1.93$\times10^{-5}$   &   183. &    5.85$\times10^{-5}$  &   1.11$\times10^{-5}$\\
\    104. &   8.60$\times10^{-5}$   & 1.73$\times10^{-5}$   &   184. &    6.13$\times10^{-5}$  &   1.02$\times10^{-5}$\\
\    105. &   7.28$\times10^{-5}$   & 1.43$\times10^{-5}$   &   185. &    5.55$\times10^{-5}$  &   8.35$\times10^{-6}$\\
\    106. &   1.08$\times10^{-4}$   & 1.43$\times10^{-5}$   &   186. &    5.72$\times10^{-5}$  &   7.67$\times10^{-6}$\\
\    107. &   7.23$\times10^{-5}$   & 1.68$\times10^{-5}$   &   187. &    6.81$\times10^{-5}$  &   3.74$\times10^{-6}$\\
\    107. &   6.72$\times10^{-5}$   & 9.40$\times10^{-6}$   &   188. &    6.96$\times10^{-5}$  &   4.33$\times10^{-6}$\\
\    108. &   6.71$\times10^{-5}$   & 1.20$\times10^{-5}$   &   197. &    7.82$\times10^{-5}$  &   1.66$\times10^{-5}$\\
\    108. &   9.21$\times10^{-5}$   & 9.25$\times10^{-6}$   &   198. &    8.02$\times10^{-5}$  &   1.44$\times10^{-5}$\\
\    109. &   9.89$\times10^{-5}$   & 1.36$\times10^{-5}$   &   199. &    5.36$\times10^{-5}$  &   1.19$\times10^{-5}$\\
\    110. &   8.17$\times10^{-5}$   & 9.04$\times10^{-6}$   &   200. &    5.21$\times10^{-5}$  &   8.33$\times10^{-6}$\\
\    111. &   7.00$\times10^{-5}$   & 1.38$\times10^{-5}$   &   201. &    5.96$\times10^{-5}$  &   1.04$\times10^{-5}$\\
\    115. &   9.10$\times10^{-5}$   & 1.43$\times10^{-5}$   &   202. &    6.01$\times10^{-5}$  &   9.77$\times10^{-6}$ \\
\    116. &   1.31$\times10^{-4}$   & 1.43$\times10^{-5}$   &   203. &    6.13$\times10^{-5}$  &   5.78$\times10^{-6}$\\
\    117. &   1.46$\times10^{-4}$   & 1.57$\times10^{-5}$   &   213. &    6.37$\times10^{-5}$  &   2.03$\times10^{-5}$\\
\    118. &   1.36$\times10^{-4}$   & 1.77$\times10^{-5}$   &   214. &    1.43$\times10^{-4}$  &   2.14$\times10^{-5}$\\
\    119. &   1.29$\times10^{-4}$   & 1.79$\times10^{-5}$   &   215. &    8.90$\times10^{-5}$  &   1.53$\times10^{-5}$\\
\    120. &   9.38$\times10^{-5}$   & 1.52$\times10^{-5}$    &  216. &    5.76$\times10^{-5}$ &    9.79$\times10^{-6}$\\
\    121. &   1.00$\times10^{-4}$   & 1.07$\times10^{-5}$    &  217. &    9.17$\times10^{-5}$ &    1.39$\times10^{-5}$\\
\    122. &   1.45$\times10^{-4}$   & 1.30$\times10^{-5}$    &  218. &    9.57$\times10^{-5}$  &   1.53$\times10^{-5}$\\
\    123. &   7.28$\times10^{-5}$   & 1.21$\times10^{-5}$    &  219. &    6.61$\times10^{-5}$  &   6.97$\times10^{-6}$\\ \hline
\end{tabular}
\tablefoot{
N/H(solar)=9.1$\times10^{-5}$; S/H(solar)= 1.6$\times10^{-5}$.}

\end{center}
\end{table}

\begin{figure}
\includegraphics[width=0.45\textwidth]{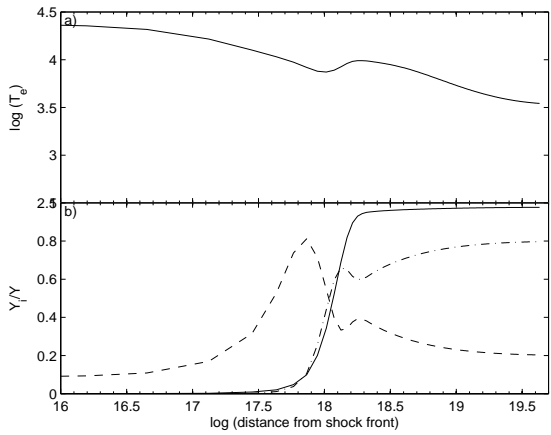}
\caption {The profile of HI (solid line), HeI (dot-dashed) and HeII (dashed)
fractional abundances as a function of distance from the shock front for position 53.
}
\label{fig9b}
\end{figure}

\subsubsection{HeI/\Hb line ratios} 

The HeI/\Hb diagram in Fig.\,\ref{fig4a} and the results in Table\,\ref{tab2} show that the calculated HeI/\Hb line ratios mostly range between 0.11 and 0.13, while the observations range between 0.04 and  0.4. To understand the results, we show  in Fig.\,\ref{fig9b} the profile of the temperature  and of HI, HeI and HeII fractional abundances  as a function of distance from the shock front, which corresponds to the cloud edge reached by the flux from the AGN. We have chosen  position 53 which shows a strong $F$ and a relatively low \Vs. The temperature maximum between 10$^{18}$ and $3\times10^{18}$ cm is  real and not due to the calculation approximation. It arises when hydrogen becomes neutral, followed by the decrease  of  the density of electrons, which, colliding with the heavy elements, produce cooling downstream (Williams 1967). The calculations of the He ion fractional abundances follow the temperature fluctuations by small variations, while the HI fluctuation is imperceptible, so the  HeI/\Hb  
calculated range is small. On the other hand, the observational error of HeI reaches 79\%.

\subsubsection{[\ion{Fe}{II}]}\label{fe2}
Rodr{\'{\i}}guez-Ardila et al. (2004) studied the [\ion{Fe}{II}] line emission in AGN and found 
that [\ion{Fe}{II}]/Pa$\beta$ increases from pure photoionization to a shock excitation regime. 
For instance, [\ion{Fe}{II}] emission and double profiles are due to jet-gas interaction 
in \object{Mrk 78} (Whittle et al. \cite{Wh88}; Ramos Almeida et al. \cite{R06}) and [\ion{Fe}{II}] lines are 
strong in the radio-jet interaction region in \object{Mrk 34} (Jackson \& Beswick \cite{J07}). 
This could indicate the importance of the shocks for strong [\ion{Fe}{II}] lines.
Iron first ionization potential (7.9 eV) which corresponds to the [\ion{Fe}{II}] lines is lower than that 
of O$^+$  and  H$^+$. Therefore [\ion{Fe}{II}] lines are mainly emitted from cool recombined gas. 
Collisional ionization coefficients are low at those low temperatures ($\leq$ 10$^4$ K), so 
strong [\ion{Fe}{II}] lines in the shock-dominated case (as for supernovae) are due to diffuse secondary radiation, which in turn is high as it is emitted from shock-heated gas slabs. 

\begin{figure*}
\includegraphics[width=0.49\textwidth]{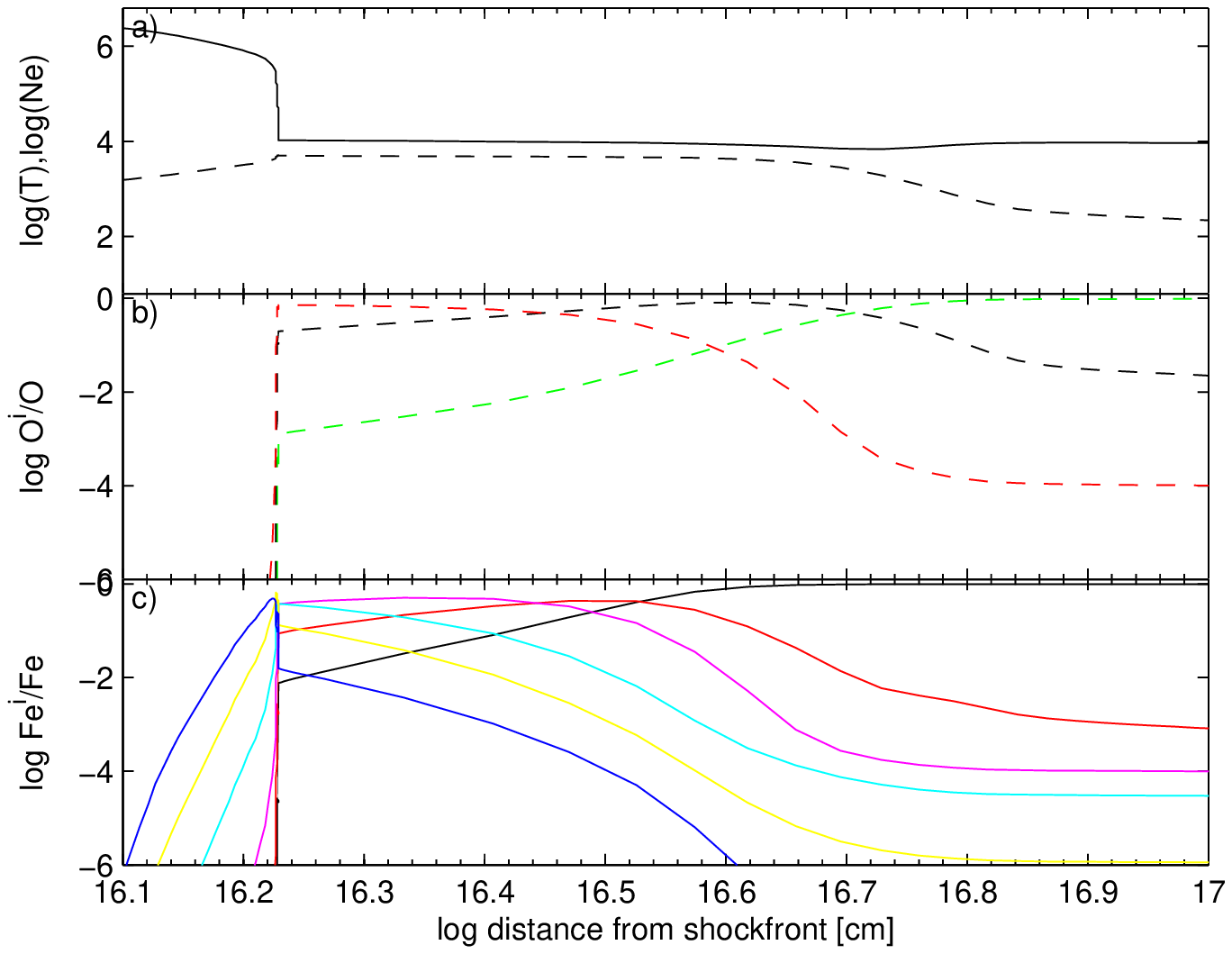}
\includegraphics[width=0.49\textwidth]{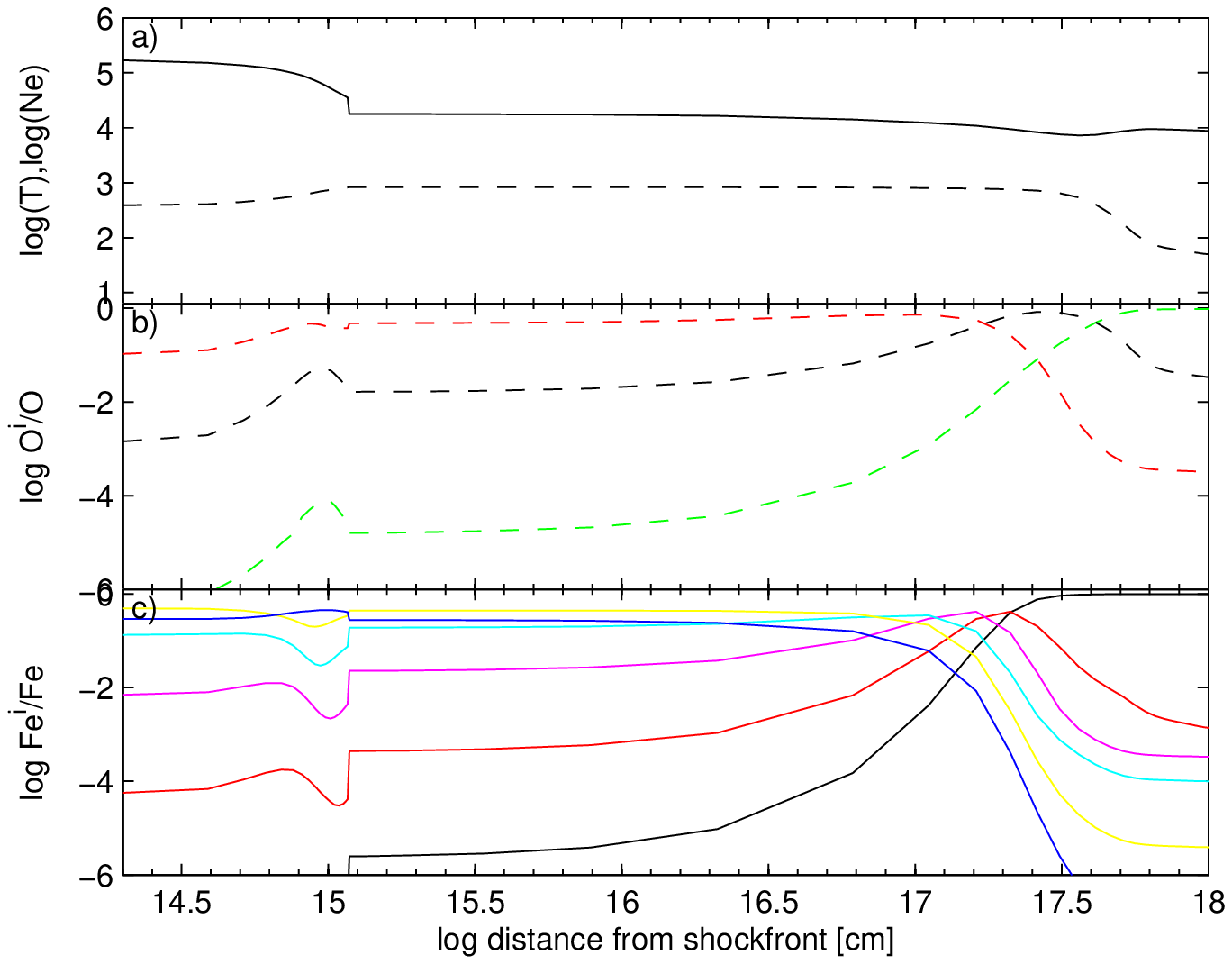}
\caption{pos. 140 (left), pos. 149 (right); top diagram: the electron temperature (solid line) and electron density (dashed line) downstream; middle diagram: the fractional abundances of O$^0$ (green), O$^+$ (black), O$^{2+}$ (red);
bottom diagram: the fractional abundances of Fe$^+$ (black), Fe$^{2+}$ (red), Fe$^{3+}$ (magenta), Fe$^{4+}$ (cyan), Fe$^{5+}$ (yellow), Fe$^{6+}$ (blue).}
\label{fig6}
\end{figure*}

In HII regions, on the other hand, the black body flux is rapidly reduced by radiation transfer throughout the gas and the temperature drop, so the [\ion{Fe}{II}] lines are weak.
In Fig.\,\ref{fig6} (top diagram) we illustrate the temperature and electron density profiles downstream in positions 140 (left) and 149 (right). 
We have chosen such positions since they have very different shock velocities and are reached by different flux intensities. Nevertheless, the spectra are sufficiently well reproduced by the models.
The $\rm O^0/O$, $\rm O^+/O$ and $\rm O^{2+}/O$  fractional abundances and the fractional abundances of significant Fe ions are described, respectively, in the middle and the bottom diagrams.

%%%%%%%%%%%%%%%%%%%%%%5

\section{The continuum SED}\label{sed}

\begin{figure}
\includegraphics[width=0.50\textwidth]{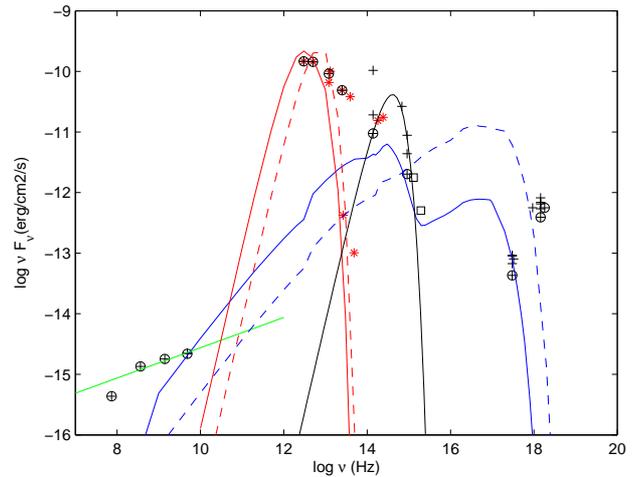}
\caption{Comparison of the continuum SED with the model results.
Black crosses: data from NED; black encircled crosses : the data from the NED believed to be dominated by the AGN;
 red asterisks: data from  Sargsyan (2011); black squares: data from GALEX archive;
green: synchrotron radiation; red: the dust reprocessed radiation flux
calculated by models 71 (solid line) and 107 (dashed line); blue: the bremsstrahlung
 for position 71 (solid line)  and for position 107 (dashed line).
Black line: the black body flux from the background stars.}
\label{fig9}
\end{figure}

We checked the models by comparing them with the observed spectral energy distribution (SED) of the continuum, shown in Fig.\,\ref{fig9}. The data (represented by black crosses) were extracted from the NASA/IPAC 
extragalactic database (NED) and are detailed in Table 4. They are taken with different apertures, and therefore they cover different regions of the galaxy. However, for apertures of few arcsec, the emission is very likely dominated by the AGN. 
We expected the same, in case of the hard X-ray emission, even if the aperture is large. 
Some problems can arise in the soft X-ray and infrared domains, where the fluxes are emitted by a large fraction 
of the galaxy and additional contributions, e.g. supernova remnants and dust, could not be negligible.

Most of the AGN emission in the optical-UV range is hidden by the molecular dusty torus, 
which makes visible the black-body flux from the old stellar population, here reproduced by black-body radiation at $T=5000$ K. The archival GALEX data in the UV follow the black-body old star radiation.  
Nevertheless, the power-law flux from the AGN and the shocks heat the gas which contributes to the SED 
by its reprocessed bremsstrahlung emission. 

Fig.\,\ref{fig9} shows that the data cover the X-ray domain, indicating the presence of high temperature regions. 
The bremsstrahlung at high frequencies depends on the  temperature of the gas in the downstream region behind  
the shock front. The higher the shock velocity, the higher the temperature of the gas downstream. 
Thus, the bremsstrahlung peaks at high frequencies for models calculated with high \Vs. 
To reproduce soft X-ray emission (0.5--2 keV) we have adopted the result of the model calculated with a 
shock velocity \Vs= 550 \kms, corresponding to position 71. 
NGC 7212 is a Compton thick source (Bennert et al. 2006) with reflected hard X-rays (2--10 keV) 
coming directly from the AGN. Yet, hard X-rays could be contributed by shocks corresponding 
to \Vs = 800 \kms~ as found in position 107. Models with lower velocities do not contribute to the continuum SED 
at high frequencies.

The IR shows a  composite black-body radiation from dust grains at different temperatures.
Dust grains are heated radiatively by the flux from the active center and collisionally by the gas. 
The  data in the near IR domain are reproduced by dust reprocessed radiation downstream of the highest 
velocity models adopting a dust-to-gas ratio $d/g \sim 2\times10^{-4}$ and $4\times10^{-5}$ (by mass) 
in positions 71 and 107, respectively. We stress, for comparison, that in 
the ISM $d/g  \sim 4 \times 10^{-3}$--$4\times10^{-4}$ (Contini \& Contini \cite{Con07}).

The data in the radio range show the characteristic power-law of the synchrotron
radiation created by the Fermi mechanism at the shock front.
The free-free and free-bound continuum radiation flux  are calculated consistently with the line emission fluxes for each model. The bremsstrahlung emitted from the gas has the same frequency distribution in the IR-radio range for different models  (see Contini et al. 2004) when free-free self-absorption is negligible (at rather low densities).

\begin{table}
\begin{center}
\caption{List of photometric data extracted from NED.}
\label{tab44}
\begin{tabular}{llllll} \\ \hline \hline
band& $\log\nu$ & $\log\nu$ $F\nu$& Instrument & Aperture  &  Ref. \\ \hline
5--10 keV   &18.257  &-12.251\tablefootmark{a}    & XMM & 30\arcsec &  1 \\
2--10 keV   &18.161  &-12.161    & XMM & 30\arcsec & 1 \\
2--10 keV   &18.161  &-12.408    & Chandra  & 1.5\arcsec  &2  \\
2--10 keV   &18.161  &-12.267\tablefootmark{a}    & Chandra & $3.5\arcsec\times2$\arcsec& 2 \\
2--10 keV   &18.161  &-12.187      & Chandra & 1\arcmin & 2 \\
2--10 keV   &18.161  &-12.086      & ASCA   & 1\arcmin & 3 \\
0.7--7 keV  &17.968  &-12.251      &   ASCA & 1\arcmin & 3 \\
0.7--2 keV  &17.513  &-13.096      &    ASCA & 1\arcmin & 3 \\
0.5--2 keV  &17.480  &-13.045      &    XMM  & 30\arcsec & 1 \\
0.5--2 kev  &17.480  &-13.173\tablefootmark{a}   & Chandra &$3.5\arcsec\times2$\arcsec& 2 \\
FUV         &15.29   &-12.299\tablefootmark{a}   & GALEX & 4.2\arcsec& \\
NUV         &15.112  &-11.753\tablefootmark{a}  & GALEX & 5.3\arcsec& \\
F330W       &14.958  &-11.694\tablefootmark{a}  &  HST/ACS & 0.3\arcsec & 4 \\
F330W       &14.958  &-11.357 &  HST/ACS & 1\arcsec & 4 \\
F330W       &14.958  &-11.054 &  HST/ACS & 2.54\arcsec& 4 \\
m$\rm_p$    &14.833  &-10.591 &     & total mag  & 5 \\
$K_{\rm s}$  &14.139  &-10.71  & 2MASS & Bulge mag & 6  \\
$K_{\rm s}$  &14.139  &-11.024\tablefootmark{a}  & 2MASS & AGN mag  &6  \\
12 $\mu$m   &13.397  &-10.311\tablefootmark{a}  & IRAS & 30\arcsec & 7 \\
25 $\mu$m   &13.079  &-10.035\tablefootmark{a}   & IRAS & 1\arcmin &7  \\
60 $\mu$m   &12.698  &-9.841\tablefootmark{a}  & IRAS & 1\arcmin  &7 \\
100 $\mu$m  &12.477  &-9.832\tablefootmark{a}  & IRAS & 2\arcmin  &7 \\
4.85 GHz    &9.685   &-14.623\tablefootmark{a}     & NRAO  & 3.5\arcmin    & 8 \\
4.85 GHz    &9.685   &-14.651    & NRAO & $3.7\arcmin\times3.3$\arcmin & 9 \\
1.40 GHz    &9.146   &-14.747\tablefootmark{a}    & NRAO & 700\arcsec& 10  \\
365 MHz     &8.562   &-14.869\tablefootmark{a}   & UTRAO  & 30\arcsec & 11 \\
73.8 MHz    &7.868   &-15.361\tablefootmark{a}   & VLA   & 80\arcsec & 12 \\ \hline
\end{tabular}
\tablebib{
(1) Guainazzi et al. (2005); (2) Levenson et al. (2006); (3) Ueda et al. (2001); (4) Mu{\~n}oz Mar{\'{\i}}n et al. (2007);
(5) Zwicky et al. (1965); (6) Peng et al. (2006); (7) Moshir et al. (1990); (8) Becker et al. (1991); (9) Gregory \& Condon (1991);
(10)  White \& Becker (1992); (11) Douglas et al. (1996); (12) Cohen et al. (2007).}
\tablefoot{
\tablefoottext{a}{data selected to be representative of the AGN emission.}
}
\end{center}
\end{table}

\section{Discussion}\label{disc}

\subsection{Comparison with other observations}

\begin{table}
\begin{center}
\caption{Comparison between our model and the Bennert et al. (2006) nuclear optical spectrum.}
\label{tab3}
\begin{tabular}{ l l l l l } \\ \hline \hline
\ line ratios to H$\beta$    & obs. & calc. \\ \hline
\ [\ion{O}{II}] 3727+ & 2.63 & 3.1 \\
\ [\ion{Ne}{III}] 3869+ & 1.7 & 1.3\\
\ [\ion{O}{III}] 4363   & 0.23 & 0.1\\
\ \ion{He}{II} 4686     & 0.24 & 0.2\\
\ [\ion{O}{III}] 5007+  & 12.06 & 12.02\\
\ [\ion{Fe}{VII}] 6087  & 0.04  & 0.02\\
\ [\ion{O}{I}] 6300     & 0.56  & 0.7 \\
\ [\ion{Fe}{X}] 6375    & 0.03  & 0.001\\
\ H$\alpha$           & 2.87  & 2.9\\
\ [\ion{N}{II}] 6583    & 1.99  & 2.9\\
\  [\ion{S}{II}] 6716   & 0.56  & 0.3 \\
\  [\ion{S}{II}] 6731   & 0.63  & 0.45\\ \hline
\ \n0 (\cm3)    & -       & 190  \\
\  \Vs (\kms)   & -     & 370  \\
\  \B0 ( gauss) & -     & $10^{-4}$ \\
\  $\log F$\tablefootmark{1}      & -     & 10.85  \\
\  $D$ (cm)     & -     & $2\times10^{17}$  \\
\  \Hb~ (\erg)   & -     & 0.23   \\
\  $n$ (\cm3)     &  -    & 6330   \\ \hline
\end{tabular}
\tablefoot{
\tablefoottext{1}{$F$ is in photons cm$^{-2}$ s$^{-1}$ eV$^{-1}$ at the Lyman limit.}
}
\end{center}
\end{table}

\begin{table}
\begin{center}
\caption{Comparison between our model and the Ramos Almeida et al. (\cite{R09}) IR spectrum.}
\label{tab4}
\begin{tabular}{ l l l l l } \\ \hline \hline
\ line ratios to [S\,{\sc iii}] 0.9    & obs. & calc. \\ \hline
\  [\ion{C}{I}] 0.98   &  0.02    & 0.02\\
\ [\ion{S}{VIII}]0.99  & 0.012    & 0.01\\
\ [\ion{S}{II}] 1.03   & 0.054    & 0.045 \\
\ \ion{He}{I} 1.08    & 0.35     & 0.36\\
\ [\ion{S}{IX}] 1.25   & 0.0057    & 0.06\\
\ [\ion{S}{IX}] 1.43  &  0.011     & 0.012\\
 P$_{\alpha}$ 1.96&0.158   &0.2\\
\ [\ion{S}{XI}] 1.9   & 0.003      & 0.01\\
\ [\ion{Si}{VI}] 1.96 & 0.028      &0.045\\ \hline
\ \n0 (\cm3)  &    -       & 200\\
\ \Vs (\kms) &     -       & 500\\
\ \B0 (gauss) &    -       & $10^{-4}$\\
\ $\log F$\tablefootmark{1}   &     -        & 10.\\
\ $D$  (cm) &-             & $3\times10^{15}$\\
\ H$\beta$  (\erg) &    -       & $6.3\times10^{-3}$\\
\ \Hb/[\ion{S}{III}]  &   -        & 1.96\\ 
\ $n$ (\cm3)          &   -     & 1300 \\ \hline  
\end{tabular}
\tablefoot{
\tablefoottext{1}{$F$ is in photons cm$^{-2}$ s$^{-1}$ eV$^{-1}$ at the Lyman limit.}
}
\end{center}
\end{table}

To confirm the ranges of the parameters obtained by reproducing Cracco et al. (\cite{Cr11}) data, we present in Table\,\ref{tab3} the results from the modelling of Bennert et al. (2006) nuclear optical spectrum, corresponding to an aperture of 1\arcsec.  
We find a shock velocity of 370 \kms~, slightly higher than that inferred from the average observed 
FWHM (250 \kms) provided by these authors.
Near-IR spectroscopic data  are reported by Ramos Almeida et al. (\cite{R09}) 
for five Seyfert galaxies showing jet-gas interaction, including NGC 7212. Its IR nuclear spectrum (1.5\arcsec-aperture) is presented in Table\,\ref{tab4}.
The line ratios are well reproduced with a model very similar to those which reproduce Cracco et al. (\cite{Cr11}) optical spectra. 
The contemporary fit of neutral  and coronal lines in different ranges requires composite models (see, e.g. Rodr{\'{\i}}guez-Ard\'{i}la et al. \cite{R05}).

The shock velocities adopted to fit the Bennert et al. and Ramos Almeida et al. nuclear spectra, were compared to the values obtained by the present modelling of the Cracco et al. spectra in a region of $3\arcsec\times3\arcsec$ centred on the nucleus. This region was chosen to take into account possible effects of slit misalignenment with respect to the active nucleus.
The values are similar: 370, 500, and 110--550 \kms, respectively.  
The densities are: 190, 200, and 70--120 \cm3, respectively. 
Recall that the  densities in the present modelling are preshock, corresponding to densities downstream higher by a factor of 4-10, 
depending on the shock velocity. The fluxes are also similar: $\log F= 10.85$, 10, and 10--10.8, respectively.   
 A difference of factor $\sim$100 is apparent in the geometrical thickness of the clouds.
 Those referring to IR spectra by Ramos Almeida et al. are approximately $3\times10^{15}$ cm, while the other 
are about $2\times10^{17}$ cm. 
It seems that  observable coronal lines  are emitted from very thin clouds, in order to prevent recombination of the Si, S, etc.

\subsection{The physical  picture of the  NLR}

The 2D maps in Fig.\,\ref{fig7} depict the distribution of the shock velocity \Vs (left panel), of the preshock density \n0 (central panel) and of the flux from the active nucleus $F$ (right panel). They refer to the shock regime. High velocity shocks (600--800 \kms) delineate the region where matter external to the ENLR collides with the ENLR gas. The highest velocities are found in the eastern regions, where the maps show low \n0 values. Matter is faster throughout the less dense gas regions, as expected from the mass conservation equation at the shock front.  This confirms the presence of shocks. 
The preshock densities demonstrate a narrow maximum  near the  eastern edge, which most probably obstruct the spreading of the high velocity shock. 
The histograms in Fig.\,\ref{fig7} illustrate the distributions of the values, \Vs, \n0 and $\log F$.

\begin{figure}
\includegraphics[width=0.50\textwidth]{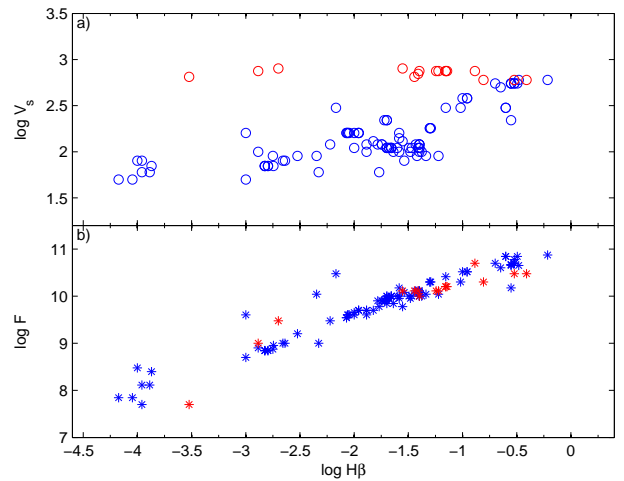}
\caption{Shock velocities vs. H$\beta$ calculated flux (top panel); fluxes from the active centre vs. H$\beta$ calculated flux (bottom panel). Red: models calculated with high \Vs ($>700$ km s$^{-1}$); blue: the other models.}
\label{fig8}
\end{figure}

Explaining the ionization of the perturbed gas in radio galaxies, Tadhunter et al. (\cite{T00}) discuss why the perturbed gas varies significantly in surface luminosity from object to object, while the quiescent gas does not. 
The correlation between $F$ and H$\beta$ absolute flux calculated at the nebula is shown in Fig.\,\ref{fig8}, as well as the correlation of the shock velocity with H$\beta$. H$\beta$ is chosen so as to represent the surface luminosity at the emitting nebula.
Extrapolating from many different objects to different regions in the same galaxy, Fig.\,\ref{fig8} shows that the luminosity of each region in NGC 7212 is strongly linked with the AGN flux reaching it.
The H$\beta$ line  emission is strong  from regions at temperatures $T\geq$10$^4$ K. 
The correlation with \Vs is less straightforward. 
Indeed, at temperatures higher than 10$^4$ K collisional parameters are much more important than the radiation ones, however, a higher \Vs generates higher densities downstream
and consequently a higher cooling rate, which results in a reduced H recombination zone.
So we can hardly recognize the shocks resulting from collisions by means of the object luminosity, but 
they can be found out by analysing line ratios and profiles.

The shock velocities  calculated in the observed positions throughout the NGC 7212 ENLR, roughly show the regions where collision of matter from the ENLR with matter from outside the cone occurred. We have found that the velocities 
are lower than 200 \kms, and the preshock densities are relatively low ($\sim100$ \cm3) in all the positions representing the ENLR matter uncontaminated by high density debris from the BLR. These velocities refer to a smooth regime. 
Tadhunter et al. (\cite{T00}) use the term ``quiescent'' to mean ``in a quasi-stable dynamical configuration in a galaxy'', 
that is, within a factor of 2 of the rotational velocity. The  high velocities, on the other hand, refer to shocks.

\begin{figure}
\begin{center}
\includegraphics[width=7cm]{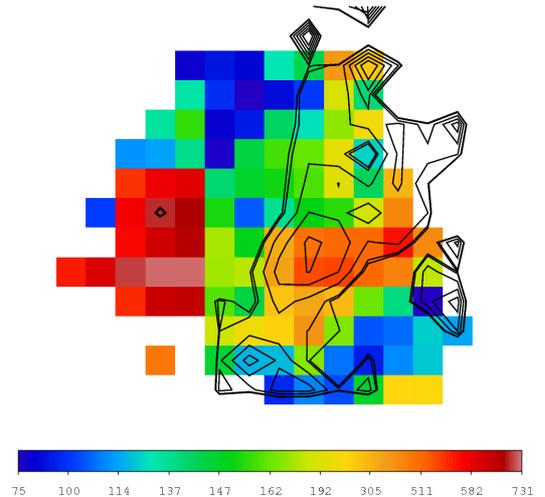}
\caption{2D map of the shock velocity in km s$^{-1}$ with the contour map of the photoionization cone ([\ion{O}{III}]/H$\beta$) overlayed. Spatial scale is 1 arcsec px$^{-1}$.}
\label{fig10}
\end{center}
\end{figure}

We confirm that shocks are strong  beyond the edges of the cone overlying in Fig.\,\ref{fig10} the shock velocities 
upon the contour map of the photoionization cone. Our calculations reveal that high velocity shocks are 
located at the edge of the ionization cone and beyond it, in agreement with the large FWHM of the line profiles, 500--800 \kms.  
Interestingly, patches of highly ionized clumps were observed also outside the ENLR of \object{NGC 3393} 
(Cooke et al. 2000) where merging is confirmed by the binary black hole system observed in the nucleus (Fabbiano et al. 2011).
Tables \ref{tab1} and \ref{tab2} show that where the highest \Vs are found (e.g. in position 108), the contribution of spectra calculated with relatively low velocity - most likely emitted from the ENLR  or even from the ISM of the host galaxy - 
cannot be excluded. The relative contribution of the two spectra to each of the observed one is revealed by the complex  
FWHM of the line profiles corresponding to the different ions, in agreement with the multiple kinematics 
found by Cracco et al. (\cite{Cr11}).
For instance, the [\ion{S}{II}]6716+ lines have a strong contribution from low velocity gas.

\subsection{The distribution of metallicities}
We have  calculated the relative abundances to H of the heavy elements throughout the galaxy, particularly those 
identified with metallicity (e.g. oxygen in the present case), because metallicity is sensitive  
not only to star formation, but is also affected by the presence of infalls and outflows, i.e. by the 
interaction between the merging galaxy and the intergalactic medium (Sommariva et al. 2011).
By adopting models which account for both shock and photoionization, we found that O/H  is solar 
in almost all positions corresponding to ``quiet'' gas, if we employ the solar 
relative abundance, O/H=$6.6\times10^{-4}$ reported by Allen (1976). 
However, O/H becomes about 0.77 solar, when we refer to the solar abundance O/H=$8.5\times10^{-4}$ calculated by Anders \& Grevesse (1989) and Feldman (1992).
Sub-solar O/H values as low as $2.6\times 10^{-4}$ (Fig.\,\ref{fig5}) were found only in some positions 
far from the centre. On the other hand, Cracco et al. (\cite{Cr11}) found sub-solar O/H throughout all 
observed positions using pure photoionization models. 
In fact, the SUMA modelling is based on both the [\ion{O}{III}] and [\ion{O}{II}] lines.
The [\ion{O}{III}] line intensity depends strongly on the flux from the active center, 
while the [\ion{O}{II}] line is affected also by the shock velocity. 
Moreover, we have improved the modelling procedure, accounting for the complete spectrum in each position.

In agreement with Cracco et al. (\cite{Cr11}) we have found that nitrogen is depleted by factors 
lower than 2 in large regions of the cone corresponding to relatively low shock velocities. 
In the regions characterised by high shock velocities (Fig.\,\ref{fig7}) N/H is  roughly solar. S/H shows 
some depletion in the same external south-east regions. N/H is generally high in AGN 
(e.g. Contini et al. \cite{Con02a}, Ciroi et al. \cite{C03}).
A low N/H may indicate inclusion of N into molecules in the ISM and/or acquisition of external gas in the merging process.
In Sect.\,\ref{sed} we have found that the dust-to-gas ratio is relatively low in NGC 7212, showing 
that dust grains and molecules can be destroyed by the itense radiation from the AGN and by relative strong shocks. We conclude that sub-solar N/H is a further indication of external gas acquisition.

It is generally known that bursts of star formation are triggered by galaxy-galaxy mergers (Somerville et al. 2001). 
As mentioned in Sect.\,\ref{intro}, Mu{\~n}oz Mar{\'{\i}}n et al. (\cite{M07}) claim that interacting galaxies can show UV-light coming from knots produced by star clusters. They identify some bright clumps at the south of NGC 7212 as  star-forming regions.
However, the contribution of the starbursts to the [\ion{O}{III}] 5007 emission line flux is less than 20\%, lower than that of the AGN, in agreement with the results for NGC 3393 (Contini 2012), another Seyfert galaxy  belonging to a merger system.
\begin{figure}
\begin{center}
\includegraphics[width=0.45\textwidth]{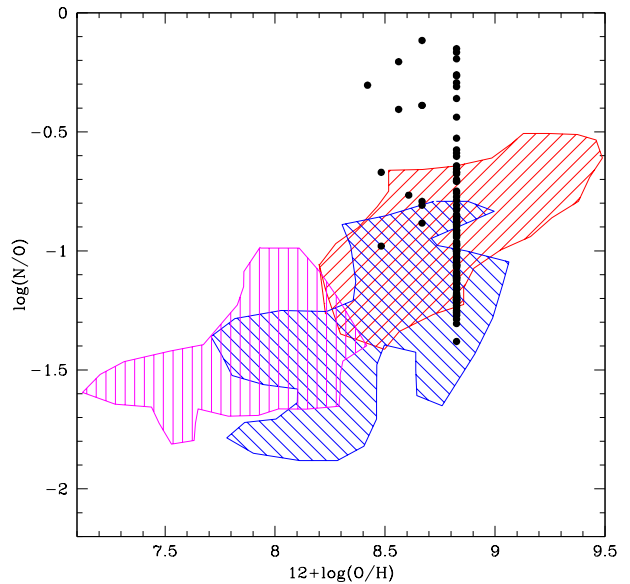}
\caption{N/O vs. O/H diagram adapted from Mouhcine \& Contini (2002), (see their fig.\,1). 
The blue area represents a sample of UV-selected galaxies, the red area includes starburst nucleus galaxies (SBNGs) selected in the optical or in the infrared, and the magenta area represents H\,{\sc ii} galaxies. The results of our model calculations are overplotted (black circles).}
\label{fig11}
\end{center}
\end{figure}
The evidence for star formation motivates us to compare the N/O relative abundance as a function of metallicity calculated throughout the galaxy with the N/O vs. O/H diagram presented by Mouhcine \& Contini (\cite{M02}) in Fig.\,\ref{fig11}. 
The calculated N/O and O/H ratios lie in the zone covered  by  UV selected galaxies and massive starbursts in the diagnostic diagram of Mouhcine \& Contini (\cite{M02}, fig.1, black squares and circles, respectively).
Figure \ref{fig11} shows a strong variation of N/O throughout NGC 7212, indicating a time delay between the release of oxygen and that of nitrogen (Contini et al. \cite{con02b}) in the regions characterised by O/H solar abundance.
Positions showing an N/O overabundance ($\log({\rm N/O}) \geq -0.86$) (see also Fig.\,\ref{fig5}) may indicate mixing with external matter as a consequence of a minor merger.

\section{Conclusions}\label{result} 

In the previous sections we presented an analysis of the  line  spectra observed by Cracco et al. (\cite{Cr11}) in the different positions throughout the ENLR of NGC 7212, adopting calculation models which account  
for the coupled effect of shocks and photoionization. The results yield new reliable estimates about the intensity and distribution of the flux from the AGN, the velocity field, the preshock density, the relative abundances, etc.

In Seyfert galaxies the radiation from the active nucleus is the main photoionization mechanism. 
The AGN  in NGC 7212 is characterised by a flux  intensity lower than 10$^{11}$ photons cm$^{-2}$ s$^{-1}$ eV$^{-1}$, at the Lyman limit, similar to  that of  low luminosity AGN (e.g. Contini \cite{Con04}), rather than to Seyfert 2 (e.g. Contini et al. \cite{Con02a}, Ciroi et al. 2003). 
The AGN flux peaks in correspondence with the maximum starlight flux. In the ENLR of NGC 7212, as in most Seyfert galaxies, the single lines show complex profiles. This indicates that the hydrodynamical field is complex. Shock waves created by winds and jets from the stars and/or by collision of matter are generally present.
Characteristic of shocked clouds is the profile of the density downstream, where the gas cools due to recombination processes. Compression speeds up recombination, enhancing the intensity of lines corresponding to low ionization levels. 
The spectra emitted by shocked gas are therefore different  from those emitted from  clouds heated and ionized by radiation from an external source.
The relative importance of photoionization and shocks in NGC 7212 throughout most of the 256 spatial elements simultaneously observed by MPFS, results from the consistent modelling of line and continuum spectra.
Each spectrum stems from a complex grid of models which account for  the parameter in various ranges.
The parameters are constrained by the \ion{He}{II}/H$\beta$ line ratios ($F$), by [\ion{S}{II}]6716/6731 (\n0), by [\ion{O}{III}]:[\ion{O}{II}]:[\ion{O}{I}] (\Vs and $F$), etc. The shock velocity is also constrained by the continuum SED.

In conclusion, the  interaction of  galaxy-galaxy or of a galaxy with matter from the interstellar or intracluster medium can  reveal  shocks by line ratios which cannot be readily explained by pure photoionization models (e.g. Villar-Mart{\'{\i}}n et al. 1999), by high electron temperatures  leading to X-ray emission and by radio synchrotron radiation.
Radiation and collisional processes are strongly intermingled in  ionizing the  gas (e.g. Contini 1997). 
The degree to which shocks contribute to the gas ionization  in the different regions of NGC 7212 can be found only by consistent modelling of line and continuum spectra. 
Our results agree with Tadhunter et al. (2000), who claim that shocks contribute significantly to the ionization of the ENLR of Seyfert galaxies, even if photoionization  contributes substantially in  most of AGNs (Robinson et al. 1987, Villar-Mart{\'{\i}}n et al. 1997).

\section*{Aknowledgements}
This research has made use of the NASA Astrophysics Data System (ADS) and the NED, operated
by the Jet Propulsion Laboratory, California Institute of Technology, under contract with NASA.
We are grateful to M. Mouhcine for allowing to reproduce  Fig.\,\ref{fig11}.

\end{document}